\documentclass{SCAEOL}%SCAEOL for online version; SCAE for publication version; SCAES for the paper dedicated to somebody.
\numberwithin{equation}{section}
\usepackage{tikz}
\usetikzlibrary{shapes,arrows}
\usetikzlibrary{arrows,shapes,chains}
\usepackage{bm}
\usepackage{graphicx} %use graph format
\usepackage{subfigure}
\usepackage{epstopdf}
\usepackage{tabularx}
\usepackage{threeparttable}

\begin{document}

%Basic Information
\Year{2020} %
\Month{Feburary}
\Vol{56} %
\No{1} %
\BeginPage{1} %
\EndPage{XX} %
%\AuthorMark{First L N {\it et al.}}
%\ReceivedDay{November 17, 2012}
%\AcceptedDay{January 22, 2013}
%\PublishedOnlineDay{; published online January 22, 2013}
%\DOI{10.1007/s11425-000-0000-0} % The author doesn't need fill in it.

% \title[short text for running head]{full title}{comments for title}
\title{Modeling the Control of COVID-19: Impact of Policy Interventions and Meteorological Factors}{}

% \author[]{Full name}{footnote}
% Remark:  One \author for one author

\author[1, 2]{JIA Jiwei}{}
\author[1]{DING Jian}{}
\author[3]{LIU Siyu}{Corresponding author}
\author[1]{LIAO Guidong}{}
\author[4]{\\[2mm]LI Jingzhi}{}
\author[5]{DUAN Ben}{}
\author[6]{WANG Guoqing}{}
\author[1, 2]{ZHANG Ran}{}

\address[{\rm1}]{School of Mathematics, Jilin University, Changchun {\rm 130012}, China;}
\address[{\rm2}]{Interdisciplinary Center of Jilin Province for Applied Mathematics, Changchun {\rm 130012}, China;}
\address[{\rm3}]{School of Public Health, Jilin University, Changchun {\rm 130021}, China;}
\address[{\rm4}]{Department of Mathematics, Southern University of Science and Technology, Shenzhen {\rm 518055}, China;}
\address[{\rm5}]{School of Mathematical Sciences, Dalian University of Technology, Dalian  {\rm 116024}, China;}
\address[{\rm6}]{School of Basic Medical Science, Jilin University, Changchun {\rm 130021}, China.}

%\address[{\rm3}]{Department of Mathematics, University3, City3 {\rm100003}, Country3;}
%\address[{\rm4}]{College of Science, University4, City4 {\rm100004}, Country4}
\Emails{ jiajiwei@jlu.edu.cn, dingjian17@mails.jlu.edu.cn,
liusiyu@jlu.edu.cn, liaogd18@mails.jlu.edu.cn, lijz@sustech.edu.cn, bduan@dlut.edu.cn, qing@jlu.edu.cn, zhangran@jlu.edu.cn}\maketitle

%     Abstract is required.

 {\begin{center}
\parbox{14.5cm}{\begin{abstract}
%In this paper, we propose a dynamical model for describing the transmission of COVID-19, which is outbreaking in China and spreading around the world. In addition to symptomatic infection, asymptomatic infection of COVID-19 is also reported. Based on this fact, we divided the population into seven categories for simulation. At the same time, in order to consider the effect of population isolation strategy, we introduce the isolation rate parameter. We employ a Least-Square procedure and the official published data to estimate the parameters and get the results of most provinces in China. We focus on the relationship between the spread rate of virus and weather factors. Numerical tests show that the proposed model can predict the transmission of COVID-19 well until now and the isolation strategy play an important role in controlling the virus transmission. The correlation between the transmission rate and the weather %condition shows that......

In this paper, we propose a dynamical model to describe the transmission of COVID-19, which is spreading in China and many other countries. To avoid a larger outbreak in the worldwide, Chinese government carried out a series of strong strategies to prevent the situation from deteriorating. Home quarantine is the most important one to prevent the spread of COVID-19. In order to estimate the effect of population quarantine, we divide the population into seven categories for simulation. Based on a Least-Squares procedure and officially published data, the estimation of parameters for the proposed model is given. Numerical simulations show that the proposed model can describe the transmission of COVID-19 accurately, the corresponding prediction of the trend of the disease is given. The home quarantine strategy plays an important role in controlling the disease spread and speeding up the decline of COVID-19. The control reproduction number of most provinces in China are analyzed and discussed adequately. We should pay attention to that, though the epidemic is in decline in China, the disease still has high risk of human-to-human transmission continuously. Once the control strategy is removed, COVID-19 may become a normal epidemic disease just like flu. Further control for the disease is still necessary, we focus on the relationship between the spread rate of the virus and the meteorological conditions. A comprehensive meteorological index is introduced to represent the impact of meteorological factors on both high and low migration groups. As the progress on the new vaccine, we design detail vaccination strategies for COVID-19 in different control phases and show the effectiveness of efficient vaccination. Once the vaccine comes into use, the numerical simulation provide a promptly prospective research.
\vspace{-3mm}
\end{abstract}}\end{center}}

%  Keyword is required.

%  \subjclass is required.

%%%%%%%%%%%%%%%%%%%%%%%%%%%%%%%%%%%%%%%%%%%%%%%%%%%%%%%%%%%%
\renewcommand{\baselinestretch}{1.2}
%\begin{center} \renewcommand{\arraystretch}{1.5}
%{\begin{tabular}{lp{0.8\textwidth}} \hline \scriptsize
%{\bf Citation:}\!\!\!\!&\scriptsize First1 L N, First2 L N, First3 L N.  SCIENCE CHINA Mathematics  journal sample. Sci China Math, 2013, 56, doi: 10.1007/s11425-000-0000-0\vspace{1mm}
%\\
%\hline
%\end{tabular}}\end{center}

%%%%%%%%%%%%%%%%%%%%%%%%%%%%%%%%%%%%%%%%%%%%%%%%%%%%%%%%%%%%
%% Text of article.
%%%%%%%%%%%%%%%%%%%%%%%%%%%%%%%%%%%%%%%%%%%%%%%%%%%%%%%%%%%%
%    Section headings
\baselineskip 11pt\parindent=10.8pt  \wuhao
\section{Introduction}

 Coronavirus is a kind of virus that causes infectious diseases in mammals and birds. Usually, the viruses cause respiratory infections among people and the first identification is in the 1960s \cite{weiji}. The main transmission of coronavirus is like other viruses: through sneezing, coughing, coming into contact with the infected people, or touching daily-used items \cite{CY7}. On December 26, 2019, the first detected novel coronary pneumonia case in China was reported as an unknown etiology pneumonia in Wuhan. Evidences pointing to the human-to-human transmission in hospitals and families are found in retrospective studies \cite{Leung2020, Phan2020, Chan2020, Rothe2020}. It takes a few days to arouse people's attention and Chinese Center for Disease Control and Prevention (China CDC) isolated the first strain of the causative virus (2019-nCoV) successfully on January 7, 2020. With Chinese New Year migration, the large epidemics occur in China and spread to many countries rapidly. The World Health Organization (WHO) declares that the pneumonia outbreak caused by 2019-nCoV as a public health emergency of international concern on January 31, 2020\cite{WHOweb}. WHO has announced the official name of the disease caused by a novel coronavirus as COVID-19 \cite{WHOweb} on February 11, 2020, it is the seventh member of the family of coronaviruses that infect humans \cite{Zhu2020}. As of February 19, 2020, there have been $74576$ confirmed COVID-19 cases, including $2118$ deaths in China, there are about 30 countries around the world reported over $1000$ diagnosed cases.

Among the seven known human coronavirus, four of them are common pathogens of human influenza. SARS-CoV, MERS-CoV and 2019-nCoV will cause fatal respiratory diseases \cite{weiji}. Although people have identified and analyzed the coronavirus for a long time, knowledge of the coronavirus is quite limited and there are no vaccines or antiviral drugs to prevent or treat human coronavirus infections. Last major outbreak in China was SARS which is an acute respiratory infectious disease with high fatality rate caused by SARS-CoV in 2003. Chinese people managed the outbreak of SARS through multiple control and prevention measures effectively. Compared with SARS, the incubation period of COVID-19 is significant and rather long. Different mean incubation period of COVID-19 is reported, $5.2$ days \cite{LiQun2020} in the early stage, $3.0$ days \cite{Guan2020.02.06.20020974} and $4.75$ days \cite{Yang2020} in recent research. The patient with up to $24$ days of incubation period is reported in \cite{Guan2020.02.06.20020974}, even $38$ days is also reported in Enshi Tujia and Miao Autonomous Prefecture in Hubei Province. Notice that there is quite a lot of people infected with asymptomatic \cite{Rothe2020}  and the fatality is much lower than SARS-CoV and MERS-CoV \cite{Guan2020.02.06.20020974}.  Genetic studies of viruses show that the homology of SARS-CoV and 2019-nCoV is $85\%$ \cite{zhenliaofangan}. But 2019-nCoV binds ACE2 with higher affinity than SARS-CoV S \cite{Science-2020}.
By the end of January 29, 2020, the confirmed cases caused by COVID-19 are surpassed SARS. The uncertain of incubation period, asymptomatic cases and super transmissibility of the virus bring great difficulties in epidemic control.

%Dynamical modeling of COVID-19 transmission were performed by many scholars. A modified SEIR model with eight components is proposed by Tang \emph{et~al.} \cite{jcm9020462}, and leads a control reproduction number of $6.47$. The travel restriction effect are also discussed by means of the simulation result for Beijing, it shows that with travel restriction (no imported exposed individuals to Beijing), the number of infected individuals in seven days will decrease by $91.14\%$ in Beijing, compared with the scenario of no travel restriction. A Bats-Hosts-Reservoir-People network is developed in \cite{Chen2020} for simulating the potential transmission from the infection source (probable be bats) to the human infection, and the analytic form of basic reproduction number for a simplified Reservoir-People network is calculated. Ming \emph{et~al.}\cite{Ming2020} applied a modified SIR model to project the actual number of infected cases and the specific burdens on isolation wards and intensive care units (ICU). The estimates suggest that assuming $50\%$ diagnosis rate if no public health interventions were implemented, the actual number of infected cases could be much higher than the reported. Chen \emph{et~al.} \cite{chen2020time} proposed a novel dynamical system with time delay to present the incubation period behavior of COVID-19, the estimated parameters show that the prediction is highly dependent on the population size and public policy carrying on  by the local governments.

Extensive research for COVID-19 with multiple perspectives are reported.  Corresponding diagnostic criteria and medication guide are designed and updated timely. Rapid detection reagent, anti-splash device for respirator and other special apparatus come into service quickly. Chinese government started first-level response of emergency health and safety, the mechanism for joint prevention and control is established in a short time. As the disease evolves, it is not only a medical problem, COVID-19 becomes a problem concern with many society questions. Collect the massive data related to COVID-19 and analyze the inherent linkage are quite important for the next step control strategy. Epidemic dynamics and population ecology are the key methods to study infectious diseases, theoretically.

Dynamical modeling of COVID-19 transmission are performed by many scholars. A modified SEIR model with eight components is proposed by Tang \emph{et~al.} \cite{jcm9020462}, the control reproduction number under their estimation is $6.47$. The travel related risks of disease spreading is evaluated in \cite{Bogoch2020}, which indicates the potential of domestic and global outbreak\cite{Leung2020}. The travel restriction effect are also discussed by means of the simulation result for Beijing, it shows that with travel restriction (no imported exposed individuals to Beijing), the number of infected individuals in seven days will decrease by $91.14\%$, compared with the scenario of no travel restriction. A Bats-Hosts-Reservoir-People network is developed in \cite{Chen2020} for simulating the potential transmission from the infection source (probable be bats) to the human infection and the analytic form of basic reproduction number for a simplified Reservoir-People network is calculated. Ming \emph{et~al.}\cite{Ming2020} apply a modified SIR model to project the actual number of infected cases and the specific burdens on isolation wards and intensive care units (ICU). The estimation suggests that assuming $50\%$ diagnosis rate if no public health interventions are implemented, the actual number of infected cases could be much higher than the reported. Chen \emph{et~al.} \cite{chen2020time} propose a novel dynamical system with time delay to present the incubation period behavior of COVID-19, the estimated parameters show that the prediction is highly dependent on the population size and public policy carrying on  by the local governments.

In this paper, we propose an extended SEIR model to describe the transmission of COVID-19 in China. In order to avoid the situation worsening, Chinese government has pursued the most strictest isolation strategy for all people throughout the country to restrict the population mobility. Traffic control, limitation of travel, extension of the Chinese  New Year vacations, delay to return to work, rigorous management of communities and even the wartime management ensure the susceptible population stay at home. At this stage, the main aim of the disease control is called 'the prevention of disease spreading inside'.
The proposed model mainly focuses on the home quarantine strategy. It requires people to stay at home for at least $14$ days, which is aiming at reduce the chance of contact with the infected people as much as possible. The asymptomatic transmission and isolation treatment policy are also taking into account in this paper. We use the official data published by China CDC and employ a Least-Squares procedure to estimate the parameters. We simulate the cases of most provinces in China, calculate the control reproduction number ($\mathcal{R}^{0}_{c}$) for each selected provinces. To capture the variation of effective control reproduction number ($\mathcal{R}_{c}(t)   $), we divide the control process into three periods, calculate the average of $\mathcal{R}_{c}(t)   $ for each stage and compare the results inside and outside Hubei province. The numerical results show that the intervention and support strategy carried out by the government decreases $\mathcal{R}_{c}(t)   $ quickly.
Chinese government provide free medical care for the diagnostic COVID-19 patient. Furthermore, as of 24:00 on February 14, 217 medical teams (military medical teams are not included) including 25633 team members have arrived Hubei from all over China. It relieved the medical pressures of Hubei Province greatly.
We estimate the disease burden by means of accumulated medical resource needed in $90$ days, the peak value and peak time of the diagnostic population are also given in the numerical simulation part.

Study shows that SARS-CoV and 2019-nCoV share high homology with genes, there should be similarity between them. Back to 2003, there is no vaccine or specific medicine for SARS, however, it seems disappear overnight. Weather factor is considered as an important reason for the vanishing of SARS \cite{tianqi1,tianqi2}. It has pointed out that high temperature will weaken the activity of SARS-CoV \cite{weaken}.
Spring is coming in China, the meteorological factors, such as temperature and humidity, is changing. To clarify the relationship between the meteorological condition and the transmission of COVID-19, we obtain the weather data from China Meteorological Data Service Center (CMDC) and define a comprehensive meteorological index $MeI$.  The outbreak of COVID-19 coincides with Chinese New Year Migration, based on the migration level, we separate all the selected provinces into two groups. The correlation analysis shows that for the spread rate are significantly associated with $MeI$ in each group. The air index plays an important but interesting role, it is a positive factor in the low-migration group, but negative in the high-migration group. The high relative humidity helps to control the spread of COVID-19 for both groups.

Some progress on the vaccines for COVID-19 are reported, we change the quarantined compartment into the vaccinated in the proposed model to describe the effect of vaccination. We simulate three scenarios to represent the vaccination starting from the three control phases, say, the first $7$ days (prophase), the second $7$ days (metaphase) and the following $14$ days (anaphase). The results show that the efficient vaccination accelerates the diagnostic population to the peak and contributes to reduce $\mathcal{R}_{c}(t)   $ effectively.

The remaining parts of the paper are organized as follows. Based on the proposed dynamical model, we analyze the current control strategy in Section 2.  We discuss the impact of meteorological factors and vaccines in Section 3. Finally, we present some concluding remarks in Section 4.

\section{Analysis for current control strategy}
\subsection{Model formulation}
\begin{figure}[h!]
\centering
\tikzstyle{startstop} = [rectangle,rounded corners, minimum width=1cm,minimum height=1cm,text centered, draw=black,fill=red!30]
\tikzstyle{style_S} = [rectangle,rounded corners, minimum width=1cm,minimum height=1cm,text centered, draw=black,fill=red!30]
\tikzstyle{style_Q} = [rectangle,rounded corners, minimum width=1cm,minimum height=1cm,text centered, draw=black,fill=red!30]
\tikzstyle{style_E} = [rectangle,rounded corners, minimum width=1cm,minimum height=1cm,text centered, draw=black,fill=red!30]
\tikzstyle{style_A} = [rectangle,rounded corners, minimum width=1cm,minimum height=1cm,text centered, draw=black,fill=red!30]
\tikzstyle{style_I} = [rectangle,rounded corners, minimum width=1cm,minimum height=1cm,text centered, draw=black,fill=red!30]
\tikzstyle{style_D} = [rectangle,rounded corners, minimum width=1cm,minimum height=1cm,text centered, draw=black,fill=red!30]
\tikzstyle{style_R} = [rectangle,rounded corners, minimum width=1cm,minimum height=1cm,text centered, draw=black,fill=red!30]
\tikzstyle{arrow} = [ultra thick,->,>=stealth]
\begin{tikzpicture}[node distance=2cm]
\node(S)[style_S]{$\bm S$};
\node(Q)[style_Q,below of=S,yshift=-1.cm]{$\bm Q$};
\node(E)[style_E,right of =S,xshift=1.cm]{$\bm E$};
\node(A)[style_A,right of =E,xshift=1.cm]{$\bm A$};
\node(I)[style_I,below of=A,yshift=-1.cm]{$\bm I$};
\node(D)[style_D,right of =A,xshift=1.cm]{$\bm D$};
\node(R)[style_R,below of=D,yshift=-1.cm]{$\bm R$};
\draw [arrow] (S.-110) -- (Q.110) node[midway,left] {$\bm p$};
\draw [arrow] (Q.70) -- (S.-70) node[midway, right] {$\bm\lambda$};
\draw [arrow] (S) -- (E) node[midway,above] {$\bm\beta$};
\draw [arrow] (E) -- (A)node[midway,above] {$\bm{\sigma(1-\rho)}$};
\draw [arrow] (E) -- (I)node[midway,left,xshift=-0.3mm] {$\bm{\sigma\rho}$};
\draw [arrow] (A) -- (D)node[midway,above] {$\bm{\epsilon_{A}}$};
\draw [arrow] (A) -- (R)node[midway,left,yshift=0.6cm,xshift=0.1cm] {$\bm{\gamma_{A}}$};
\draw [arrow] (I) -- (D)node[midway,right,yshift=-0.6cm,xshift=-0.55cm] {$\bm{\epsilon_{I}}$};
\draw [arrow] (I) -- (R)node[midway,above] {$\bm{\gamma_{I}}$};
\draw [arrow] (D) -- (R)node[midway,right] {$\bm{\gamma_{D}}$};
\draw [arrow] (6,-3.5) -- (6,-4)node[midway,right] {$\bm{d_{I}}$};
\draw [arrow] (9,0.5) -- (9,1)node[midway,right] {$\bm{d_{D}}$};
\end{tikzpicture}
\caption{Flow diagram of the compartmental model of COVID-19 in China}\label{fig1_illustration}
\end{figure}
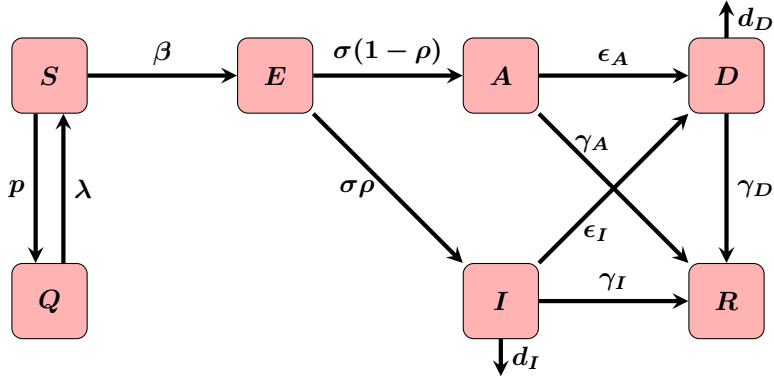

Based on the epidemiological feature of COVID-19 and the isolation strategy carrying out by the government, we extend the classical SEIR model to describe the transmission of COVID-19 in China. We use a short-term model to describe the strictest isolation strategy, the population is assumed relatively fixed. Consideration must be given to both the actual situations and theoretical analysis, some simplifications are necessary. This model satisfies the following assumptions,
\begin{itemize}
\item[(1)] All coefficients involved in the model are positive constants.

\item[(2)] It is a short-term model, so the natural birth and death rate are not considered.

\item[(3)] If one has been cured well, the immune efficacy will maintain for some time, i.e. second infection is not considered in the model.
\end{itemize}
Based on the above assumptions and actual isolation strategy, the spread of COVID-19 in the populations is shown in Figure~\ref{fig1_illustration}.

And the corresponding dynamical model is formulated as follows,
\begin{equation}\label{eq1}
\begin{array}{lcl}
\dfrac{{\rm{d}} S}{{\rm{d}} t}&=&-\beta S(I+\theta A) - pS + \lambda Q\\[3mm]
\dfrac{{\rm{d}}   Q}{{\rm{d}} t}&=& pS-\lambda Q\\[3mm]
\dfrac{{\rm{d}} E}{{\rm{d}} t}&=& \beta S(I + \theta A) - \sigma E\\[3mm]
\dfrac{{\rm{d}}  A}{{\rm{d}} t}&=&\sigma(1-\rho)E - \epsilon_{A}A-\gamma_{A}A\\[3mm]
\dfrac{{\rm{d}}  I}{{\rm{d}} t}&=&\sigma\rho E - \gamma_{I}I - d_{I}I - \epsilon_{I}I\\[3mm]
\dfrac{{\rm{d}} D}{{\rm{d}} t}&=&\epsilon_{A}A + \epsilon_{I}I - d_{D}D - \gamma_{D}D\\[3mm]
\dfrac{{\rm{d}} R}{{\rm{d}} t}&=&\gamma_{A}A + \gamma_{I}I + \gamma_{D}D
\end{array}
\end{equation}

The population is divided into seven compartments, where $S(t), E(t), I(t), R(t), Q(t), A(t)$ and $D(t)$ denote the susceptible, exposed, infectious with symptoms, recovered, home quarantined, asymptomatic infected and diagnosed individuals at time $t$, respectively. The exposed class means low-level virus carriers, which are considered to be no infectiousness. The class $Q(t)$ denotes the class in which the individual who is in the process of home quarantine, according to their travel limitation and the rigorous management of communities, we suppose that they won't contact with the infected population. Specially, in order to match the reported data better, the class $D(t)$ represents the number of medical confirmed cases which are in isolation treatment at time $t$.

In system (\ref{eq1}), we adopt bilinear incidence rates to describe the infection of disease and parameter $\beta$ denotes the contact rate. The spread ability between infectious individuals with symptoms and asymptomatic infectious class is different, parameter $\theta\in(0, 1)$ is used to describe the difference. We use parameter $p$ and $\lambda$ to present the quarantined rate and release rate of quarantined compartment $Q$, respectively. Transition rate of exposed to infected class is denoted as $\sigma$. Once infected, the proportion of becoming symptomatic is $\rho$ and asymptomatic is $1-\rho$. Diagnostic rate of asymptomatic and symptomatic infectious are $\epsilon_{A}$ and $\epsilon_{I}$. And the mean recovery period of class $A, I, D$ are $1/\gamma_{A}, 1/\gamma_{I}$ and $1/\gamma_{D}$, respectively. The parameters $d_{I}$ and $d_{D}$ represent the disease-induced death rate.

Under the isolation control strategy, we employ the next generation matrix approach \cite{reproductionnumber} to calculate the control reproduction number,

\begin{equation}
\begin{aligned}
\mathcal{R}^{0}_{c}  &= r(\mathcal{F}\cdot\mathcal{V}^{-1})\\
&=\left( \dfrac{\beta\theta(1-\rho)}{\epsilon_{A}+\gamma_{A}} + \dfrac{\beta\rho}{\gamma_{I}+d_{I}+\epsilon_{I}}  \right) S_{0}.
\end{aligned}
\end{equation}
where $r$ denotes the spectral radius and the matrices $\mathcal{F}$ and $\mathcal{V}$ are given by
$$
\mathcal{F} = \left(
\begin{array}{ccc}
0 & \beta S \theta & \beta S\\
0&0&0\\
0&0&0
\end{array}
\right),
\ \ \
\mathcal{V} = \left(
\begin{array}{ccc}
\sigma & 0 & 0\\
-\sigma(1-\rho)&\epsilon_{A}+\gamma_{A}&0\\
-\sigma\rho&0&\gamma_{I}+d_{I}+\epsilon_{I}
\end{array}
\right),
$$

And the corresponding effective control reproduction number is defined as
\begin{equation}
\begin{aligned}
\mathcal{R}_{c}(t)    =\left( \dfrac{\beta\theta(1-\rho)}{\epsilon_{A}+\gamma_{A}} + \dfrac{\beta\rho}{\gamma_{I}+d_{I}+\epsilon_{I}}  \right) S(t).
\end{aligned}
\end{equation}

It provides us a clear index to evaluate the control strategy for any time $t$.

\subsection{Data-Based Parameter Estimation}
\label{secp}

\emph{\textbf{Data preparation}}

On January 23, 2020, China CDC and each provincial CDC started to publish the epidemic data in their official websites, we collect these data up to February 19 for parameter estimation. The period we select is $28$ days, it is just twice of the least home quarantine period. According to the average incubation period, we divide the total control period into three phases: prophase (1-7 days), metaphase (8-14 days) and anaphase (15-28 days). $D(t)$ is an important variable in model (\ref{eq1}), it can be calculated by the published data, say, $D(t)$ is equal to the accumulated diagnostic cases subtract the recovered and death cases. The counting rule of $D(t)$ is subtraction, it can avoid the loss of data immensely. We take $D(t)$ as the benchmark for fitting problem. Qinghai and Tibet have only few of diagnostic cases and the epidemic is already under control by the government, so these two provinces are excluded in our research, Hong Kong, Macao and Taiwan are also not included because of their different diagnostic criteria and control strategy.

\noindent\emph{\textbf{Parameter Estimation and Prediction Procedure}}

In order to estimate the value of $\mathcal{R}^{0}_{c} $ and $\mathcal{R}_{c}(t)   $, we use the daily published data to perform  fitting. The data pre-processing is as above mentioned,  the basic time scale is day.  Some parameters are assumed as follows. According to the response of each province, $1/p$ is estimated as 3 to 5 days, $1/\lambda$ is taken as 60 days for most provinces. The mean incubation period ($1/\sigma$) is about $7$ days \cite{incubation, jcm9020462}. Based on the percentage of symptomatic infected patients reported in \cite{incubation}, we estimate the proportion of symptomatic in the infected class in $[0.7, 0.99]$.  Although the testing kits are developed and come into service quickly, the shortage of capacity delays the diagnosis. Refer to the detail of confirmed cases, the average time of diagnosis $(1/\epsilon_{I})$ for infectious with symptoms is taken in $[3, 9]$. The diagnosis of asymptomatic infected is much harder, we estimate the period $(1/\epsilon_{A})$ as 3 to 15 days. Luckily, the spread ability of asymptomatic infected is limited, we set parameter $\theta \in [0.005, 0.2]$. Back to model (\ref{eq1}), the mortality rate  ($mr$)  of disease can be written as
\begin{equation}\nonumber
mr =\dfrac{{d_{D}}}{{d_{D} + \gamma_{D}}}.
\end{equation}
By calculation, we set $mr$ as $2.1\%$ for most provinces. The death rate of patient without effective medicine care will be higher, we describe it as $d_{I} = c_I\cdot d_{D}$, where $c_{I} \in [1.1, 1.6]$. The relationship in average recover period is assumed as $\gamma_{A} = c_{A}\cdot\gamma_{I}$ and $\gamma_{D} = c_{D}\cdot\gamma_{I}$, where $c_{A}, c_{D} \in [1.1, 1.5]$. The reason is that the speed of recovery in asymptomatic infected and patients with medicine care is quicker than those infected without treatment.

We employ a Least-Squares procedure to estimate parameter $\beta$ and $\gamma_{I}$. Suppose we have a proper estimation for other parameters in (\ref{eq1}), we need to solve the following optimization problem.

\begin{equation}\label{eqbeta}
\min_{\beta, \gamma_{I}}\| D(t; \beta, \gamma_{I}) - D_{pub}\|_{2},
\end{equation}
where $D_{pub}$ is the data published by CDC. Then the estimation and prediction procedure follows,
\begin{itemize}
\item[(1)] Set the initial condition for $\{S(t_{0}), Q(t_{0}), E(t_{0}), A(t_{0}), I(t_{0}), R(t_{0}) \}$ and the proper guess for the parameters in (\ref{eq1}), except for $\beta$ and $\gamma_{I}$.
\item[(2)] Based on the official published data $D_{pub}$, solve the optimization problem (\ref{eqbeta}) to obtain the estimated $\beta^{*}$ and $\gamma_{I}^{*}$.
\item[(3)] Based on $\beta^{*}$ and $\gamma_{I}^{*}$, the initial condition and parameters set in Step (1), solve the dynamical system (\ref{eq1}) to obtain $\{S(t), Q(t), E(t), A(t), I(t), R(t), D(t)\}$.
\end{itemize}

\begin{remark}
For initial values, the total population data is based on the report published by the National Bureau of Statistics \cite{totalpopulation}. Due to the Chinese New Year, many people are in vacation and they stay at home originally. We estimate the fraction of original home quarantine as about $30\%$.
\end{remark}

\begin{remark}
Due to the change of diagnostic criteria from nucleic acid detection to clinical diagnosis, there is a jump for $D_{pub}$ of Hubei Province on February 12, 2020 (see  Figure~\ref{fig2}(b)).
\end{remark}

%We estimate $\beta, \gamma_{I}$ and calculate the control reproduction number $\mathcal{R}^{0}_{c} $ for each selected provinces.

\noindent\emph{\textbf{Accumulated Medical Resource Estimation}}

In order to avoid the delay of medical treatment caused by personal economic ability, Chinese government started to carry out free medical care strategy timely. National finance support the treatment for COVID-19 patients strongly. We introduce an index to evaluate the finance support for the disease treatment. The diagnostic compartment $D(t)$ represents the population of confirmed infected patients being treated in hospital at time $t$, we define the accumulated medical resource $(AMR)$ needed until $t_{f}$ as the integration of $D(t)$,
\begin{equation}\label{AMR}
AMR = k\int_{0}^{{t_{f}}}D(t){\rm{d}}t,
\end{equation}
where the parameter $k$ represents the average index of medical resource a patient needs daily.

\subsection{Numerical simulation}

\noindent\emph{\textbf{Prediction and Estimation}}

\begin{figure}[h!]
\centering
\subfigure[Outside Hubei Province]{
\begin{minipage}[t]{0.45\linewidth}
\centering
\includegraphics[width=8cm]{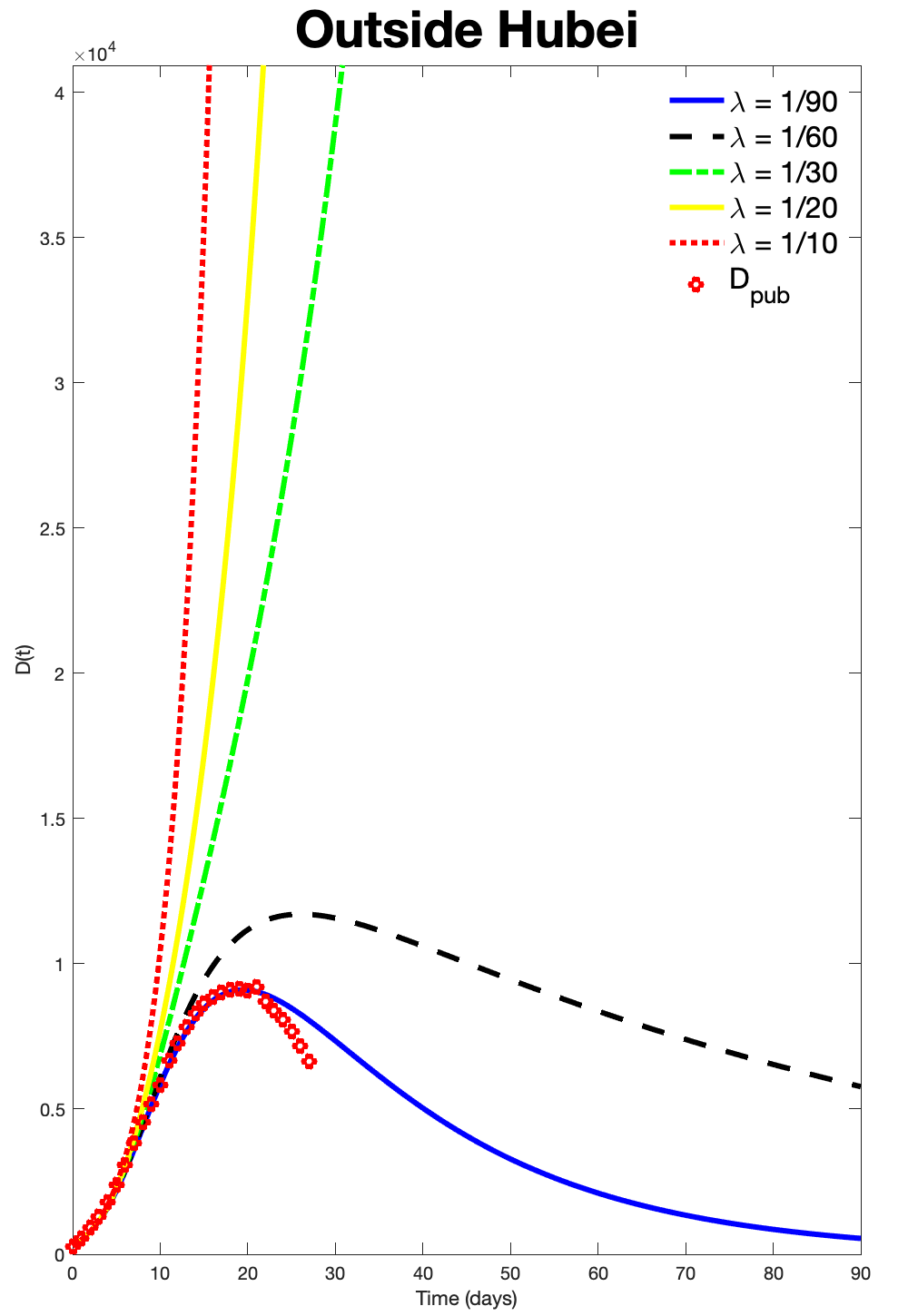}
%\caption{fig1}
\end{minipage}%
}%
\hspace{1cm}
\subfigure[Inside Hubei Province]{
\begin{minipage}[t]{0.45\linewidth}
\centering
\includegraphics[width=8cm]{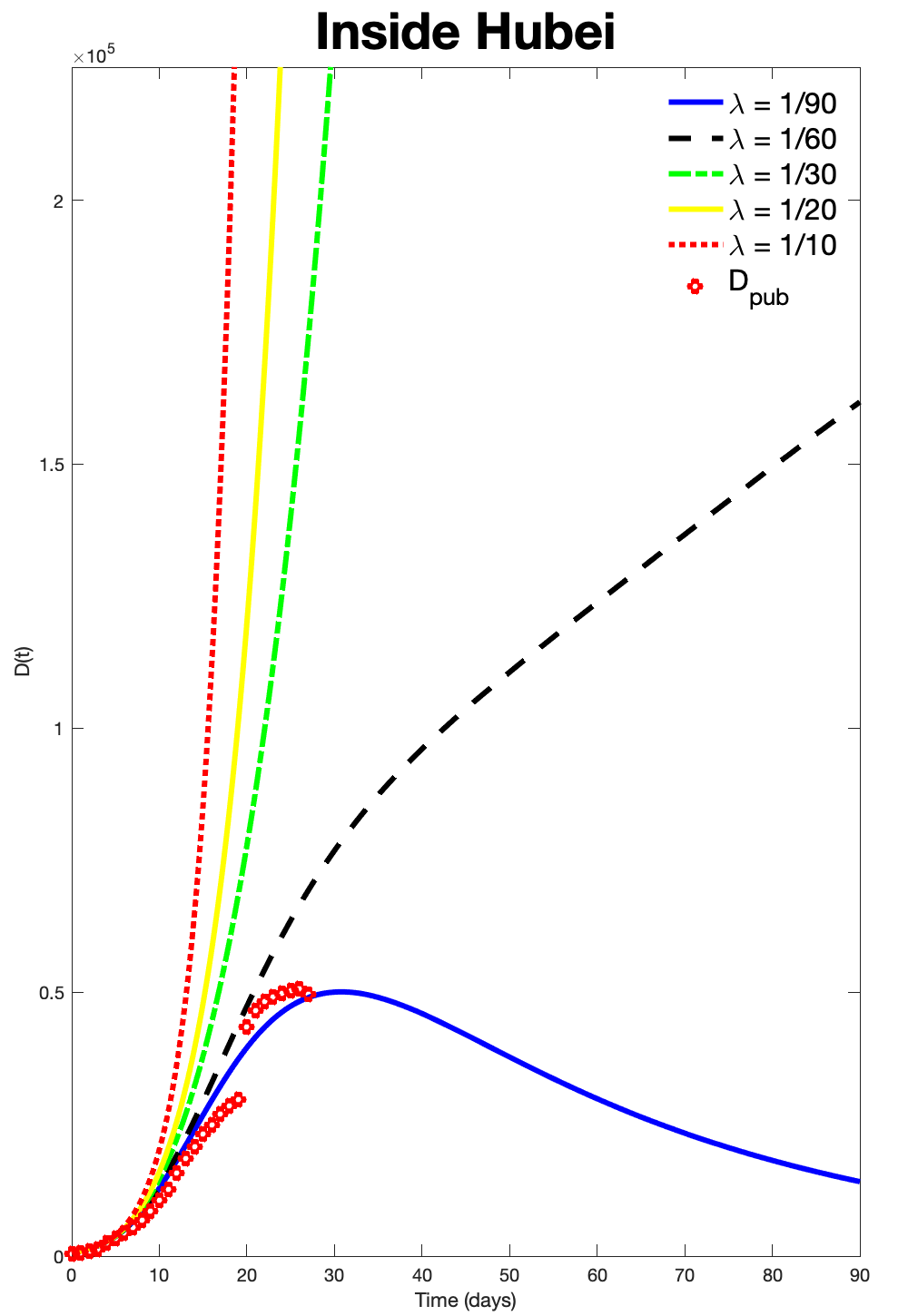}
\end{minipage}%
}%
\centering
\caption{Optimal simulation and prediction of the transmission trend}\label{fig2}
\end{figure}

The first confirmed COVID-19 patient of China is in Wuhan,  because of the limited knowledge of the disease, the control measure in Wuhan is insufficient at the beginning. It results in the outbreak of COVID-19 in Hubei Province. Due to the different circumstance inside and outside Hubei Province, the estimation and prediction are investigated for the two cases. Notice that, the spread of COVID-19 is very strong, it may cause twice outbreak without effective control. Zhong declares that, the disease may be controlled well by the end of April \cite{zhongnanshansiyue} under our powerful control strategy, we take parameter $1/\lambda$ as 90 days. Based on the procedure in Section~\ref{secp}, the corresponding parameter $\beta$ and $\gamma_{I}$ are given. The results of inside and outside Hubei province are presented in Figure~\ref{fig2}, the blue solid line shows the fit based on current circumstances and the trends until the end of April are shown. Asterisks represent the $D_{{pub}}$. All the other parameters, initial values and the corresponding $\mathcal{R}^{0}_{c} $ for inside and outside of Hubei Province can be seen in Tabel~\ref{table_para_inout}. We find that the $\mathcal{R}^{0}_{c} $  outside Hubei is much higher than it in inside Hubei. This is mainly caused by the huge total size of Chinese population, it represent the initial situation in a way. To evaluate the strictest isolation strategy, we calculate the average $\mathcal{R}_{c}(t)   $ for three stages and show them in Table \ref{table_averageRc}. We can clearly see that the values of the average $\mathcal{R}_{c}(t)   $ decrease quickly under current control strategy. It almost down to $1$ in metaphase outside Hubei Province and below $1$ now. The situation in Hubei Province is more complex, though the average $\mathcal{R}_{c}(t)   $ decreases sharply, it is still greater than $1$ in anaphase. The disease isn't under control completely, it still has a high risk of sustainable spread. Specific medicine and effective vaccine are absent till now, the strictest isolation strategy makes great contribution to the prevention of the disease spread, it needs to persist in.

%\begin{table}[h!]
%\centering
%\begin{threeparttable}
%\caption{Parameter estimation and initial value}\label{table_para_inout}
%\begin{tabular}{cll|cll}
%\hline
%\hline
%Parameter & Outside Hubei & Inside Hubei & Parameter & Outside Hubei & Inside Hubei\\
%\hline
%$\beta$  & $5.5010\times10^{-9}$& $1.0014\times10^{-7}$ & $\epsilon_{A}$  &$1/5$ & $1/10$\\
%$\theta$  & $0.1000$ & $0.1600$  & $\epsilon_{I}$ & $1/4$ & $1/3$\\
%$p$    &  $1/3$ & $1/6.2$ & $\gamma_{A}$ & $0.1496$ & $0.1500$\\
%$\lambda$ & $1/90$ & $1/90$ & $\gamma_{I}$ & $0.0998$ & $0.1000$\\
%$\sigma$  & $1/7$  & $1/7$ & $\gamma_{D}$ & $0.1496$  &$0.1400$\\
%$\rho$   &  $0.8800$ & $0.8800$ & $d_{I}$ & $0.0046$ & $0.0105$\\
%$\mathcal{R}^{0}_{c} $ & $12.7700$ & $8.5423$ & $d_{D}$  & $0.0031$ & $0.0030$\\
%\hline
%$S(0)$ & $921984900$ & $41419000$ & $I(0)$ & $563$ & $1206$ \\
%$Q(0)$ & $414225100$ & $17751000$ & $D(0)$  & $227$ & $494$\\
%$E(0)$ & $3207$  & $2280$ & $R(0)$ & $3$ & $31$\\
%$A(0)$ & $595$ & $1450$\\
%\hline
%\hline
%\end{tabular}
%\end{threeparttable}
%\end{table}

\begin{table}[h!]
\centering
\begin{threeparttable}
\caption{Parameter estimation and initial value}\label{table_para_inout}
\begin{tabular}{cll|cll}
\hline
\hline
Parameter & Outside Hubei & Inside Hubei & Parameter & Outside Hubei & Inside Hubei\\
\hline
$\beta$  & $5.5010\times10^{-9}$& $1.0014\times10^{-7}$ & $\epsilon_{I}$ & $1/4$ & $1/3$\\
$\theta$  & $0.1000$ & $0.1600$  & $\gamma_{A}$ & $0.1496$ & $0.1500$\\
$p$    &  $1/3$ & $1/6.2$ & $\gamma_{I}$ & $0.0998$ & $0.1000$\\
$\lambda$ & $1/90$ & $1/90$ & $\gamma_{D}$ & $0.1496$  &$0.1400$\\
$\sigma$  & $1/7$  & $1/7$ & $d_{I}$ & $0.0046$ & $0.0105$\\
$\rho$   &  $0.8800$ & $0.8800$ & $d_{D}$  & $0.0031$ & $0.0030$\\
$\epsilon_{A}$  &$1/5$ & $1/10$&  & &\\
\hline
$\mathcal{R}^{0}_{c} $ & $12.7700$ & $8.5423$ && \\
\hline
$S(0)$ & $921984900$ & $41419000$ & $I(0)$ & $563$ & $1206$ \\
$Q(0)$ & $414225100$ & $17751000$ & $D(0)$  & $227$ & $494$\\
$E(0)$ & $3207$  & $2280$ & $R(0)$ & $3$ & $31$\\
$A(0)$ & $595$ & $1450$\\
\hline
\hline
\end{tabular}
\end{threeparttable}
\end{table}

\begin{table}[h!]
\centering
\caption{Average $\mathcal{R}_{c}(t)   $}\label{table_averageRc}
\begin{tabular}{c|ccc}
\hline
\hline
&Prophase&Metaphase&Anaphase\\
\hline
Outside Hubei & 6.0295 &  1.0843 &  0.6208\\
Inside Hubei & 5.6870 & 2.2426 &  1.0560\\
\hline
\hline
\end{tabular}
\end{table}

To make a better illustration of quarantine strategy, we test different home quarantine period ($1/\lambda$) in Figure~\ref{fig2}. The corresponding peak value, peak time of $D(t)$ and the accumulated medical resource needed in $90$ days by $AMR$ which defined in (\ref{AMR}) are listed in Table~\ref{table_peak}. In Figure~\ref{fig2}, the colorful dashed lines show that if the quarantine period isn't long enough, the isolation strategy can not work well. There needs a longer quarantine period in Hubei Province than outside. The longer quarantine period, the earlier peak time of $D(t)$ comes. Unlike normal infectious disease, the ward for COVID-19 is particular. To avoid nosocomial infection, it requires maximal barrier precautions in the hospital and there are quite a few critically ill patient in ICU. Our medical system is facing the huge challenges caused by patients shoot up rapidly, especially the peak value of $D(t)$. The study of the maximum capacity to deal with the emergency is necessary for disease control. In the simulation, the peak time of $D(t)$ in outside Hubei is on February 12, 2020 and inside Hubei is February 24, 2020, which is in line with the actual data (the last line in Table~\ref{table_peak}). By adjust the parameter $\lambda$, we find that, if we fix the quarantine period as $30$ days, both the peak value and $AMR$ are more than triple as now inside Hubei. Particularly, the peak time won't reach in three months in Hubei. The situation outside Hubei is not much better than that in Hubei. The peak time delays a week, though the peak value isn't increase much, the $AMR$ almost doubled. More details of shorter quarantine period can be seen in Table~\ref{table_peak}. Notice that, in shorter quarantine period (5, 10 and 20 days), $T_{{peak}}$ seems  proportional to $1/\lambda$. It is mainly caused by a large amount of infected people, that is, the disease almost infected every susceptible person. The medical burden in shorter quarantine period situation is horrible.
\begin{table}[htp]
\centering
\begin{threeparttable}
\caption{Peak value, peak time of $D(t)$ and $AMR$ during 90 days.}\label{table_peak}
%Estimation of peak value, peak time and medical resource needed in 90 days. (* do not reach peak in 90 days, - not applicable.)
\begin{tabular}{c|ccc|ccc}
\hline
\hline
&\multicolumn{3}{c|}{Outside Hubei} & \multicolumn{3}{c}{Inside Hubei}\\
\hline
$1/\lambda$	&	$Peak$	&	$T_{peak}$	&	$AMR$	&	$Peak$	&	$T_{peak}$	&	$AMR$	\\
\hline
90	&	9094	&	20	&	354017$k$	&	50041	&	32	&	2649863$k$	\\
30	&	11706	&	27	&	743494$k$	&	161780	&	*	&	8342449$k$	\\
20	&	1540394	&	*	&	25210395$k$	&	2503084	&	78	&	100643586$k$	\\
10	&	47382685	&	*	&	618221390$k$	&	4372193	&	60	&	185123007$k$	\\
5	&	142619535	&	67	&	4664793399$k$	&	7810630	&	45	&	264760516$k$	\\
\hline
Data	&	9211	&	22	&	-	&	50633	&	27	&	-	\\

\hline
\hline
\end{tabular}
\begin{tablenotes}
\item[1]* do not reach peak in 90 days.
\item[2]- not applicable.
\end{tablenotes}
\end{threeparttable}
\end{table}

\begin{figure}[h!]
\centering
\subfigure{
\begin{minipage}[t]{0.25\linewidth}
\centering
\includegraphics[width=4cm]{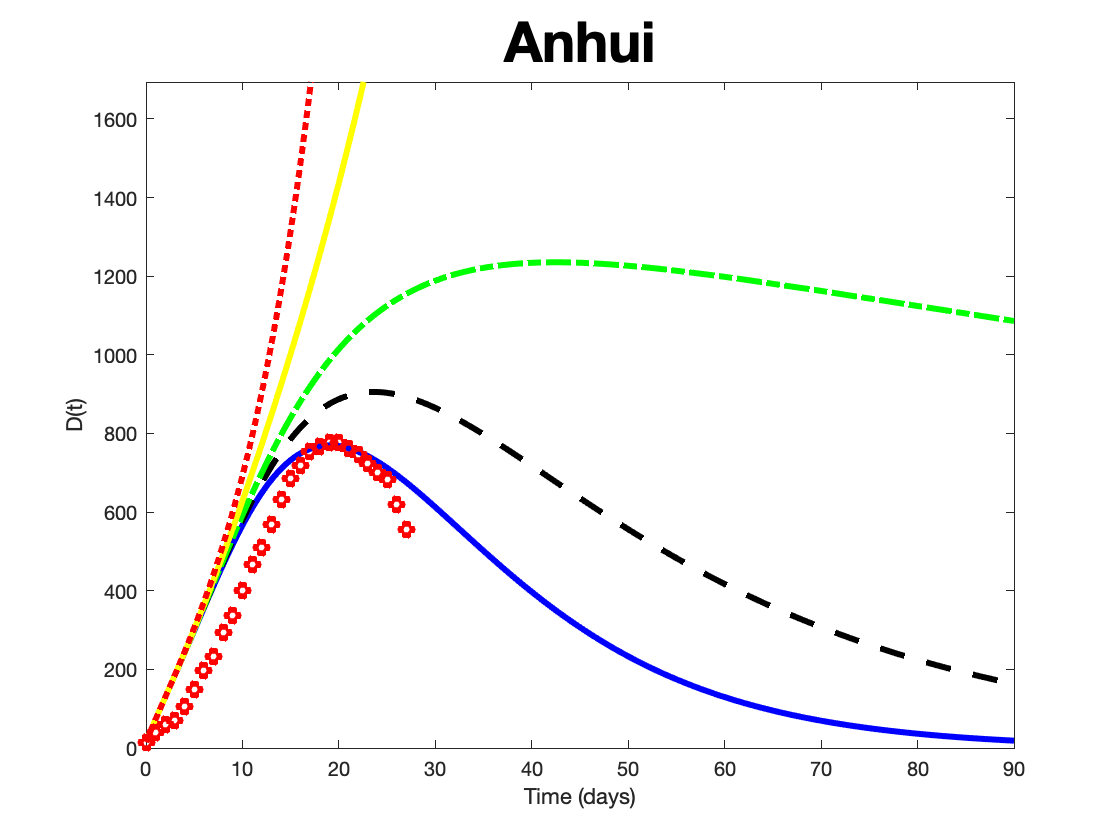}
%\caption{fig1}
\end{minipage}%
}%
%\hspace{1cm}
\subfigure{
\begin{minipage}[t]{0.25\linewidth}
\centering
\includegraphics[width=4cm]{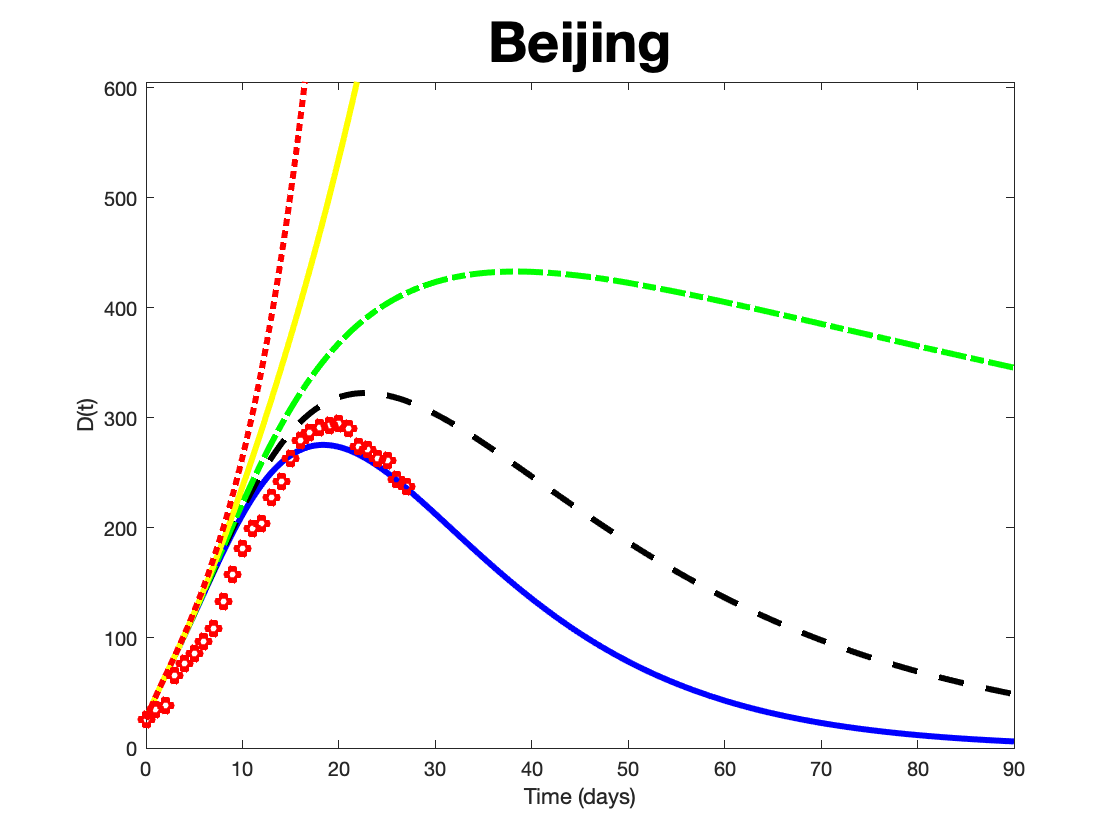}
%\caption{fig1}
\end{minipage}%
}%
\subfigure{
\begin{minipage}[t]{0.25\linewidth}
\centering
\includegraphics[width=4cm]{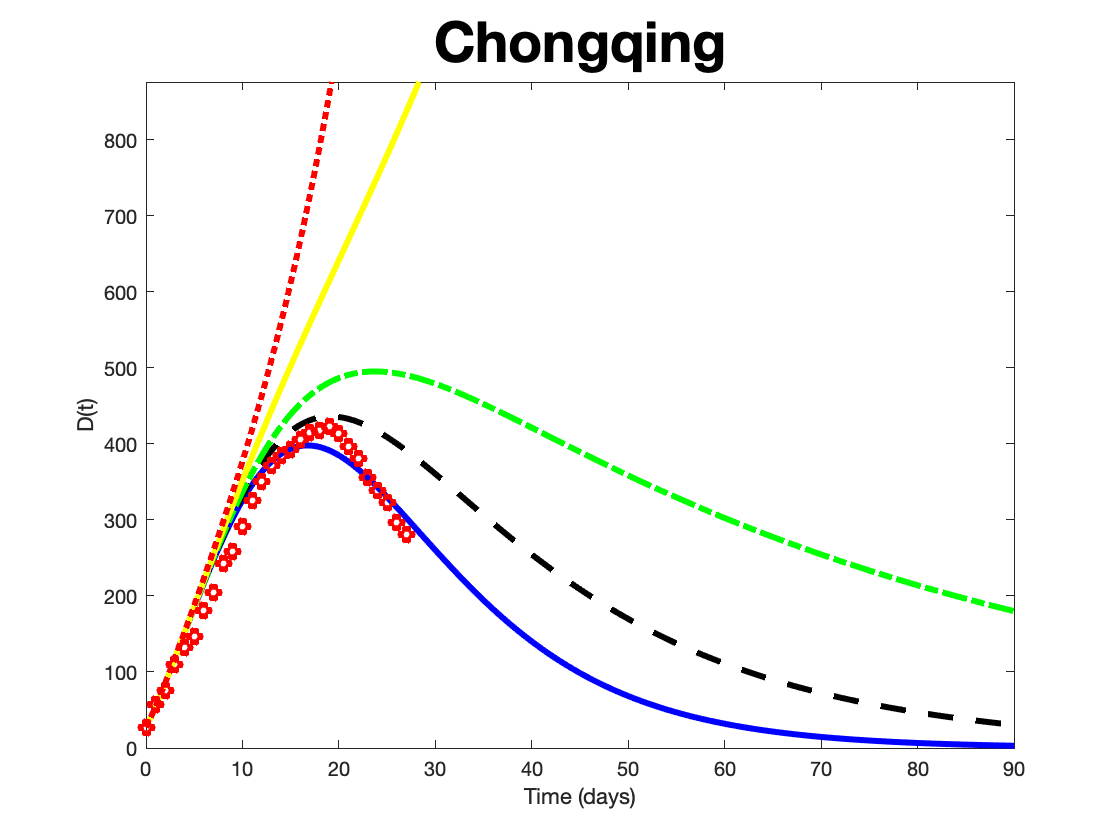}
%\caption{fig1}
\end{minipage}%
}%
\subfigure{
\begin{minipage}[t]{0.25\linewidth}
\centering
\includegraphics[width=4cm]{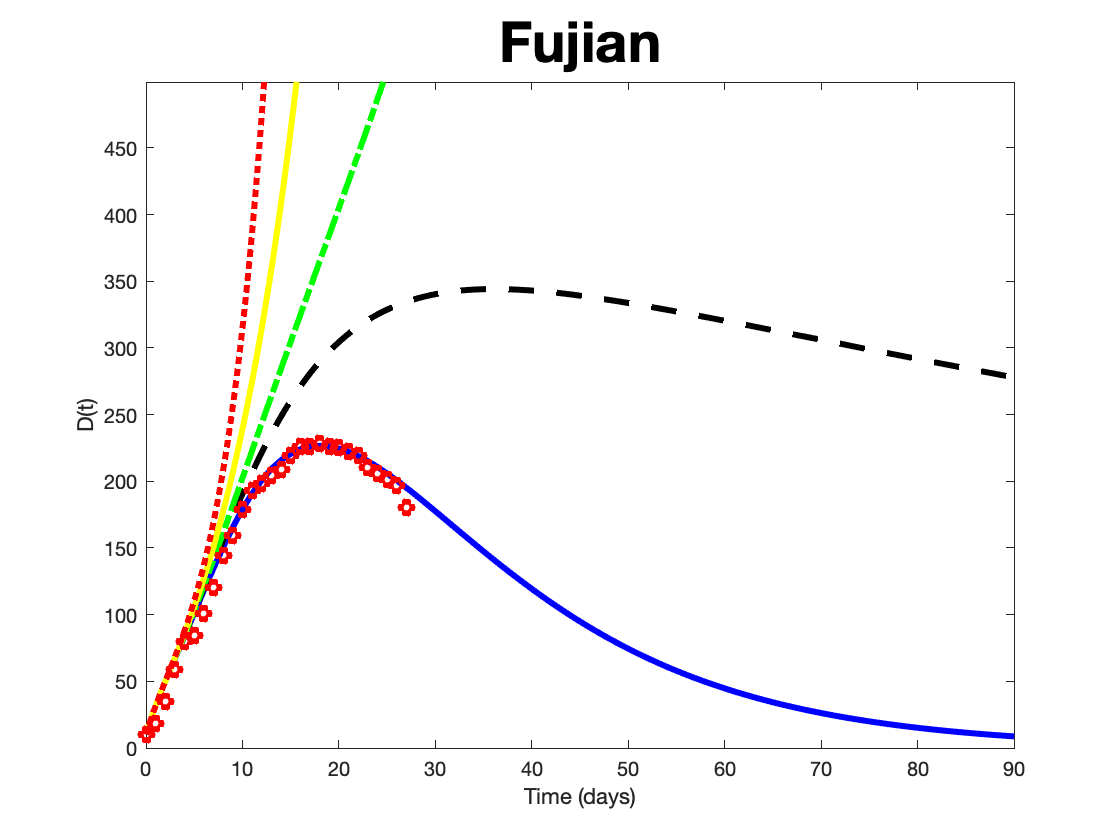}
%\caption{fig1}
\end{minipage}%
}%
%---------------------------------------------------------------
\vskip -8pt

\subfigure{
\begin{minipage}[t]{0.25\linewidth}
\centering
\includegraphics[width=4cm]{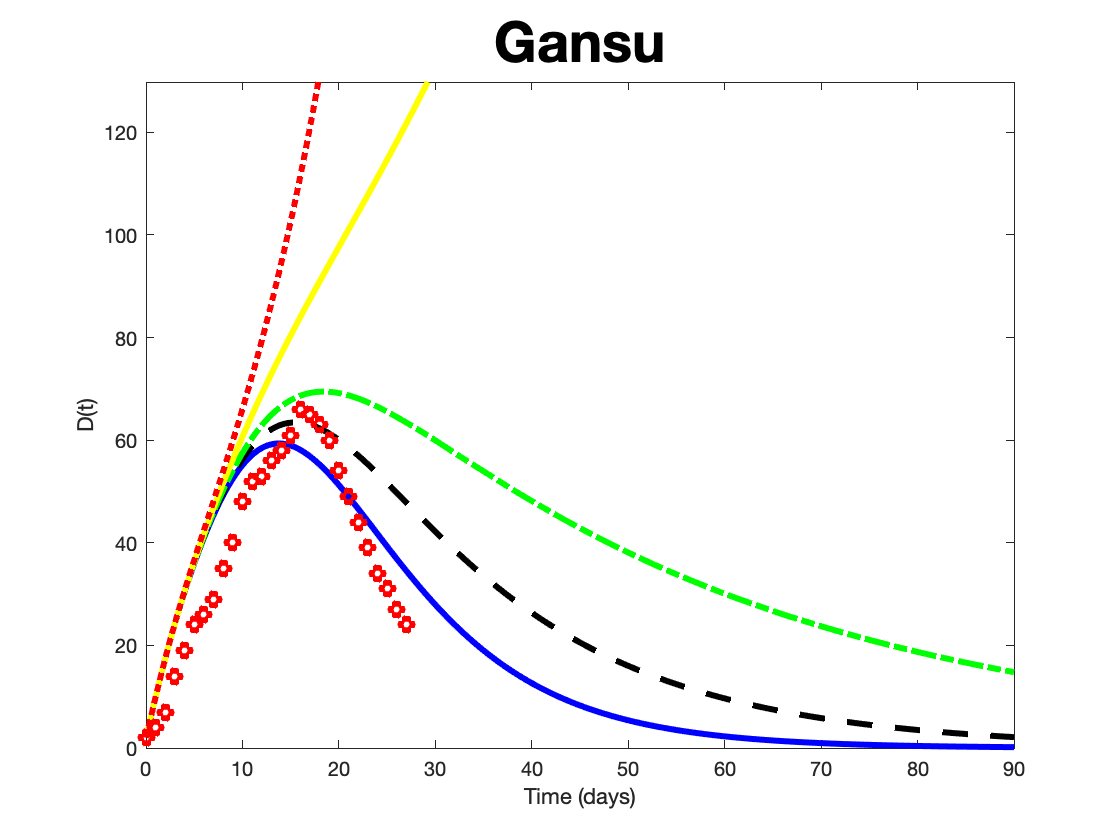}
%\caption{fig1}
\end{minipage}%
}%
%\hspace{1cm}
\subfigure{
\begin{minipage}[t]{0.25\linewidth}
\centering
\includegraphics[width=4cm]{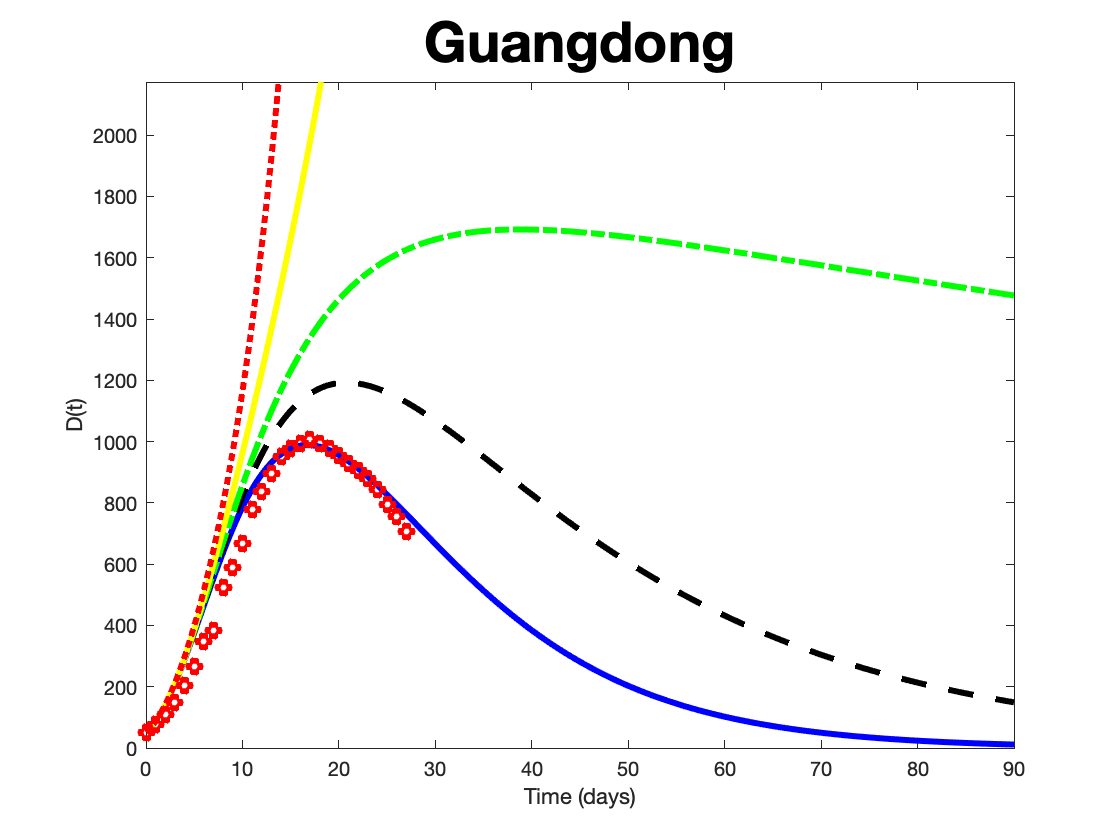}
%\caption{fig1}
\end{minipage}%
}%
\subfigure{
\begin{minipage}[t]{0.25\linewidth}
\centering
\includegraphics[width=4cm]{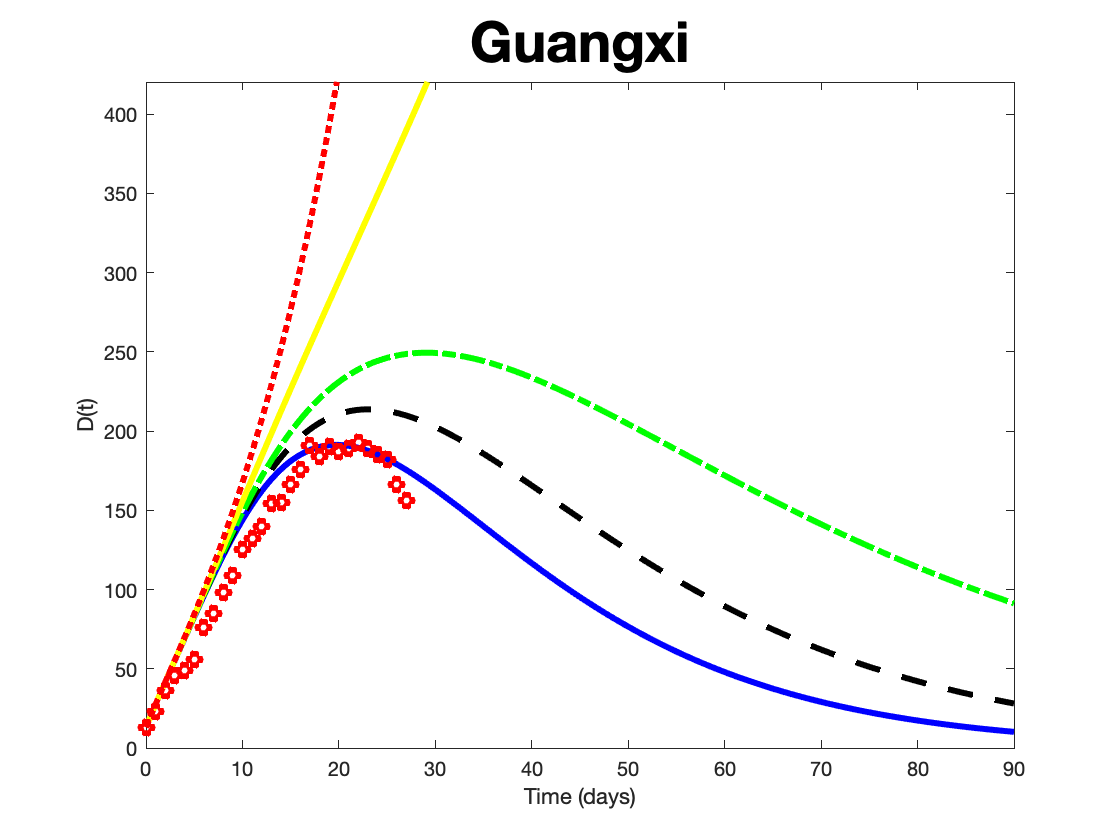}
%\caption{fig1}
\end{minipage}%
}%
\subfigure{
\begin{minipage}[t]{0.25\linewidth}
\centering
\includegraphics[width=4cm]{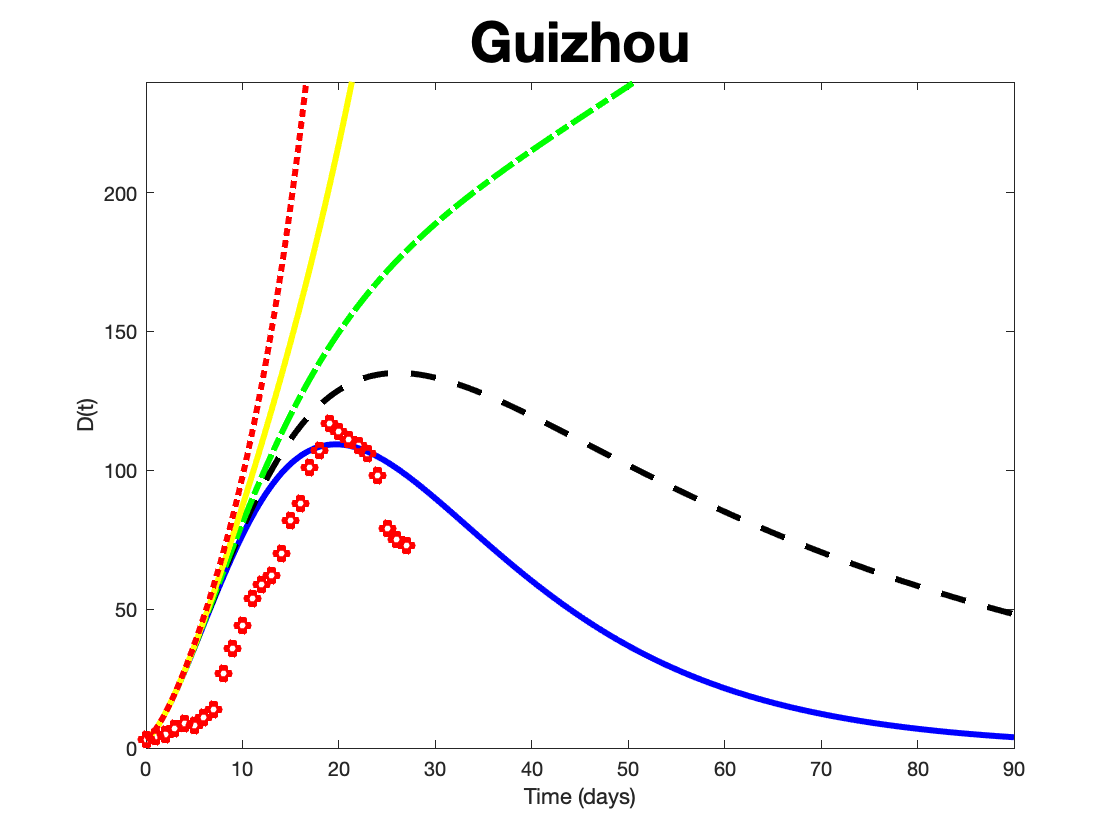}
%\caption{fig1}
\end{minipage}%
}%
%---------------------------------------------------------------
\vskip -8pt

\subfigure{
\begin{minipage}[t]{0.25\linewidth}
\centering
\includegraphics[width=4cm]{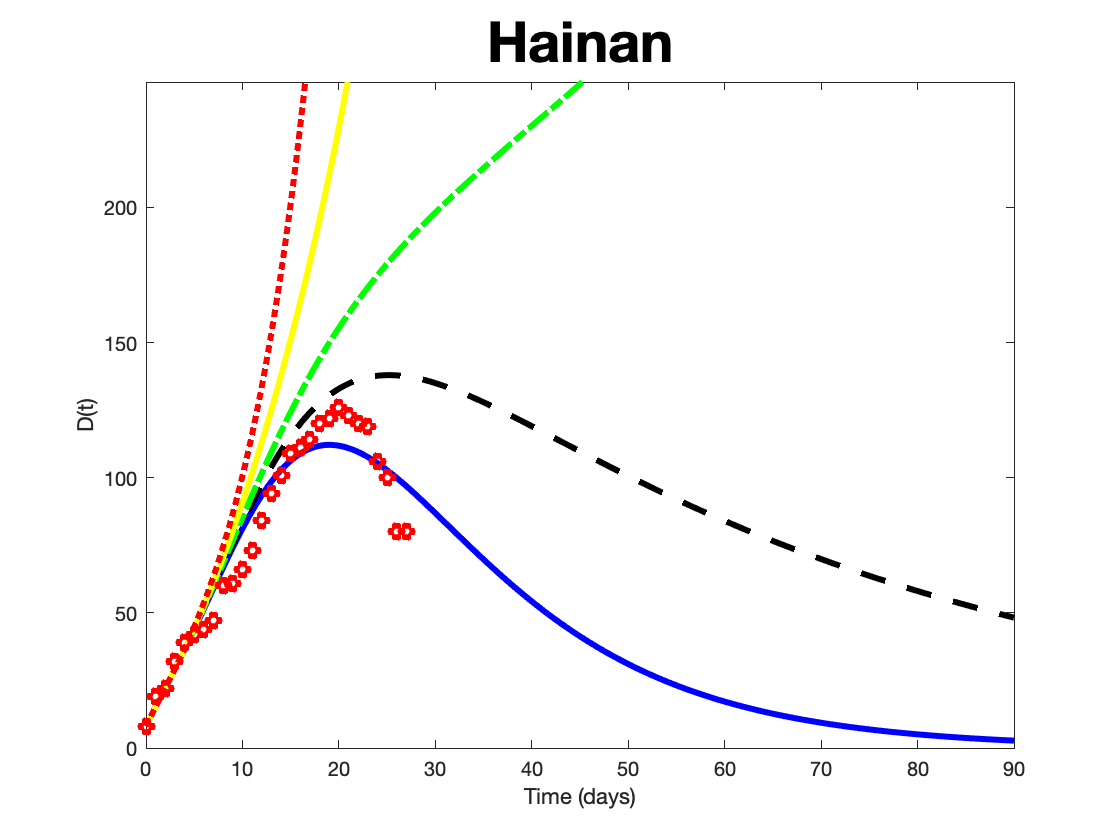}
%\caption{fig1}
\end{minipage}%
}%
%\hspace{1cm}
\subfigure{
\begin{minipage}[t]{0.25\linewidth}
\centering
\includegraphics[width=4cm]{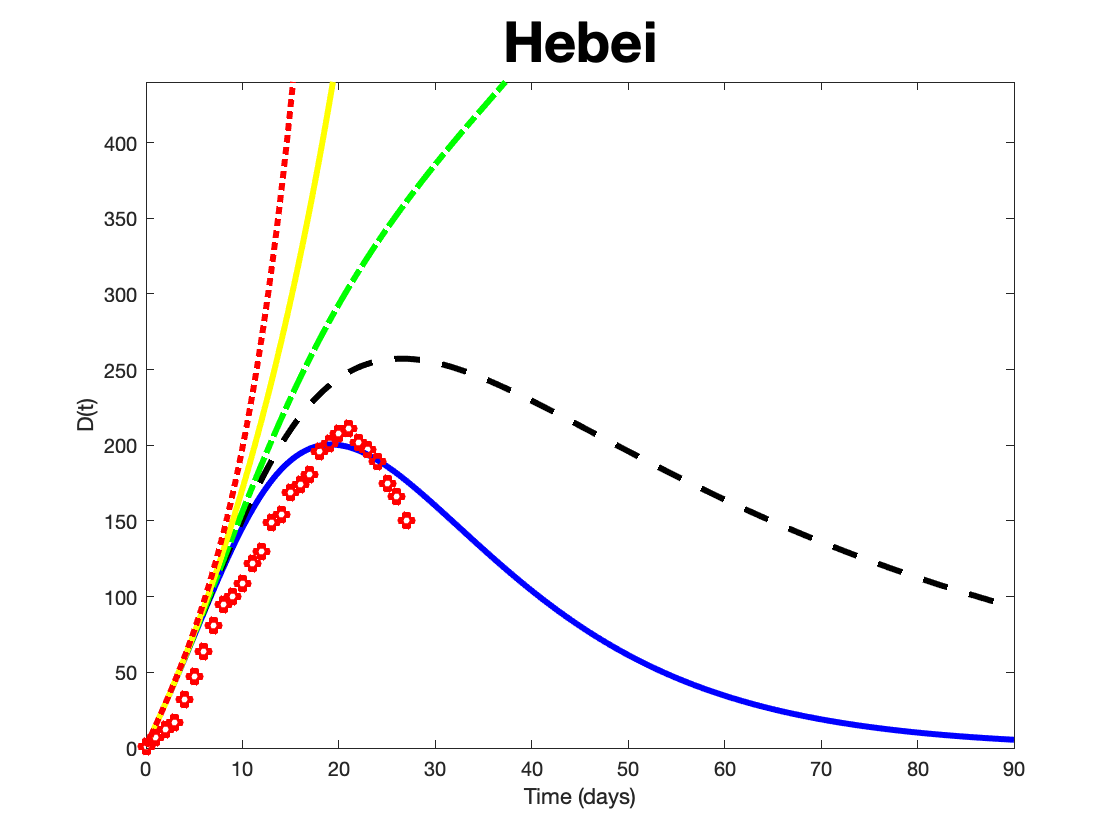}
%\caption{fig1}
\end{minipage}%
}%
\subfigure{
\begin{minipage}[t]{0.25\linewidth}
\centering
\includegraphics[width=4cm]{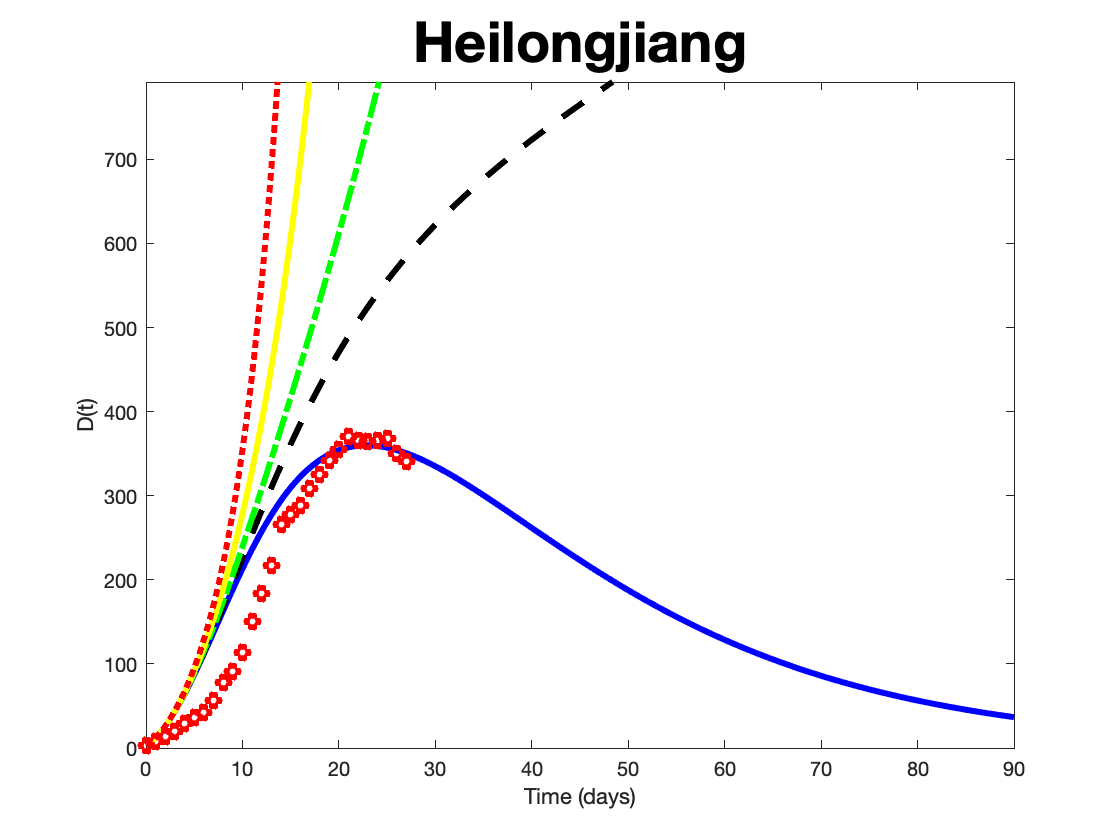}
%\caption{fig1}
\end{minipage}%
}%
\subfigure{
\begin{minipage}[t]{0.25\linewidth}
\centering
\includegraphics[width=4cm]{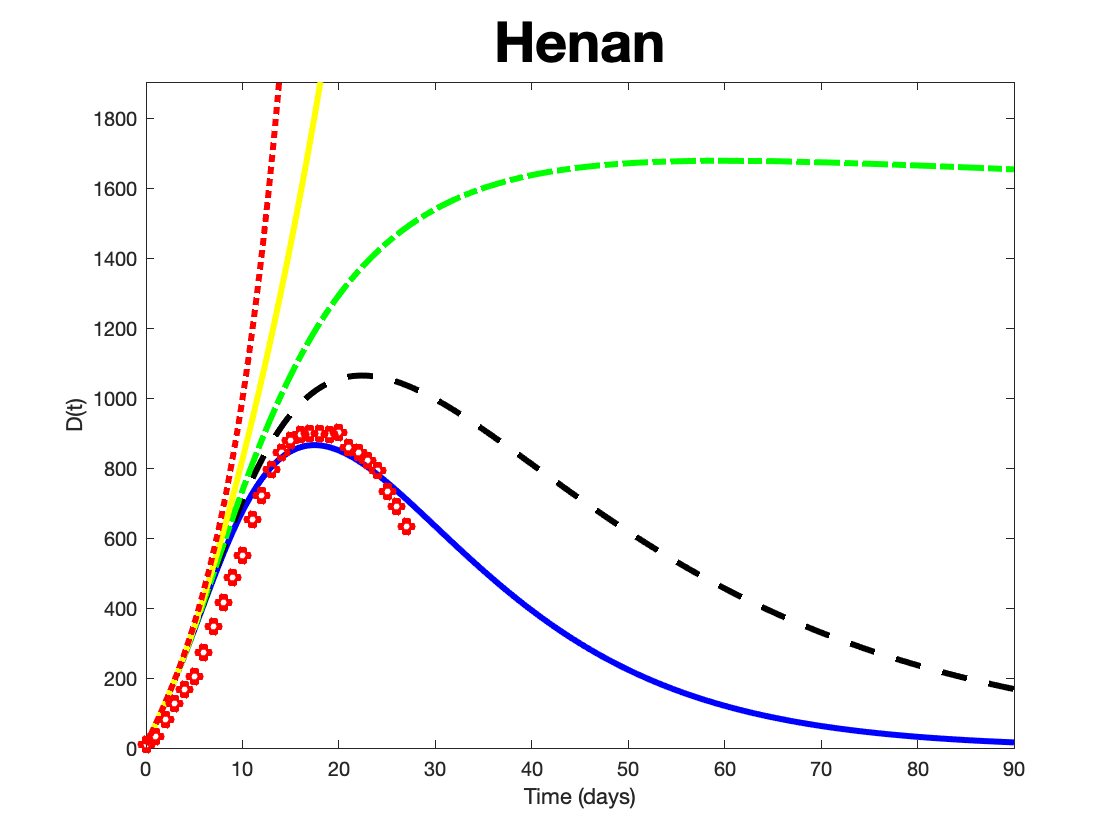}
%\caption{fig1}
\end{minipage}%
}%
%---------------------------------------------------------------
\vskip -8pt

%\hspace{1cm}
\subfigure{
\begin{minipage}[t]{0.25\linewidth}
\centering
\includegraphics[width=4cm]{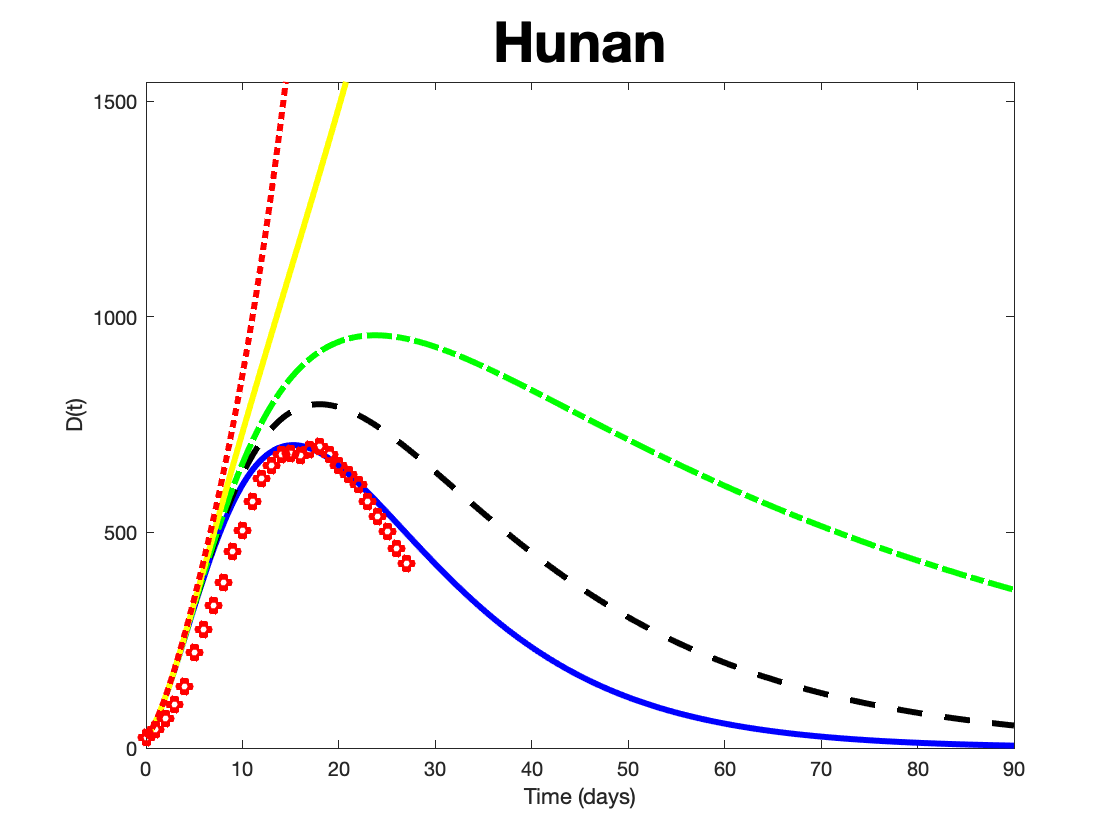}
%\caption{fig1}
\end{minipage}%
}%
\subfigure{
\begin{minipage}[t]{0.25\linewidth}
\centering
\includegraphics[width=4cm]{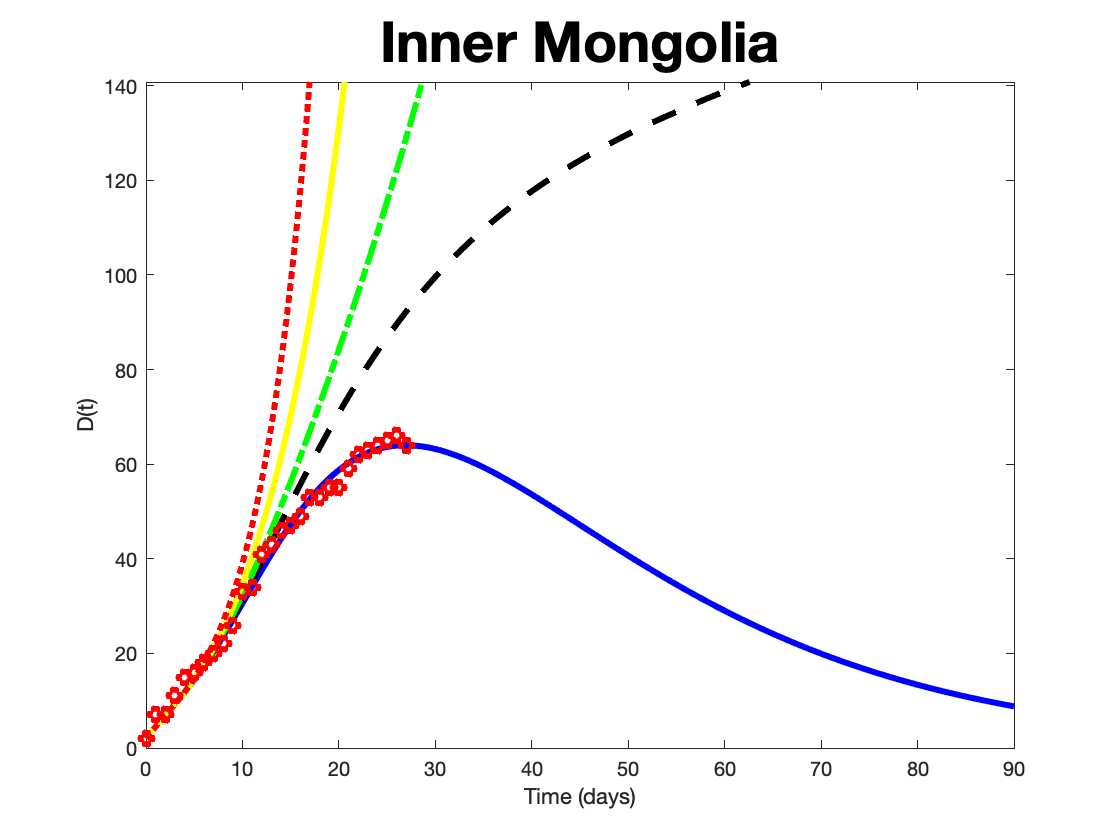}
%\caption{fig1}
\end{minipage}%
}%
\subfigure{
\begin{minipage}[t]{0.25\linewidth}
\centering
\includegraphics[width=4cm]{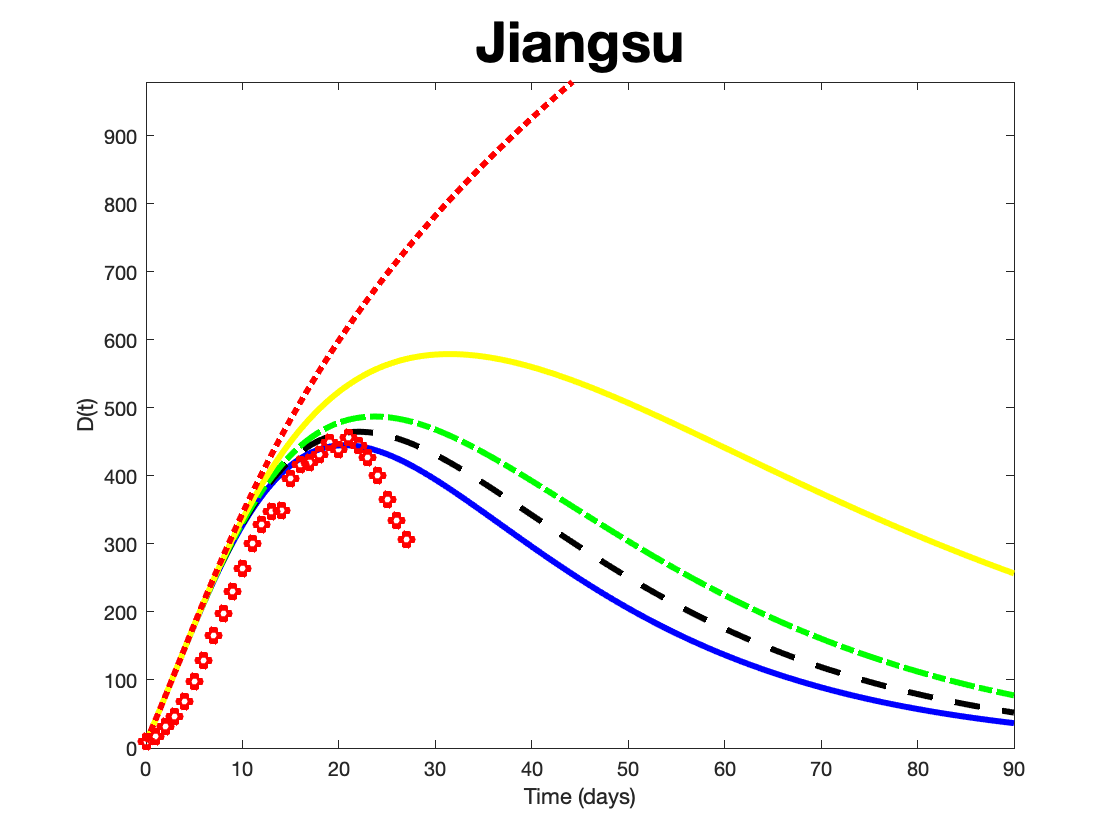}
%\caption{fig1}
\end{minipage}%
}%
\subfigure{
\begin{minipage}[t]{0.25\linewidth}
\centering
\includegraphics[width=4cm]{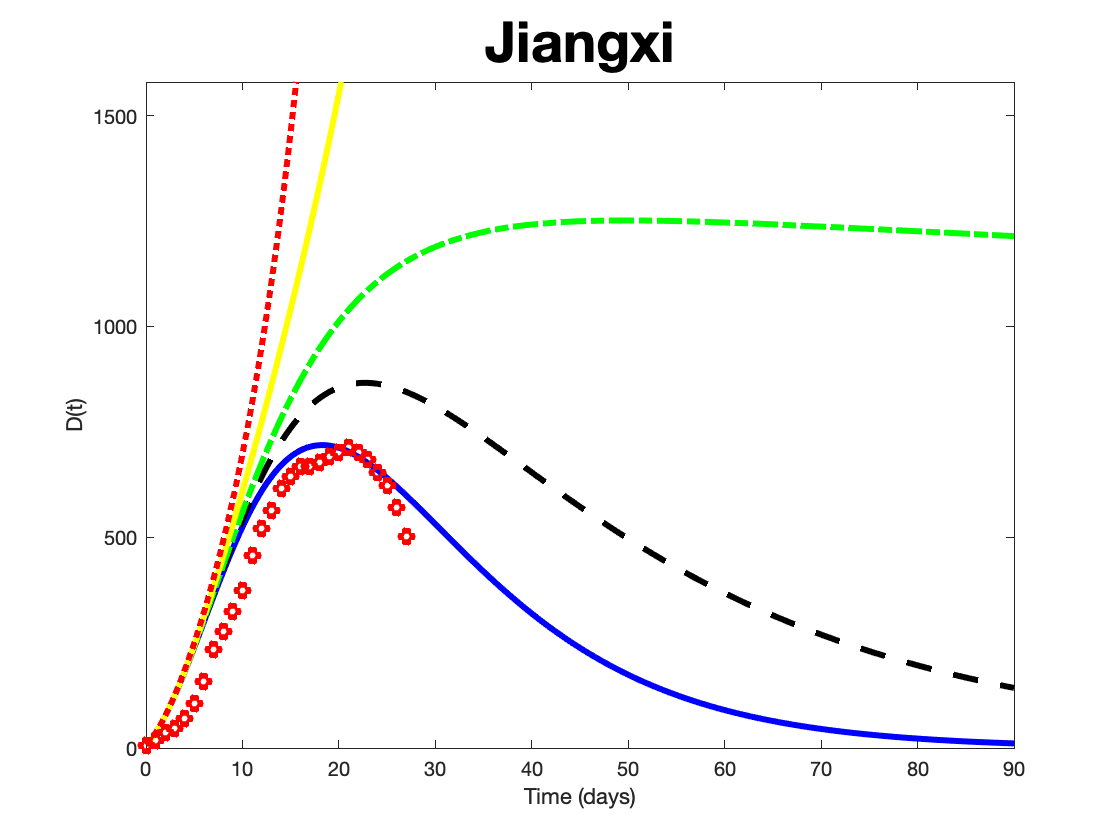}
%\caption{fig1}
\end{minipage}%
}%
%---------------------------------------------------------------
\vskip -8pt

\subfigure{
\begin{minipage}[t]{0.25\linewidth}
\centering
\includegraphics[width=4cm]{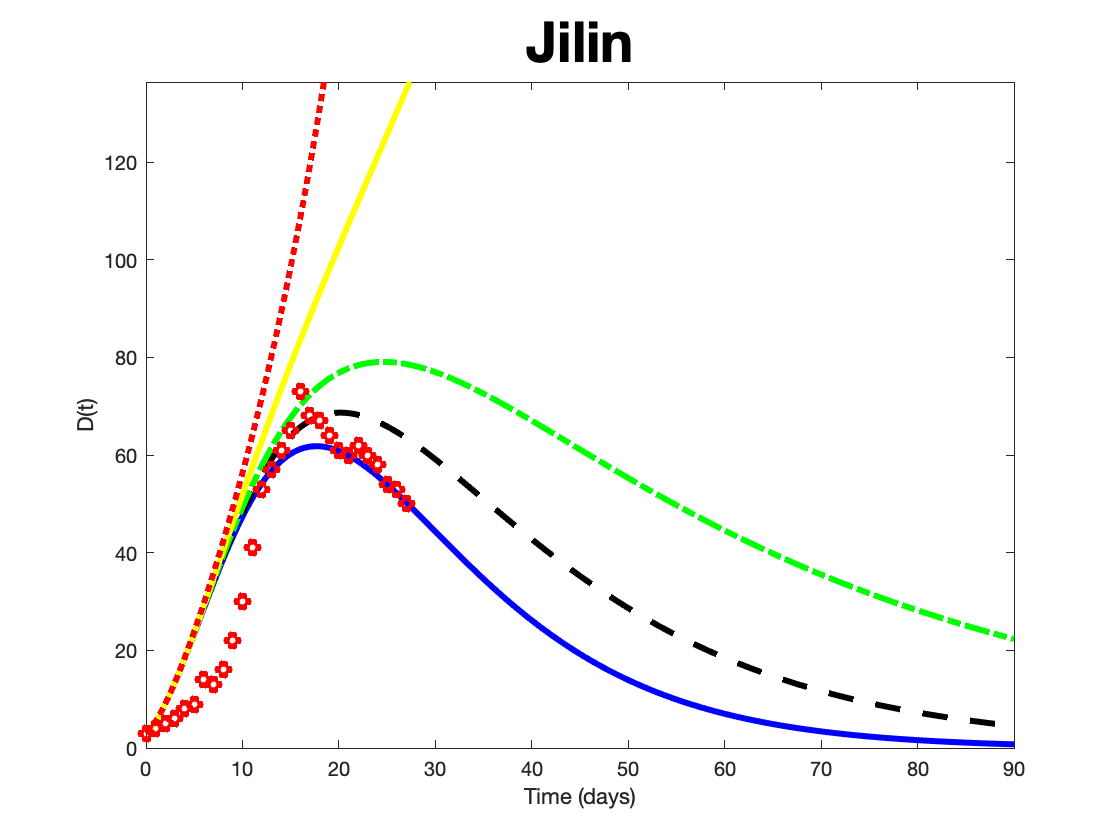}
%\caption{fig1}
\end{minipage}%
}%
\subfigure{
\begin{minipage}[t]{0.25\linewidth}
\centering
\includegraphics[width=4cm]{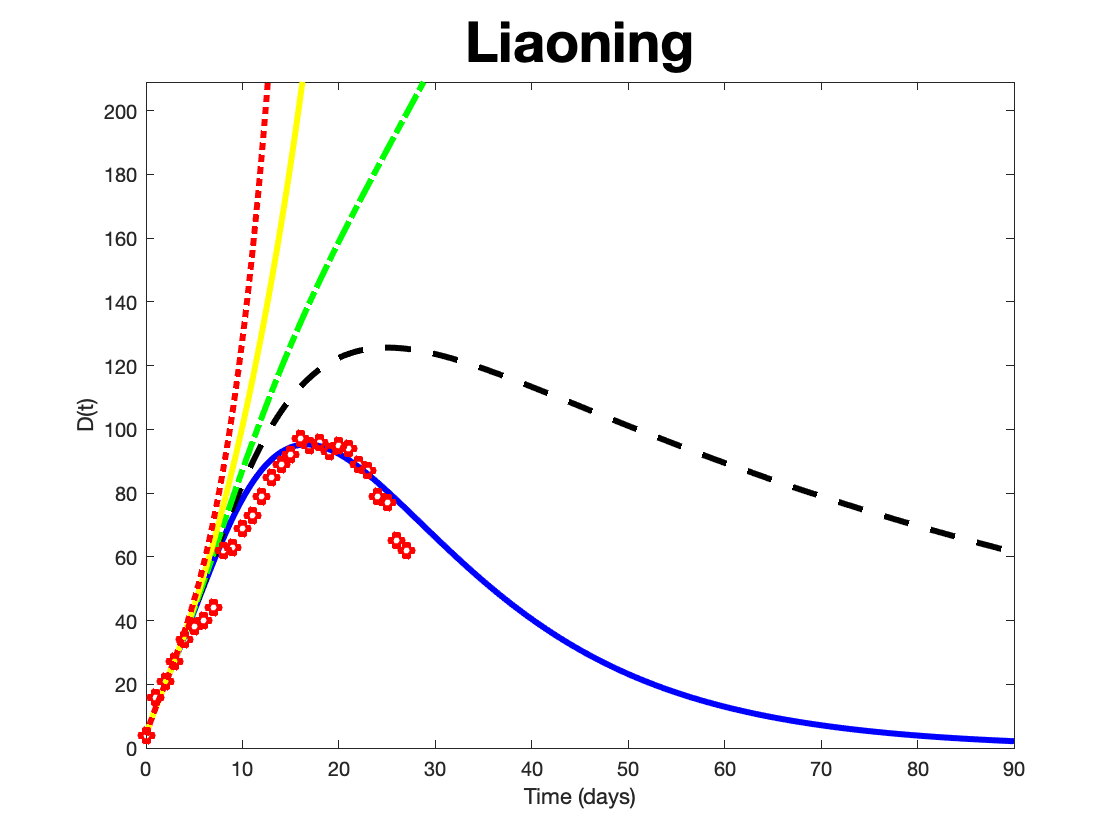}
%\caption{fig1}
\end{minipage}%
}%
\subfigure{
\begin{minipage}[t]{0.25\linewidth}
\centering
\includegraphics[width=4cm]{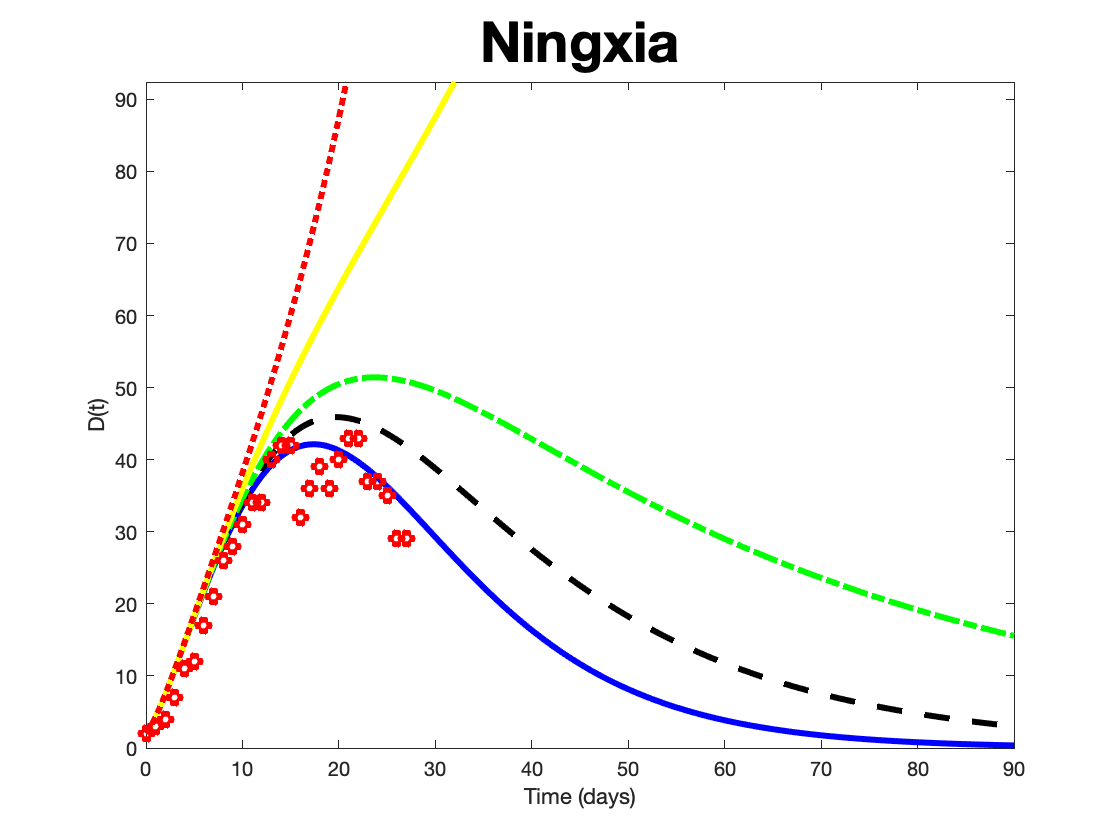}
%\caption{fig1}
\end{minipage}%
}%
%\hspace{1cm}
\subfigure{
\begin{minipage}[t]{0.25\linewidth}
\centering
\includegraphics[width=4cm]{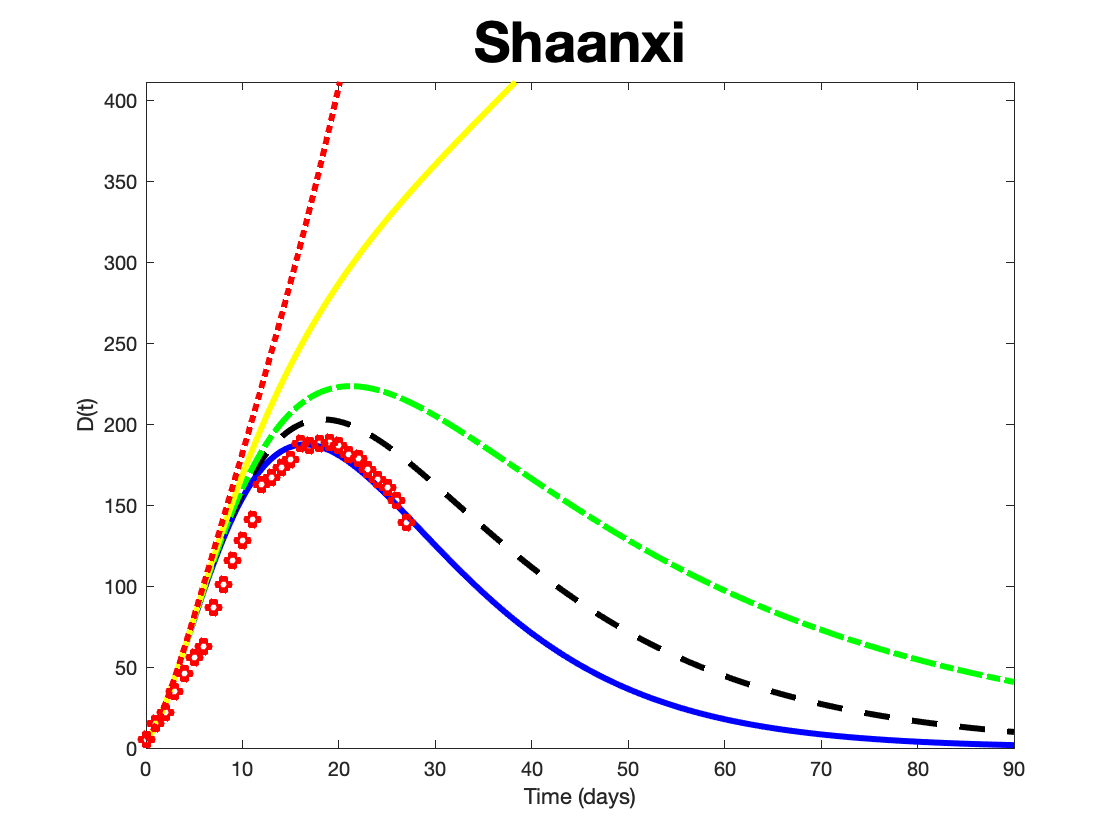}
%\caption{fig1}
\end{minipage}%
}%
%---------------------------------------------------------------
\vskip -8pt

\subfigure{
\begin{minipage}[t]{0.25\linewidth}
\centering
\includegraphics[width=4cm]{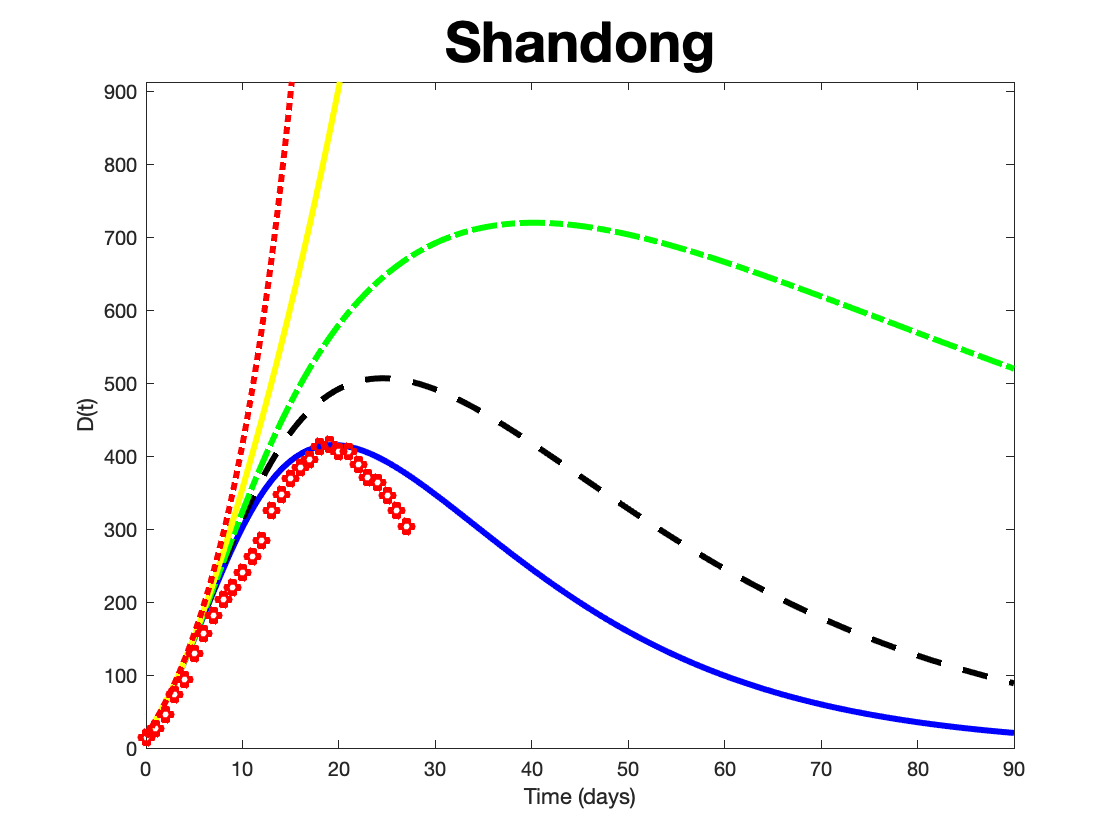}
%\caption{fig1}
\end{minipage}%
}%
\subfigure{
\begin{minipage}[t]{0.25\linewidth}
\centering
\includegraphics[width=4cm]{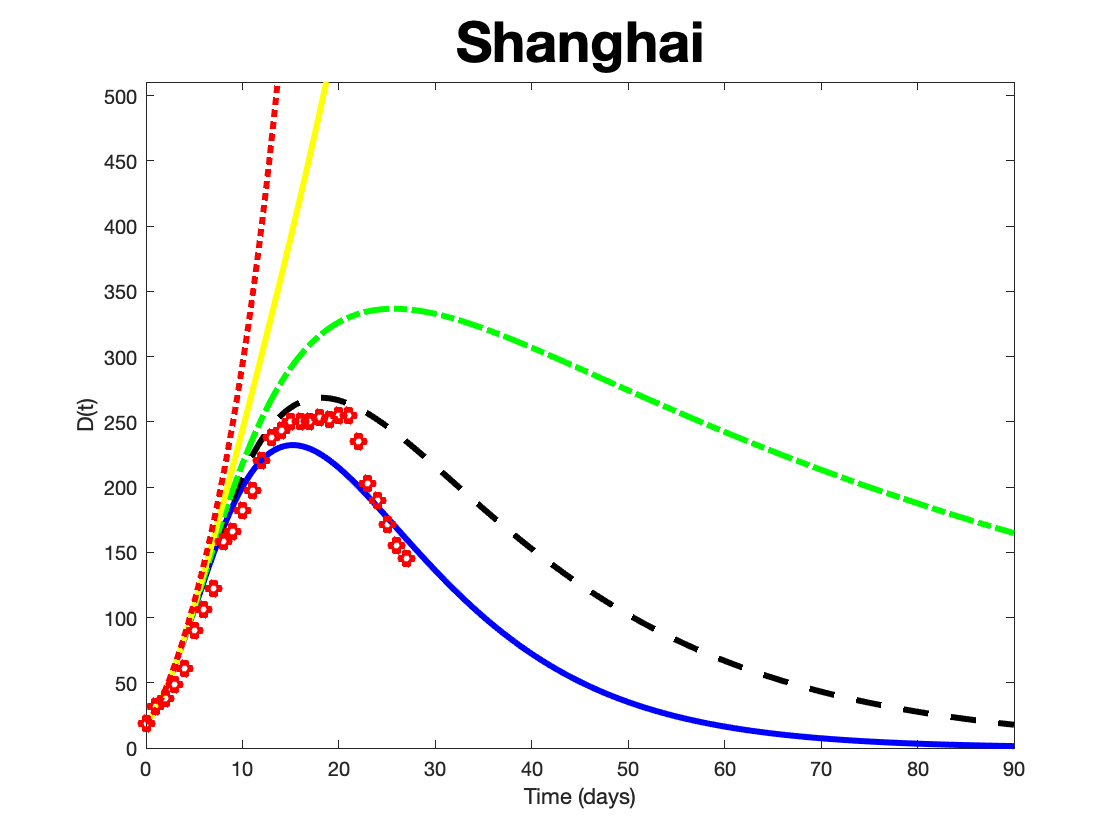}
%\caption{fig1}
\end{minipage}%
}%
%\hspace{1cm}
\subfigure{
\begin{minipage}[t]{0.25\linewidth}
\centering
\includegraphics[width=4cm]{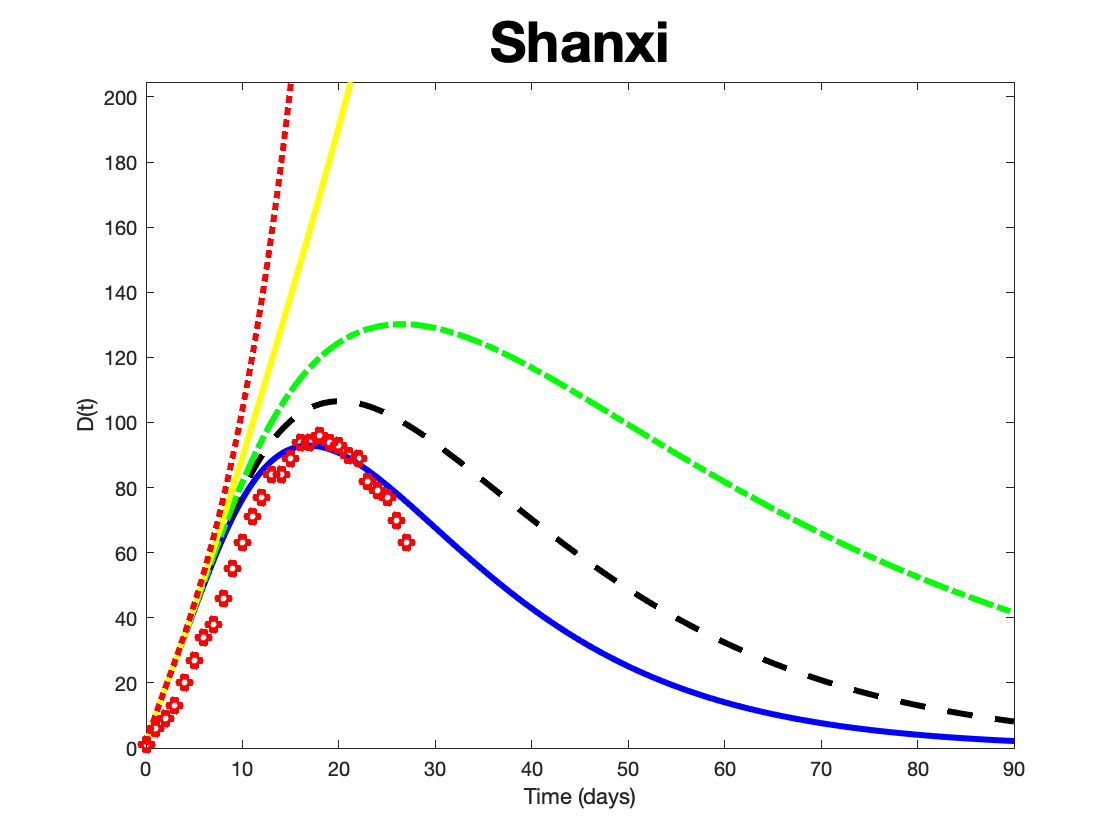}
%\caption{fig1}
\end{minipage}%
}%
\subfigure{
\begin{minipage}[t]{0.25\linewidth}
\centering
\includegraphics[width=4cm]{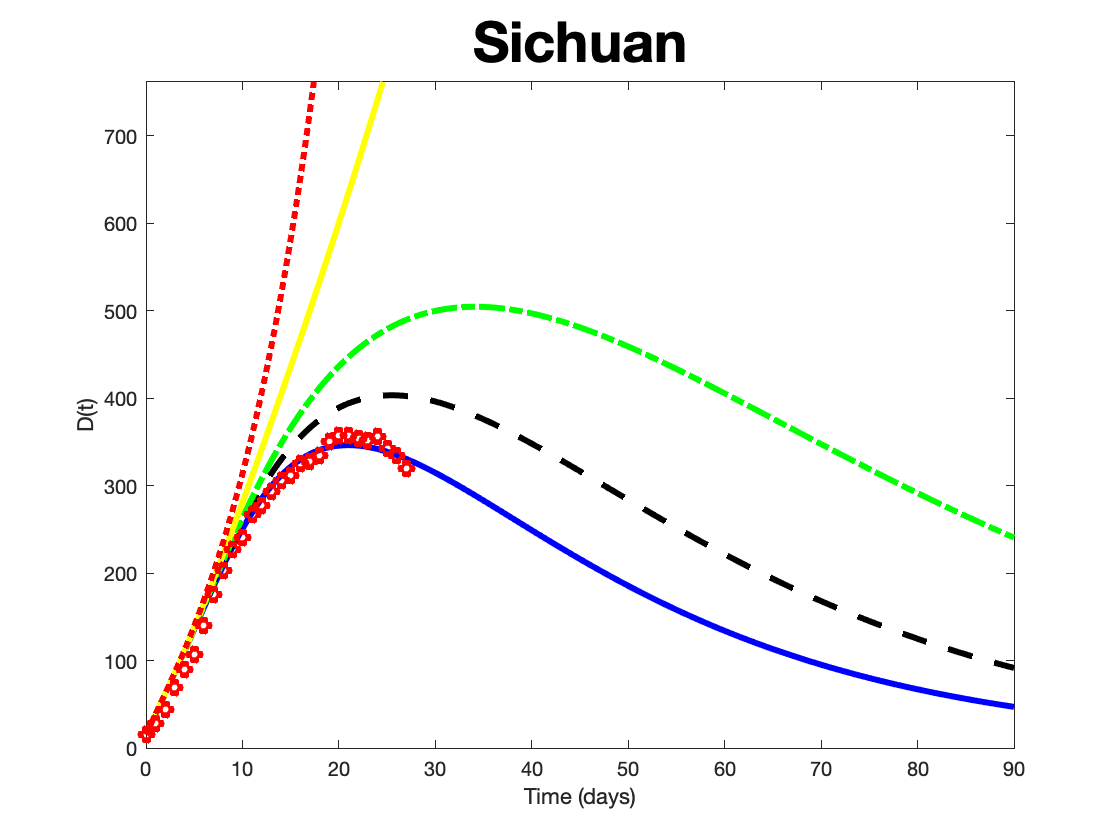}
%\caption{fig1}
\end{minipage}%
}%
%---------------------------------------------------------------
\vskip -8pt

\subfigure{
\begin{minipage}[t]{0.25\linewidth}
\centering
\includegraphics[width=4cm]{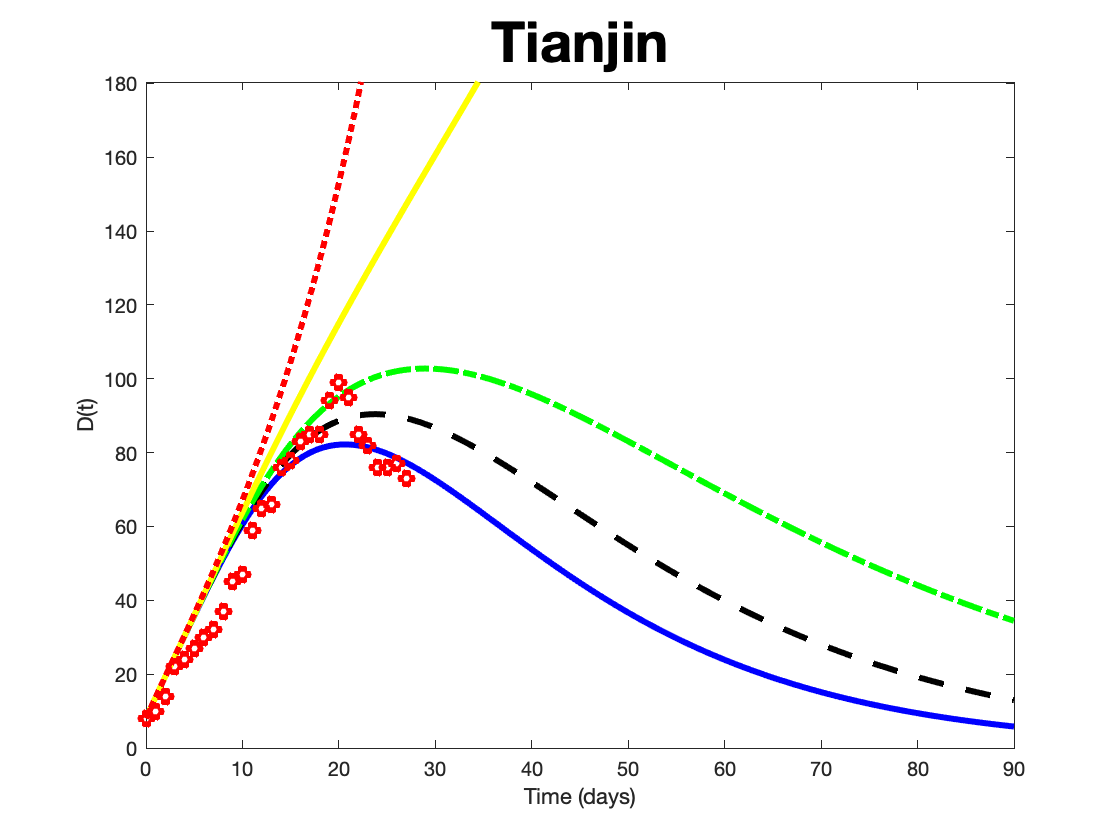}
%\caption{fig1}
\end{minipage}%
}%
\subfigure{
\begin{minipage}[t]{0.25\linewidth}
\centering
\includegraphics[width=4cm]{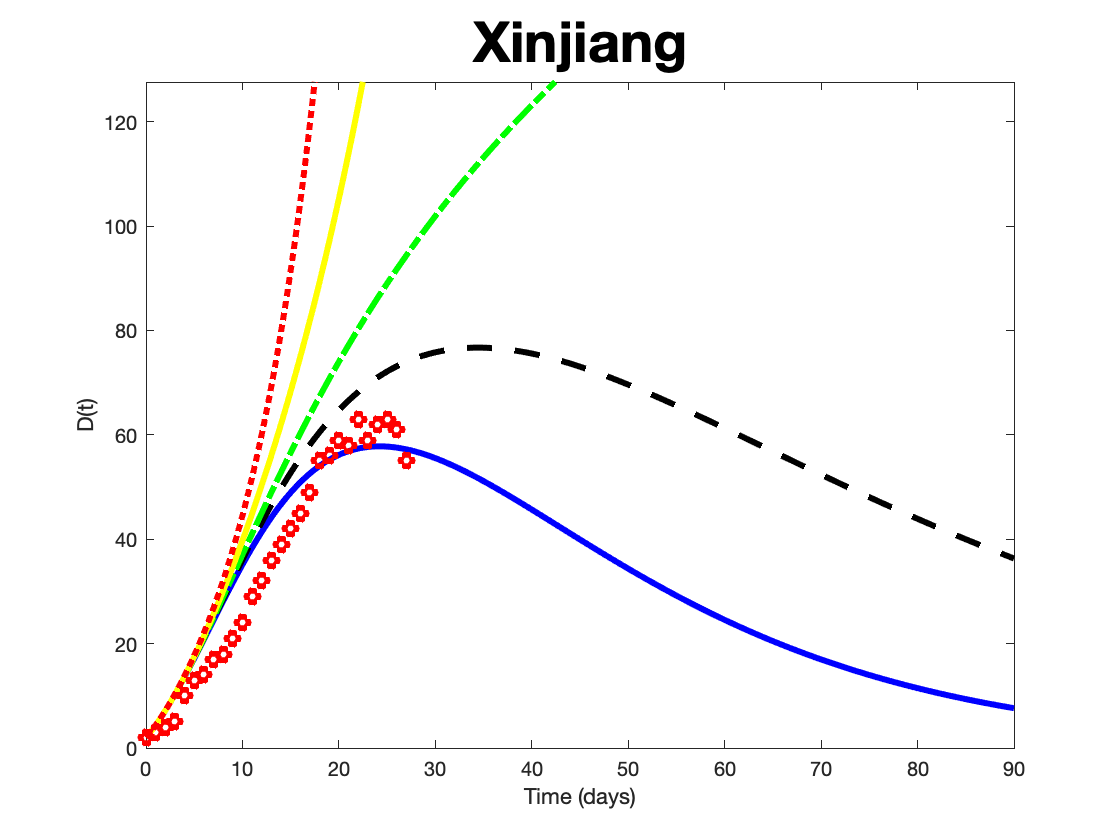}
%\caption{fig1}
\end{minipage}%
}%
%\hspace{1cm}
\subfigure{
\begin{minipage}[t]{0.25\linewidth}
\centering
\includegraphics[width=4cm]{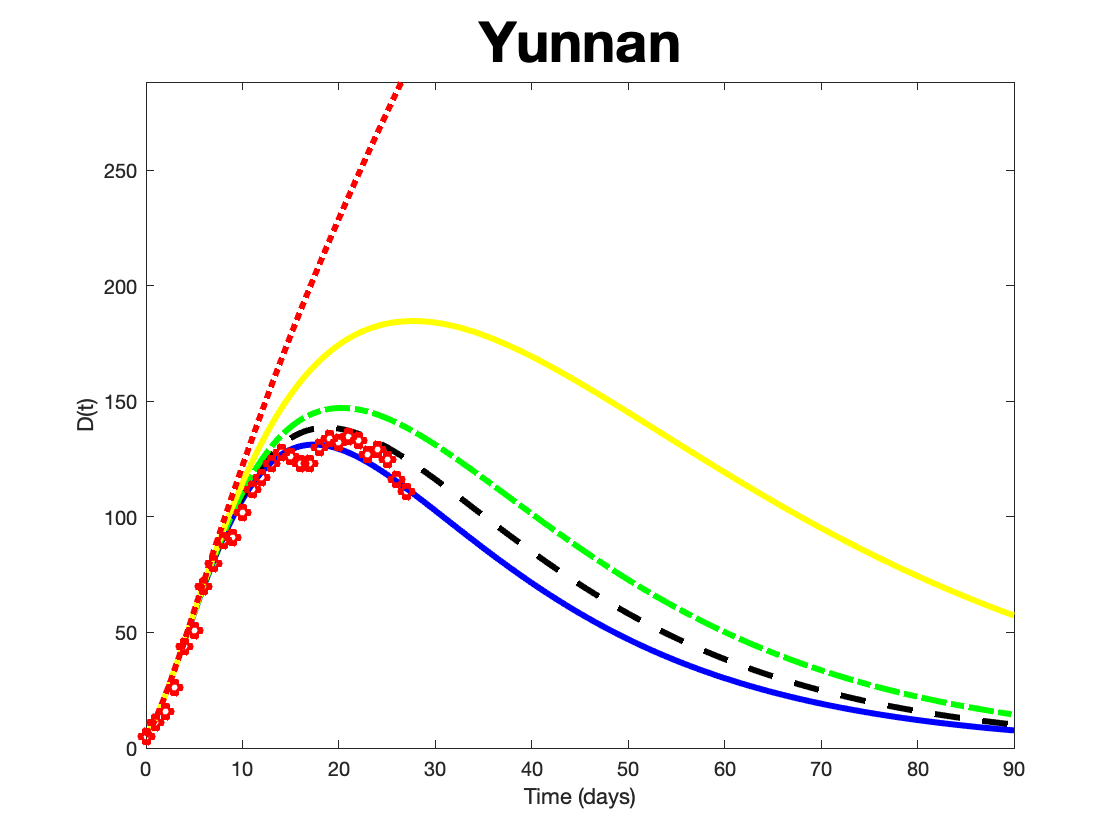}
%\caption{fig1}
\end{minipage}%
}%
\subfigure{
\begin{minipage}[t]{0.25\linewidth}
\centering
\includegraphics[width=4cm]{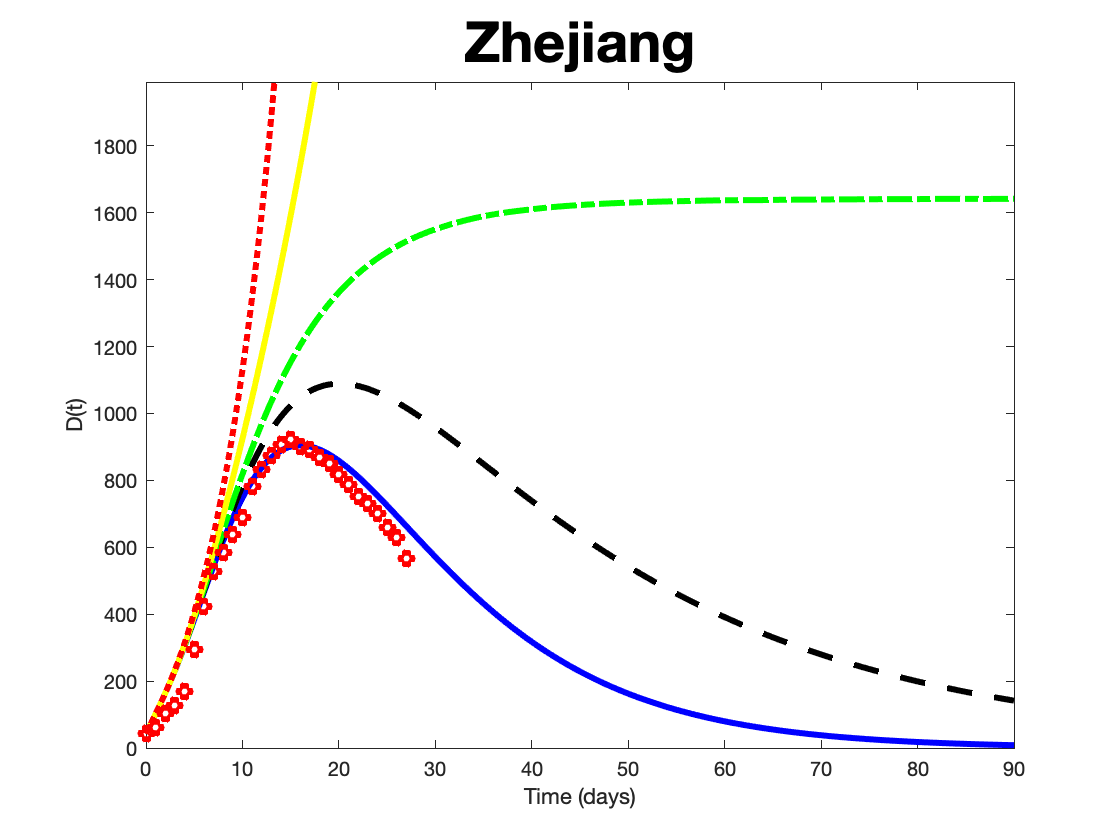}
%\caption{fig1}
\end{minipage}%
}%
%---------------------------------------------------------------
%\vskip -8pt
%
%\subfigure{
%\begin{minipage}[t]{0.25\linewidth}
%\centering
%\includegraphics[width=4cm]{hubei.png}
%%\caption{fig1}
%\end{minipage}%
%}%
\centering
\caption{Simulation results of $D(t)$ (Solid blue line: $\lambda = 1/60$, dashed dark line: $\lambda = 1/30$, dash-dot green line: $\lambda = 1/20$, solid yellow line: $\lambda = 1/10$, dotted red line: $\lambda = 1/5$) and published data $D_{pub}$ (Red asterisk).}\label{fig3}
\end{figure}

\begin{figure}[h!]
\centering
\includegraphics[width=15cm]{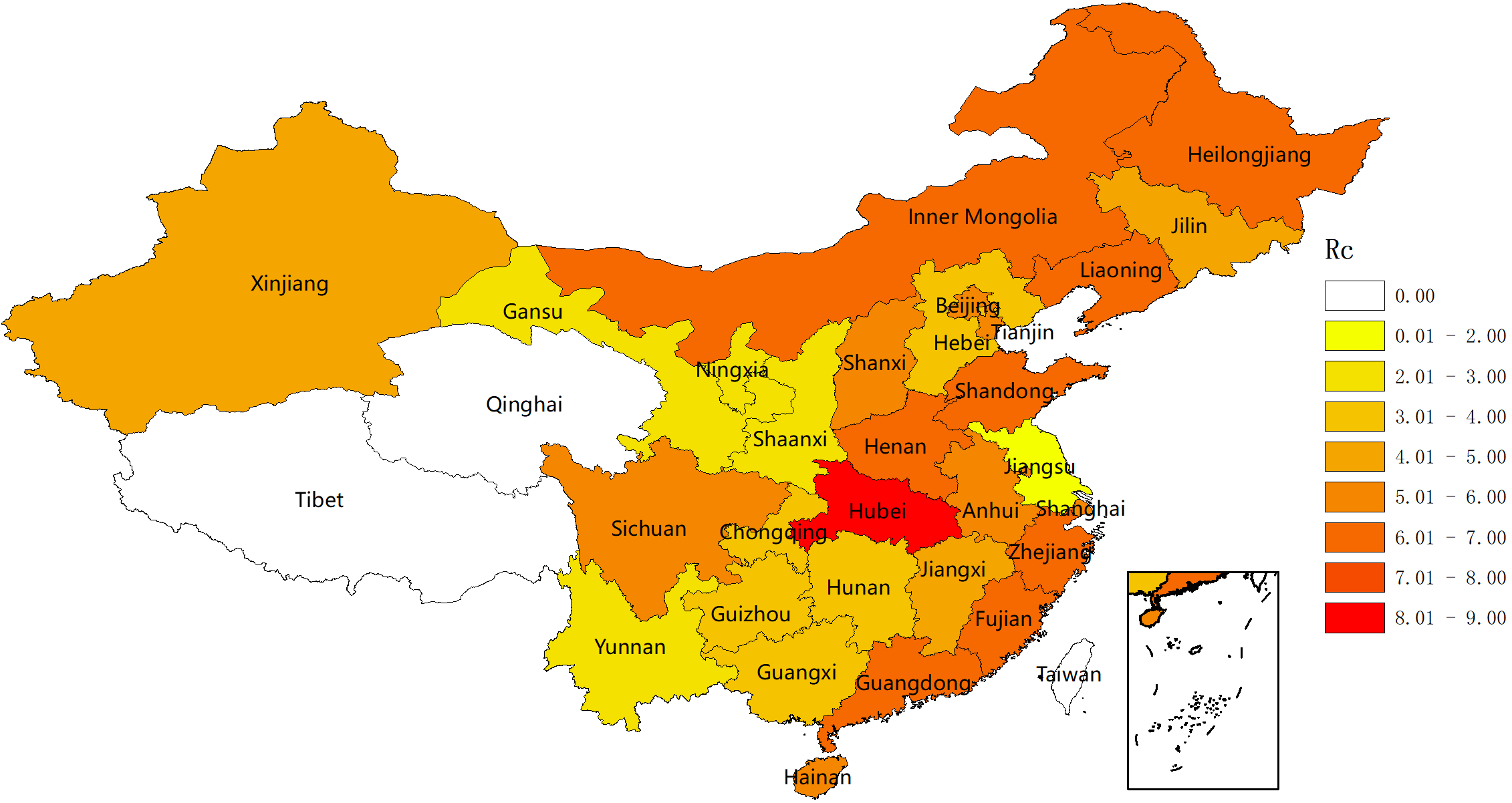}
\centering
\caption{ $\mathcal{R}^{0}_{c} $ distribution heat map}\label{fig_Rc_map}
\end{figure}

Above study gives an intuitionistic understanding of COVID-19 in China. According to the different characters in each province, we study the transmission and control strategy of COVID-19 more intensively. The simulation for other provinces are shown in Figure~\ref{fig3} with blue solid lines. Colorful dashed lines represent the trends of $D(t)$ under different quarantine period. All the parameters are shown in Appendix Table~\ref{table_para_I} and Table~\ref{table_para_II}. The peak value, peak time of $D(t)$ for each province under different $\lambda$ can be found in Appendix Table~\ref{table_peak_all}. Our model fit the published data accurately for most of the provinces. We estimate the $\mathcal{R}^{0}_{c} $ for each province and plot the distribution heat map in Figure~\ref{fig_Rc_map}, the initial situation for each province is serious. If the control strategy is insufficient, the disease will out of control in each province. Back to the strictest isolation strategy, up to February 19, 2020, the disease is in the decline phase for most provinces, it is effective for controlling the transmission of COVID-19. Inner Mongolia, Xinjiang and Heilongjiang should pay more attention, they are in the key period of the disease control. The home quarantine period in most province can be much shorter than Hubei and the disease can also be controlled well. From Figure~\ref{fig3}, we find that, the trends of $D(t)$ between 30 days and 60 days quarantine period are roughly the same, such as GanSu, Yunnan, Tianjin and so on. Considering the need of daily production, the workers of these provinces can return to work actively and orderly.

\begin{table}[htp!]
\centering
\caption{ Accumulated medical resource, 3A and designated hospital comparison}\label{table_AMR_3ch}
\resizebox{0.9\textwidth}{!}{
\begin{tabular}{c|cccccccc}
\hline\hline
Province & Anhui & Beijing & Chongqing & Fujian & Gansu & Guangdong & Guangxi & Guizhou \\
\hline
$AMR$ & 28759 & 10216 & 12705 & 8866 & 1639 & 32278 & 8093 & 4203 \\
$N_{3}$ & 20 & 30 & 11 & 24 & 12 & 66 & 25 & 23 \\
$N_{d}$ & 271 & 90 & 104 & 765 & 99 & 886 & 48 & 1143 \\
$MB_{3}$ & 1438 & 341 & 1155 & 369 & 137 & 489 & 324 & 183 \\
$MB_{W}$ & 92	&	68	&	101	&	11	&	13	&	32	&	83	&	4 \\
\hline
Province & Hainan & Hebei & Heilongjiang & Henan & Hubei & Hunan & Inner Mongolia & Jiangsu \\
\hline
$AMR$ & 4078 & 7495 & 16502 & 30486 & 2649863 & 21867 & 3167 & 20035 \\
$N_{3}$ & 5 & 32 & 31 & 24 & 36 & 20 & 13 & 38 \\
$N_{d}$ & 317 & 394 & 493 & 508 & 678 & 325 & 213 & 769 \\
$MB_{3}$ & 816 & 234 & 532 & 1270 & 73607 & 1093 & 244 & 527 \\
$MB_{W}$ & 12	&	16	&	30	&	55	&	3533	&	60	&	13	&	24\\
\hline
Province & Jiangxi & Jilin & Liaoning & Ningxia & Shaanxi & Shandong & Shanghai & Shanxi \\
\hline
$AMR$ & 24493 & 2086 & 3336 & 1384 & 6093 & 16992 & 7036 & 3371 \\
$N_{3}$ & 33 & 20 & 36 & 3 & 25 & 21 & 24 & 32 \\
$N_{d}$ & 307 & 133 & 228 & 73 & 354 & 541 & 29 & 205 \\
$MB_{3}$ & 742 & 104 & 93 & 461 & 244 & 809 & 293 & 105 \\
$MB_{W}$ & 66	&	12	&	11	&	18	&	15	&	29	&	91	&	13\\
\hline
Province & Sichuan & Tianjin & Xinjiang & Yunnan & Zhejiang &  &  &  \\
\hline
$AMR$ & 16921 & 3660 & 2868 & 5330 & 28601 &  &  &  \\
$N_{3}$ & 36 & 17 & 9 & 5 & 26 &  &  &  \\
$N_{d}$ & 1910 & 27 & 176 & 350 & 323 &  &  &  \\
$MB_{3}$ & 470 & 215 & 319 & 1066 & 1100 &  &  &  \\
$MB_{W}$ & 9	&	60	&	15	&	15	&	76 &  &  &  \\
\hline\hline
\end{tabular}}
\end{table}
\newpage

The medical resources in each province are distributed unevenly. After the disease outbreak, the government deployed  medical resources to rescue Hubei Province. We estimate the $AMR$ needed for each provinces by integrating $D(t)$ from $0$ (Jan 23) to $90$ (Apr 22) days under different quarantine period ($5-60$ days). The detail parameter values and corresponding $AMR$ in 90 days are shown in Table~\ref{table_peak_all}. We collect the number of 3A hospital  in each province, which can reflect the level of medical resources. In Table \ref{table_AMR_3ch}, we show the 3A hospital ($N_{3}$) number in each province, the medical burden ($MB$) is measured by
\begin{equation}\nonumber
MB_{3}=\dfrac{{AMR}}{N_{3}}.
\end{equation}
The value of $AMR$ is taken as the optimal simulation results, i.e., $\lambda=1/60$. The median of $MB_{3}$ among 29 provinces is 461, then for which province the $MB_{3}$ Index is more than $461$, it should be paid more attention. However, after the disease outbreak, The government established many designated hospitals and fever clinics to relieve the stress of medical resources. Denote $N_{d}$ as the number of designated hospital, we set a weight between the 3A hospital and designated hospital. The weighted medical burden is described as
\begin{equation}\nonumber
MB_{W}=\dfrac{{AMR}}{{3*N_{3}+(N_{d}-N_{3})}}.
\end{equation}

Comparing the value of $MB_{3}$ with $MB_{W}$ in each province, we can clearly see that, thanks to the adjoint of designated hospital, it has decreased the medical burden sharply. It is benefit for early diagnosis, early isolation and better treatment. But the disease burden in Hubei is still huge, it is almost 60 times greater than other province. On one hand, we have taken full advantage of current medical resources; on the other hand, the additional medical workers and materials from other province are supported to Hubei Province spare no effort. Sufficient and effective medical support is the basic foundation to control COVID-19.

%---------------------------------------------------------------
% provinces
%---------------------------------------------------------------

%---------------------------------------------------------------

%\begin{table}[htbp]
%\centering
  %  \resizebox{\textwidth}{!}{
%	\begin{tabular}{c|ccccccc}
%\hline\hline
%		&	Anhui	&	Beijing	&	Chongqing	&	Fujian	&	Gansu	&	Guangdong	&	Guangxi	\\
%\hline	$R_{c}$	&	5.09	&	5.15	&	3.09 	&	6.49 	&	3.00 	&	6.31	&	3.72	\\
%\hline
%\hline
%		&	Guizhou	&	Hainan	&	Hebei	&	Heilongjiang	&	Henan	&	Hunan	&	Inner Mongolia	\\

%\hline	$R_{c}$	&	3.92	&	5.06	&	3.88	&	6.23 	&	6.62	&	3.50 	&	6.57	\\
%\hline\hline
%		&	Jiangsu	&	Jiangxi	&	Jilin	&	Liaoning	&	Ningxia	&	Shaanxi	&	Shandong	\\

%\hline	$R_{c}$	&	1.90	&	4.32	&	4.21	&	6.73	&	2.44	&	2.24 	&	6.25	\\
%\hline\hline	
	%	&	Shanghai	&	Shanxi	&	Sichuan	&	Tianjin	&	Xinjiang	&	Yunnan	&	Zhejiang	\\
	
%\hline	$R_{c}$	&	4.39	&	5.41	&	5.15	&	5.29	&	4.25	&	2.62	&	6.98	\\

%\hline\hline
%\end{tabular}}
%\caption{$\mathcal{R}^{0}_{c} $ estimation}\label{table_Rc}
%\end{table}

%----------------------------

\section{Further trends for the control of COVID-19}
 In previous section, the transmission of COVID-19 inside and outside Hubei is discussed.  Simulations for most provinces can conduce to understand the effect of control strategy in China. And the corresponding analysis for medical resources are given. Isolation strategy plays an important role in the prevention of disease spreading. But the strictest isolation strategy brings great inconvenient to people's daily life. Now the disease spread is in decline, more and more people will return to their normal life paths. In this section, we will explore further control strategy from different perspectives for the following control phase.

\subsection{The impact of Meteorological Index ($MeI$)}

Seventeen years ago, SARS outbroke in China, it spread quickly and disappeared suddenly. There is no specific medicine and vaccine, the medical level and control strategy are not as complete as today. Meteorological factor is regarded as a critical role for the vanishing of SARS \cite{tianqi1,tianqi2}. Due to the high genetic homology of SARS-CoV and 2019-nCoV, we want to clarify the meteorological influence in the spreading of COVID-19. We collect the official published the average meteorological data during the simulation period, including air index $(AI)$, temperature $(TE)$,  precipitation $(PR)$, relative humidity $(RH)$ and wind power $(WP)$ from China Meteorological Data Service Center (CMDC) website \cite{CMDCweb}  for correlation analysis.

\begin{figure}[h!]
\centering
\subfigure[Group I. (Left, $\mathcal{R}^{0}_{c} $, $r = 0.69, p = 0.0015$; Right, $\beta$, $r = 0.70, p = 0.0012$.)]{
\begin{minipage}[t]{0.9\linewidth}
\centering
\includegraphics[height=7cm]{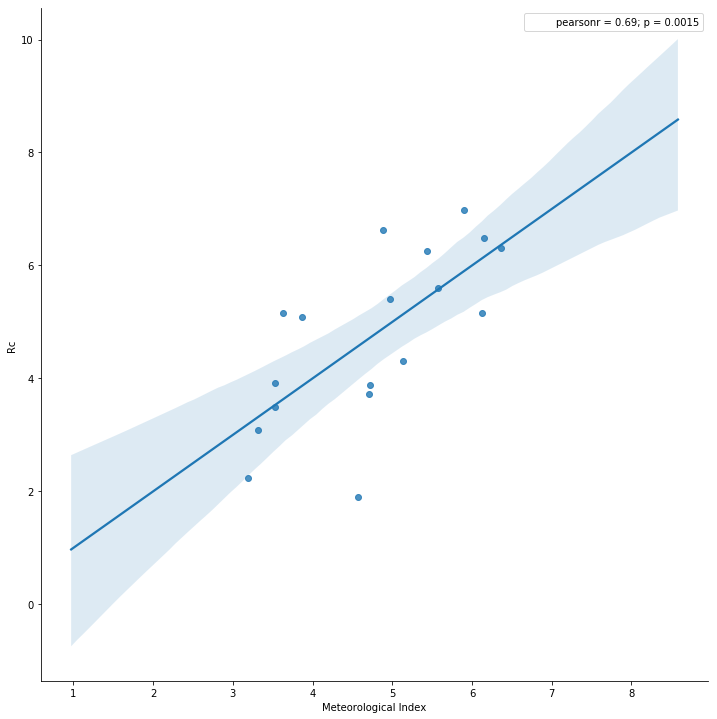}
\includegraphics[height=7cm]{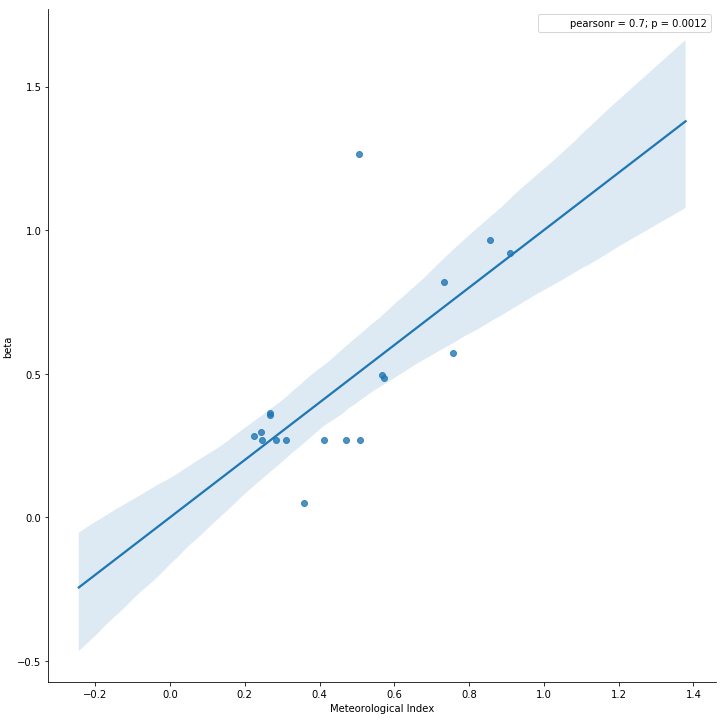}
%\caption{fig2}
\end{minipage}%
}%
%\hspace{5mm}

\subfigure[Group II. (Left, $\mathcal{R}^{0}_{c} $, $r = 0.84, p = 0.0021$; Right, $\beta$, $r = 0.63, p = 0.049$.)]{
\begin{minipage}[t]{0.9\linewidth}
\centering
\includegraphics[height=7cm]{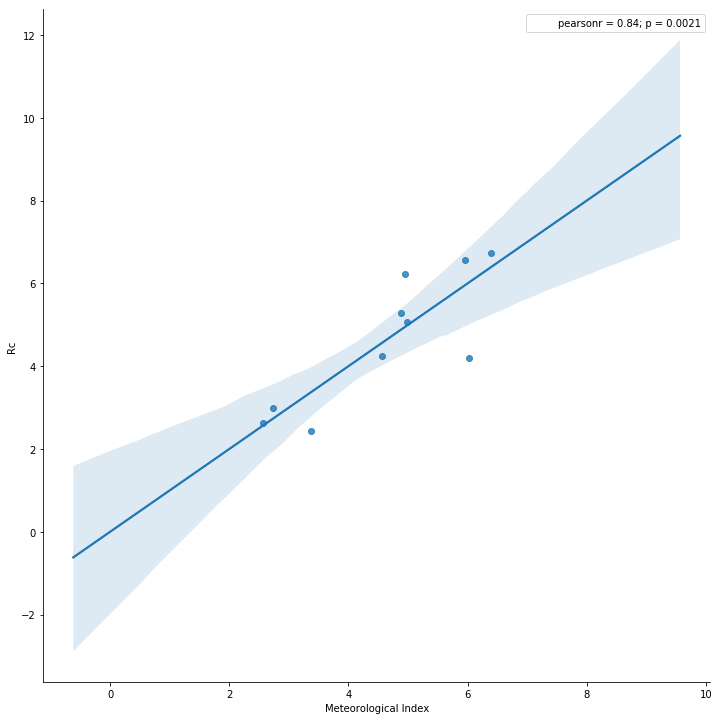}
\includegraphics[height=7cm]{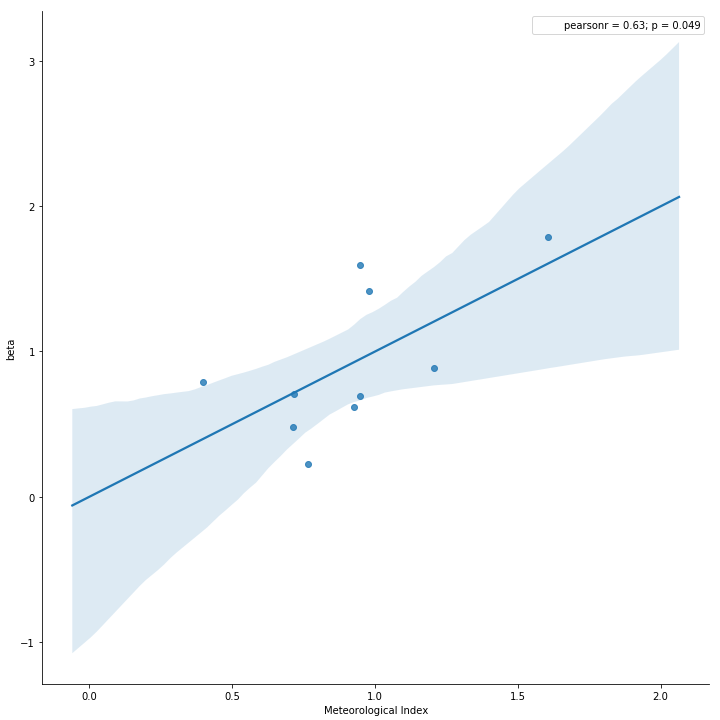}
%\caption{fig1}
\end{minipage}%
}%
\centering
\caption{ Correlation analysis of $\mathcal{R}^{0}_{c} $, $\beta$ and $MeI$.}\label{fig_Rc_beta_MI}
\end{figure}

The outbreak of COVID-19 coincided with Chinese New Year, the population migration is busy at that time. Wuhan is the towngate of nine provinces, a large number of people moved out or passed by Wuhan. Population migration is a key factor in the spread process and this factor cannot be ignored in the analysis. We define an index $MiI$ for each provinces to reflect the population migrate from Hubei Province, the geographical distance from Wuhan is also considered.

$$
MiI = C\dfrac{PE}{DIS^{2}},
$$
where $PE$ is the percentage of population moving in from Hubei, $DIS$ is the distance between the capital of destination province and Wuhan, $C=10^{10}$ is an adjustment constant. The parameter $PE$ and $DIS$ are from Baidu Migration \cite{baiduqianxi} and Baidu Map \cite{baidumap}.

Based on the descending order of $MiI$ for provinces outside Hubei (see Table~\ref{table_dis} for details),  we separate them into two groups to represent different migration levels. We set critical value as $MiI_{y}=100$. If the value of $MiI$ is greater than $MiI_{y}$, it is in the high level group called Group~I, others provinces are in low level group called Group~II. From Table \ref{table_dis}, the high $MiI$ areas are almost the province surround Hubei Province.

\begin{table}[h!]
\caption{Migration Index Computation}\label{table_dis}
\centering
\resizebox{\textwidth}{!}{
\begin{tabular}{c|ccccccc}
%\begin{tabularx}{16cm}{p{0.6cm}<{\centering}|p{1.8cm}<{\centering}p{1.8cm}<{\centering}p{1.9cm}<{\centering}p{1.8cm}<{\centering}p{1.8cm}<{\centering}p{1.8cm}<{\centering}p{1.8cm}<{\centering}}
\hline
\hline
Province	&	Hunan	&	Jiangxi	&	Henan	&	Anhui	&	Jiangsu	&	Chongqing	&	Guangdong	\\
\hline
$PE$	&	15.65\%	&	7.75\%	&	18.66\%	&	6.58\%	&	3.86\%	&	8.23\%	&	7.27\%	\\
$DIS$	&	284.4	&	254.5	&	468.2	&	304.3	&	452.3	&	759.1	&	835.9	\\
$MiI$	&	19349	&	11965	&	8512	&	7106	&	1887	&	1428	&	1041	\\
\hline
Province	&	Zhejiang	&	Shaanxi	&	Shandong	&	Fujian	&	Sichuan	&	Hebei	&	Shanghai	\\
\hline
$PE$	&	3.1\%	&	3.59\%	&	2.74\%	&	2.47\%	&	4.49\%	&	1.84\%	&	1.13\%	\\
$DIS$	&	558.3	&	662.5	&	726.1	&	690.3	&	985.4	&	836.6	&	684.7	\\
$MiI$	&	995	&	818	&	520	&	518	&	462	&	320	&	241	\\
\hline
Province	&	Guizhou	&	Shanxi	&	Guangxi	&	Beijing	&	Yunnan	&	Gansu	&	Hainan	\\
\hline
$PE$	&	1.68\%	&	1.25\%	&	1.98\%	&	1.44\%	&	1.24\%	&	0.84\%	&	0.86\%	\\
$DIS$	&	869.5	&	825.8	&	1048.1	&	1054.7	&	1295.3	&	1155	&	1242.1	\\
$MiI$	&	222	&	183	&	180	&	130	&74&	63	&	56	\\
\hline
Province	&	Liaoning	&	Tianjin	&	InnerMongolia	&	Heilongjiang	&	Jilin	&	Ningxia	&	Xinjiang	\\
\hline
$PE$	&	0.67\%	&	0.28\%	&	0.32\%	&	0.52\%	&	0.36\%	&	0.11\%	&	0.26\%	\\
$DIS$	&	1480.4	&	976.6	&	1161.4	&	2000.2	&	1760.4	&	1145.6	&	2770.4	\\
$MiI$	&	31 &	29	&	24	&	13	&	12	&	8	&	3	\\
\hline
\hline
\end{tabular}}
\end{table}

Based on above grouping criteria, we use the following formula and linear regression procedure to calculate a comprehensive meteorological index $MeI$ and apply the correlation analysis to $\mathcal{R}^{0}_{c} , \beta$ and $MeI$ for each group.
\begin{equation}
MeI =  c_{1}\ln{(AI)} + c_{2}TE + c_{3}PR^{2} + c_{4}RH + c_{5}WP + c_{6}.
\end{equation}
The coefficients and intercepts are listed in Table~\ref{table_lr}. For each group, we calculate $MeI_{\mathcal{R}^{0}_{c} } $ and $MeI_{\beta}$ and perform the correlation analysis with $\mathcal{R}^{0}_{c} $ and $\beta$, respectively. The results are presented in Figure~\ref{fig_Rc_beta_MI}, it shows that $\mathcal{R}^{0}_{c} $ and $\beta$ are significantly associated with $MeI$.
%We plot the heat map of $\mathcal{R}^{0}_{c} $ in Figure~\ref{fig_Rc_beta_MI}.

%MI_Rc_g1
%'空气质量指数的对数','平均温度','平均降水量的平方','平均相对湿度','风力']
%[-1.99870830e+00  1.19125639e-01  1.63122398e-04 -1.38258705e-01 -3.77764827e-02] 21.316708877160814
%[-2.05977488e+00  1.19290870e-01  1.68842767e-04 -1.43886187e-01 1.84268575e-01] 21.57572433288313

%MI_beta_g1
%[-6.49653164e-08 -2.99572293e-09  3.79725972e-12 -1.67574464e-09 -2.77071869e-08]  4.957934716015854e-07
%[-6.66713636e-08 -2.99110682e-09  3.95707252e-12 -1.83296239e-09 -2.15038036e-08]  5.030297153574646e-07

%MI_Rc_g2
%[ 1.88016164 -0.17992474  0.01322516 -0.13903845  1.16559193] -0.32147814593176616
%[ 1.88016164 -0.17992474  0.01322516 -0.13903845  1.16559193] -0.32147814593176616

%MI_beta_g2
%[ 2.58544311e-08  1.87778667e-10  2.68979834e-10 -1.98361985e-09 -2.18822404e-08] 9.746778571551124e-08
%[ 2.58544311e-08  1.87778667e-10  2.68979834e-10 -1.98361985e-09 -2.18822404e-08] 9.746778571551124e-08

\begin{table}[h!]
\caption{Linear regression coefficients and intercept for $\mathcal{R}^{0}_{c} , \beta$ and $MeI$.}\label{table_lr}
\centering
\begin{tabular}{c|cccccc}
\hline
\hline
 & $c_{1}$ & $c_{2}$ & $c_{3}$ & $c_{4}$ & $c_{5}$ & $c_{6}$ \\
\hline
\multicolumn{7}{c}{Group I}\\
\hline
$MeI_{\mathcal{R}^{0}_{c} } $  &  $ -2.0598$ & $ 0.1193$ &  $1.6884\times10^{-4}$ & $ -0.1439$ & $0.1843 $ & 21.5757 \\
$MeI_{\beta} (\times10^{-8})$ & $-6.6671$  &  $  -0.2991 $  &  $ 3.9571\times10^{-4} $  &  $  -0.1833  $  &  $ -2.1504$  &  $  50.3030$\\
\hline
\hline
\multicolumn{7}{c}{Group II}\\
\hline
$MeI_{\mathcal{R}^{0}_{c} } $  &  $  1.8802 $  &  $  -0.1799$  &  $   0.0132$  &  $  -0.1390 $  &  $   1.1656 $  &  $  -0.3215 $ \\
$MeI_{\beta} (\times10^{-8})$ & $ 2.5854  $ & $ 0.0188$ & $ 0.0269$ & $ -0.1984 $ & $ -2.1882  $ & $  9.7468$\\
\hline
\hline
\end{tabular}
\end{table}

%\begin{figure}[htp!]
%\begin{center}
%\subfigure{
%  % Requires \usepackage{graphicx}
%  % replace aims_logo.eps by your figure file name
%  \includegraphics[width=0.28\textheight]{MI_Rc_g1.png}
%  \includegraphics[width=0.285\textheight]{MI_beta_g1.png}}\\
%\hspace{0.01in}
%\subfigure{
%  % Requires \usepackage{graphicx}
%  % replace aims_logo.eps by your figure file name
%  {\includegraphics[width=0.28\textheight]{MI_Rc_g2.png}}
%  \includegraphics[width=0.28\textheight]{MI_beta_g2.png}}
%  \caption{Linear regression of $\mathcal{R}^{0}_{c} $, $\beta$ and $MI$. Group I. (Left, $\mathcal{R}^{0}_{c} $, $r = 0.66, p = 0.0027$; Right, $\beta$, $r = 0.80, p = 5.8\times10^{-5}$.) Group II. (Left, $\mathcal{R}^{0}_{c} $, $r = 0.84, p = 0.0021$; Right, $\beta$, $r = 0.63, p = 0.049$.)}\label{fig_Rc_beta_MI}
%  \end{center}
%\end{figure}

The coefficients in Table \ref{table_lr} remind us that the effort of meteorology factors on the disease spread is quite different between the two groups. For example, air index is the most important meteorology factor in our study, in high $MiI$ group, $\ln(AI)$ is inversely proportional to $MeI_{{\mathcal{R}^{0}_{c} }}$, however, in low $MiI$ group, the result is inverse. The reason is that, the fraction of imported cases is high in high $MiI$ group, if the value of $AI$ is small, it represents good air condition, the social activities will increase. In this case, the probability of  contact with infected people is increased. In low $MiI$ group, the bad air condition will aggravate the spread. The impact of wind power in both group has similar pattern. Bad air condition and strong wind suggests that, people should pay more attention to the personal protection. In high $MiI$ group, higher temperature will cause undesirable impact on the control of disease. However, in low $MiI$ group, the result shows that higher temperature will reduce the spread. Precipitation shows low influence on COVID-19 spread. Notice that, the results in both groups show that higher relative humidity is the protection factor for the disease control.

\subsection{Further control with new vaccine}

On February 22, 2020, Zhejiang Government reported some progress on the vaccine of 2019-nCoV. It is said that the first vaccine has produced antibodies and the process is in the animal experiment \cite{zhejiangvaccine}. The director-general of WHO said that there are more than 20 COVID-19 vaccines candidates are currently in development phase and some new treatment are in clinical trials, the results will be expected within a few weeks \cite{tandesai}. The progress of the development of vaccines is much quicker than expected. If the new vaccine of COVID-19 comes into service, it will be great benefit for the disease control. Considering the strictest isolation strategy, from the mathematical point of view, the process of the strategy is just like a short-time vaccine for susceptible population. The effect of stay away from the source of infection is equal to contact with infected people but doesn't get infected. After a little modification on model (\ref{eq1}), it can describe the impact of the vaccine, by changing the quarantine compartment $Q$ into the vaccine compartment $V$ and setting parameter $p$ as the vaccination rate, $1/\lambda$ as the mean protection period of the new vaccine.

\begin{figure}[htp!]
\centering
\subfigure[]{
\begin{minipage}[t]{0.45\linewidth}
\centering
\includegraphics[width=\linewidth]{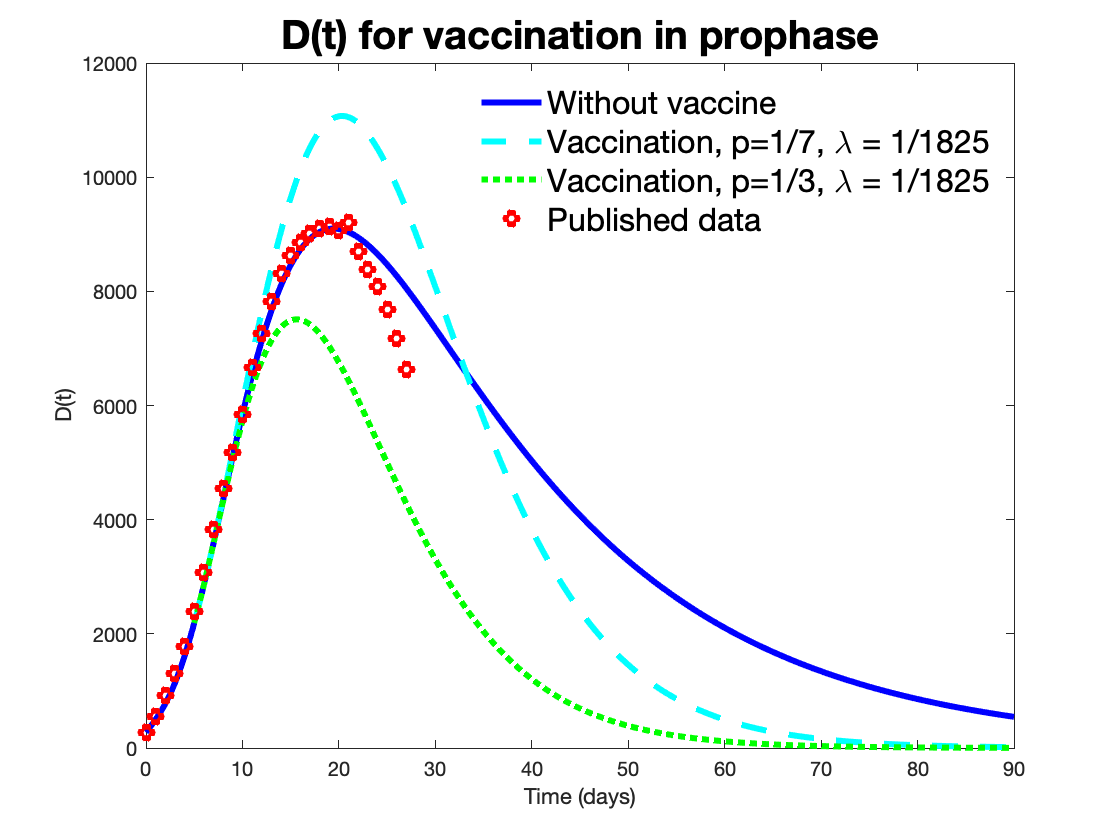}
\end{minipage}}
\subfigure[]{
\begin{minipage}[t]{0.45\linewidth}
\centering
\includegraphics[width=\linewidth]{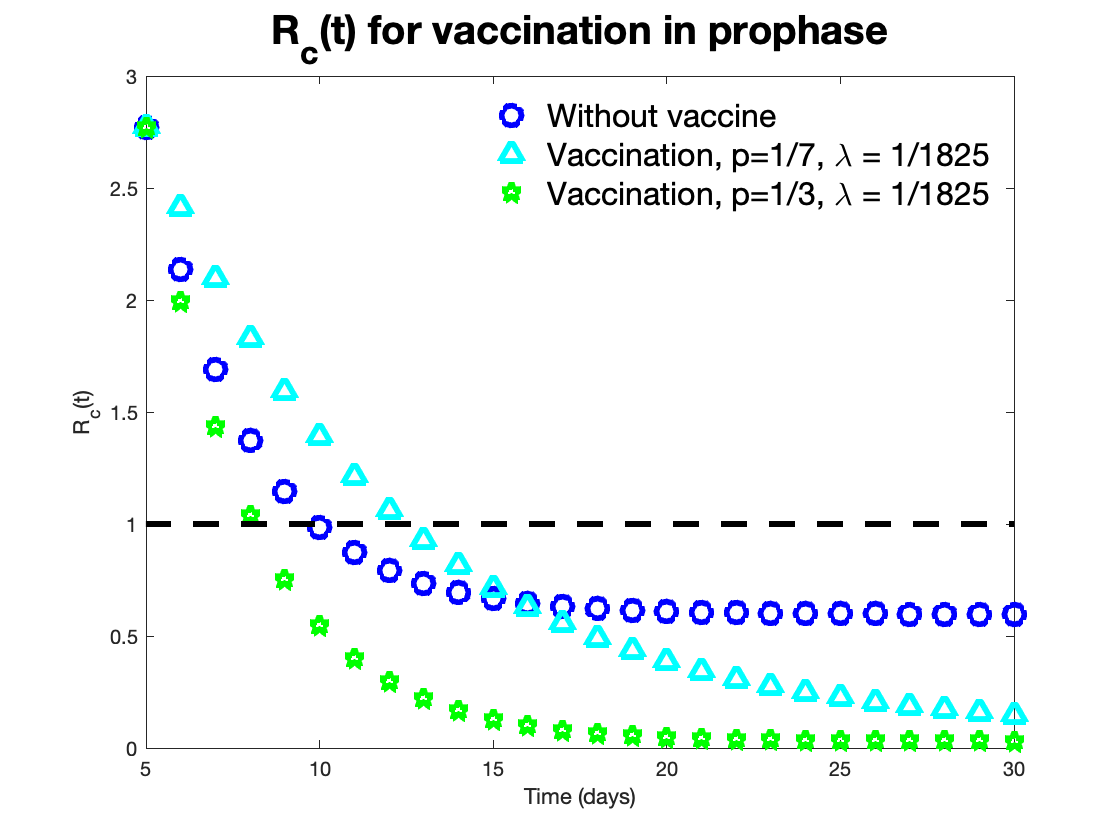}
\end{minipage}}

\subfigure[]{
\begin{minipage}[t]{0.45\linewidth}
\centering
\includegraphics[width=\linewidth]{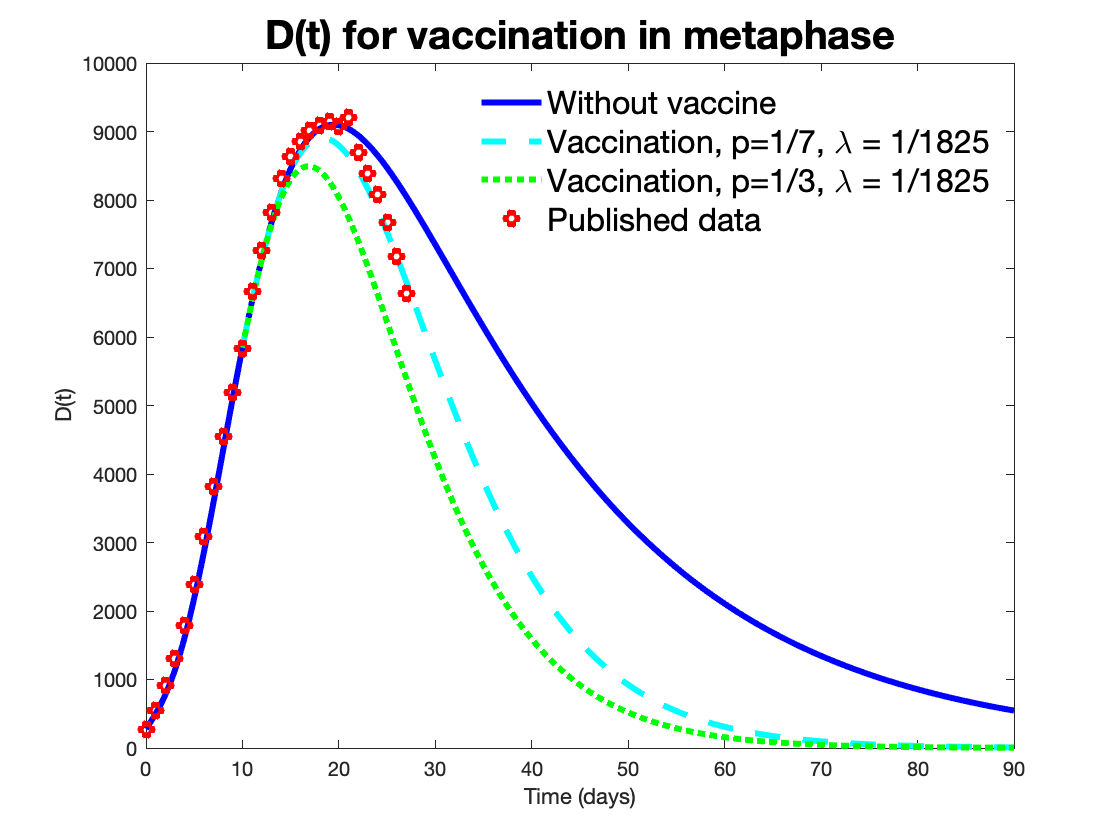}
\end{minipage}}
\subfigure[]{
\begin{minipage}[t]{0.45\linewidth}
\centering
\includegraphics[width=\linewidth]{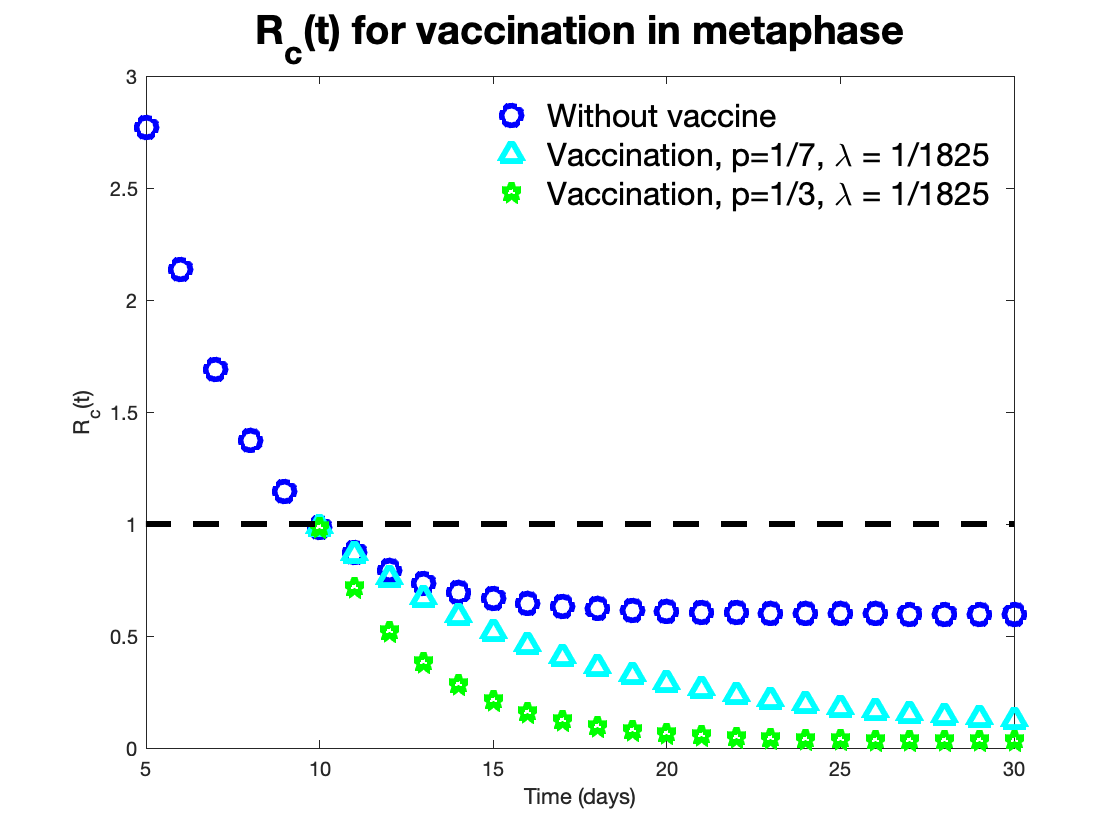}
\end{minipage}}

\subfigure[]{
\begin{minipage}[t]{0.45\linewidth}
\centering
\includegraphics[width=\linewidth]{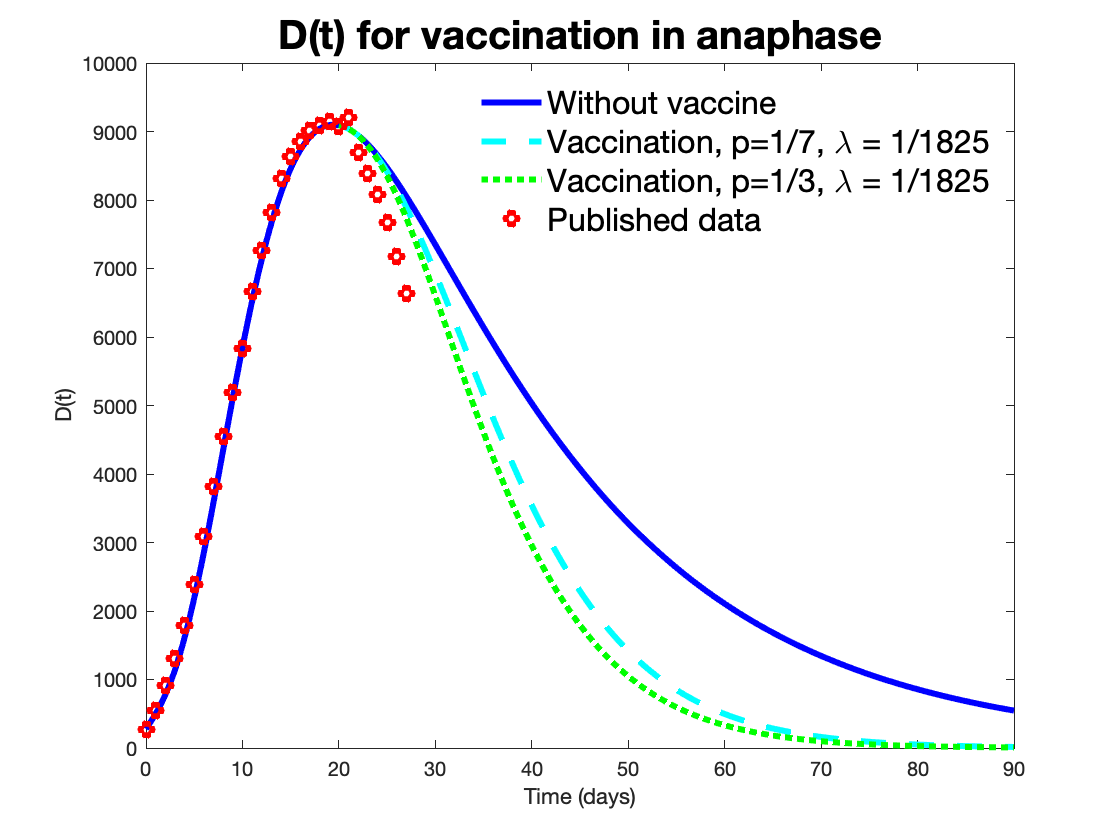}
\end{minipage}}
\subfigure[]{
\begin{minipage}[t]{0.45\linewidth}
\centering
\includegraphics[width=\linewidth]{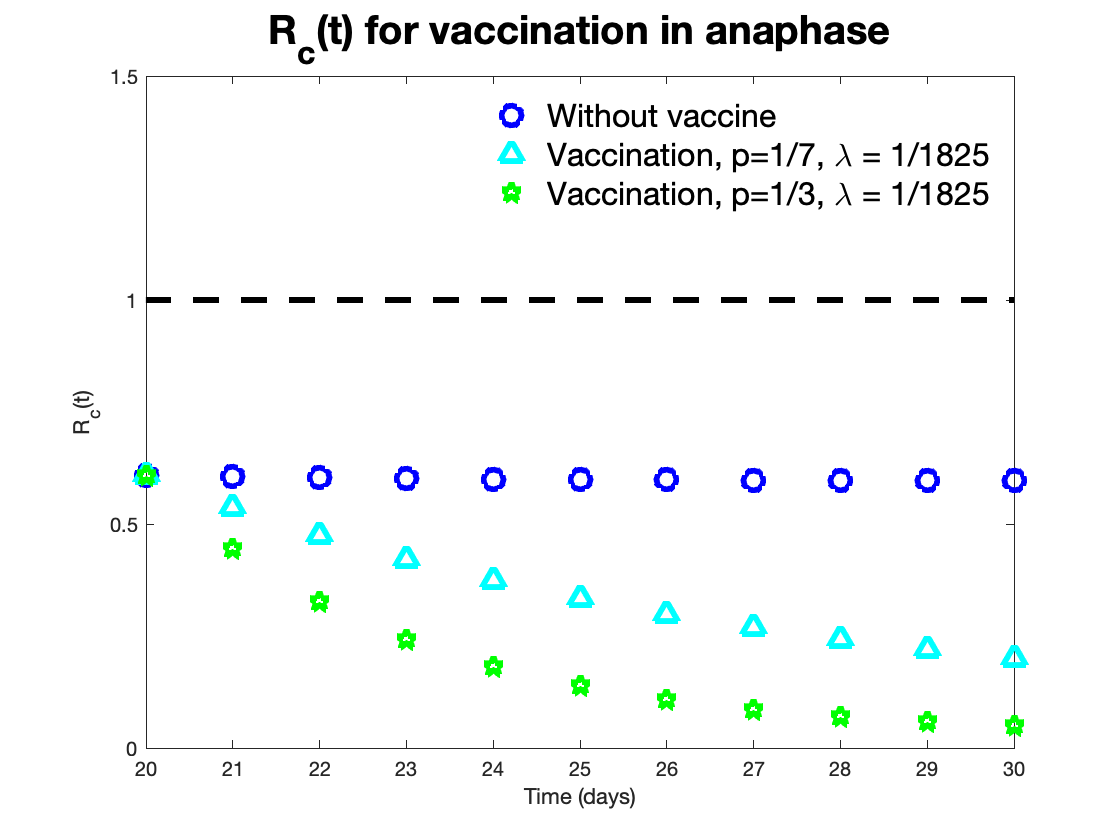}
\end{minipage}}
\caption{Impact of vaccination strategy outside Hubei Province}\label{fig6_out}
\end{figure}

\begin{figure}[htp!]
\centering
\subfigure[]{
\begin{minipage}[t]{0.45\linewidth}
\centering
\includegraphics[width=\linewidth]{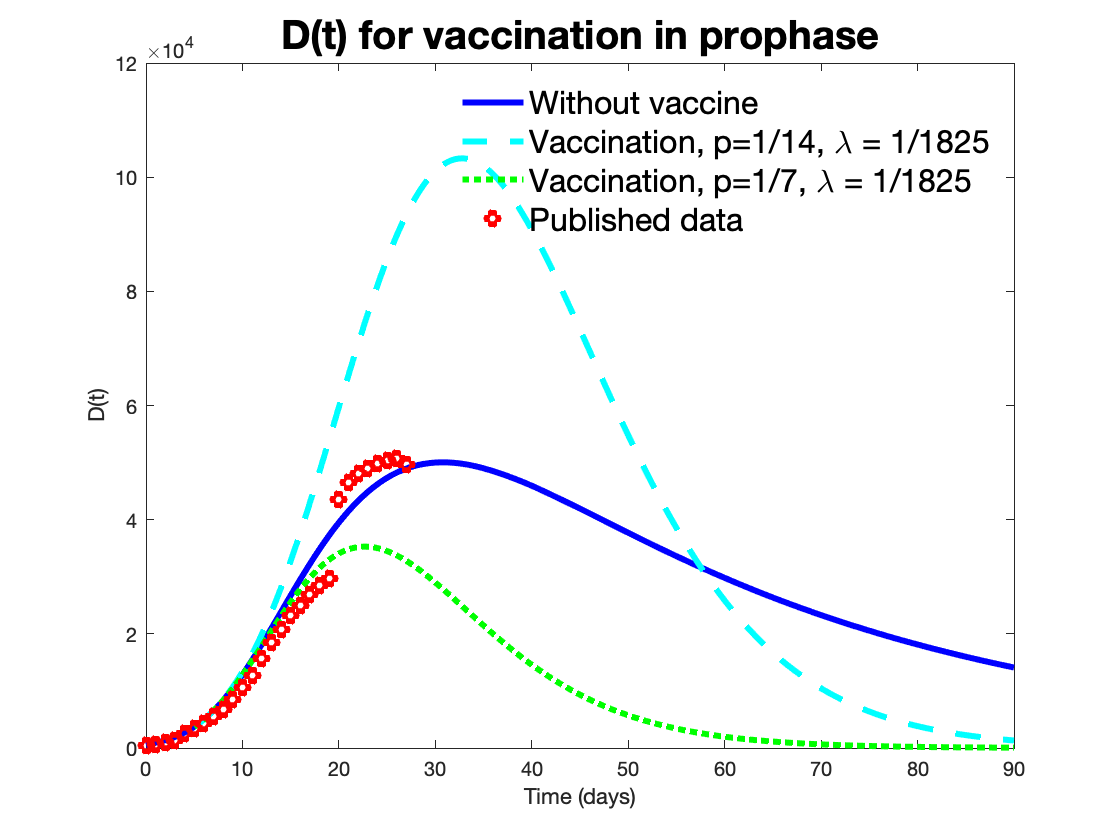}
\end{minipage}}
\subfigure[]{
\begin{minipage}[t]{0.45\linewidth}
\centering
\includegraphics[width=\linewidth]{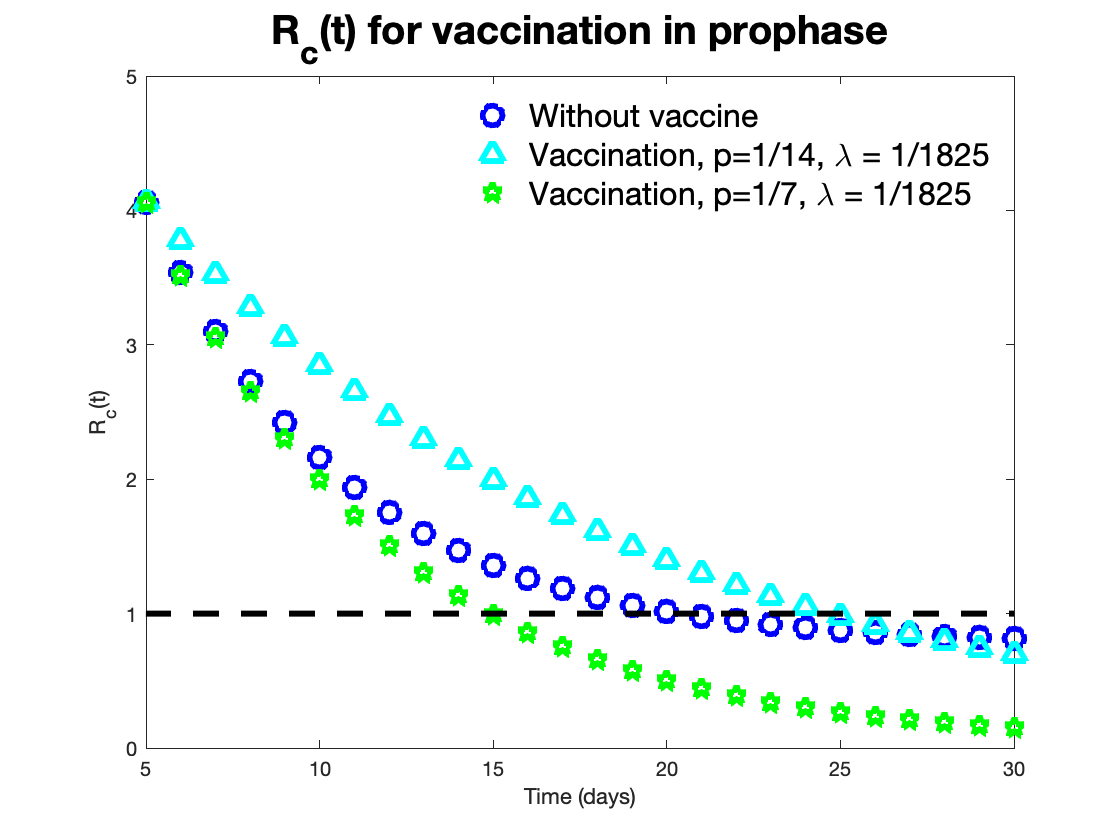}
\end{minipage}}

\subfigure[]{
\begin{minipage}[t]{0.45\linewidth}
\centering
\includegraphics[width=\linewidth]{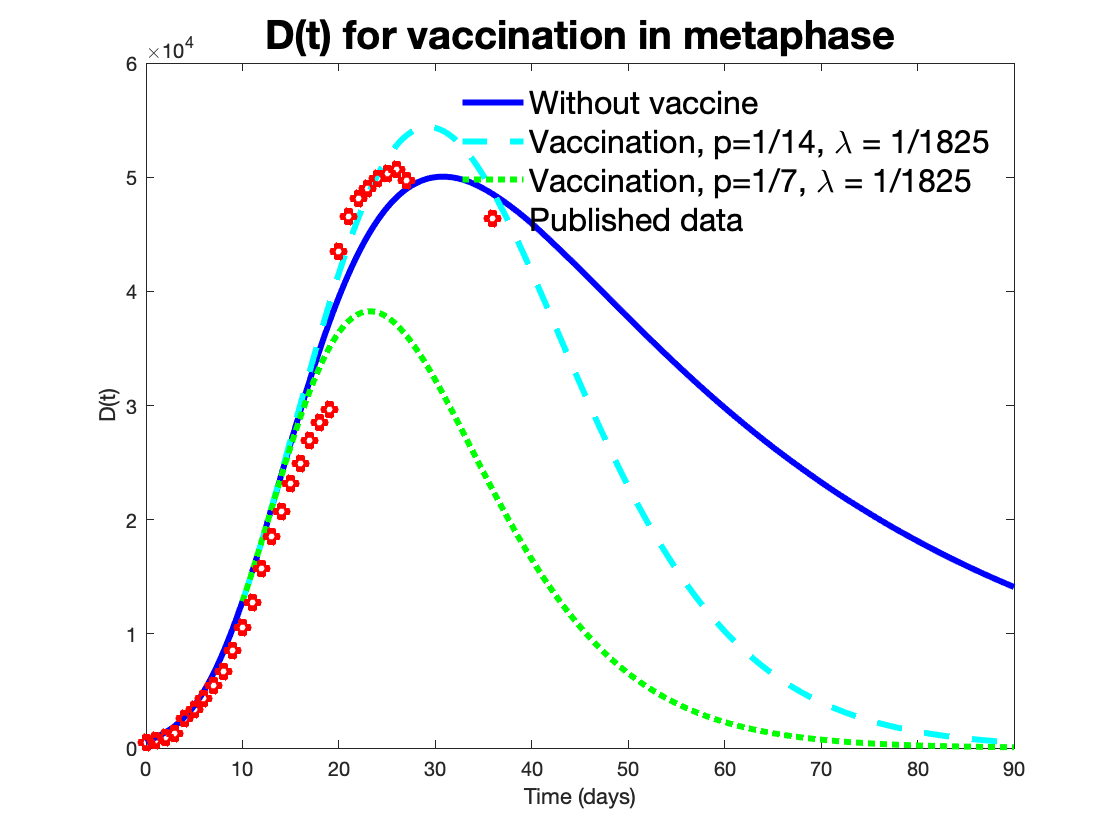}
\end{minipage}}
\subfigure[]{
\begin{minipage}[t]{0.45\linewidth}
\centering
\includegraphics[width=\linewidth]{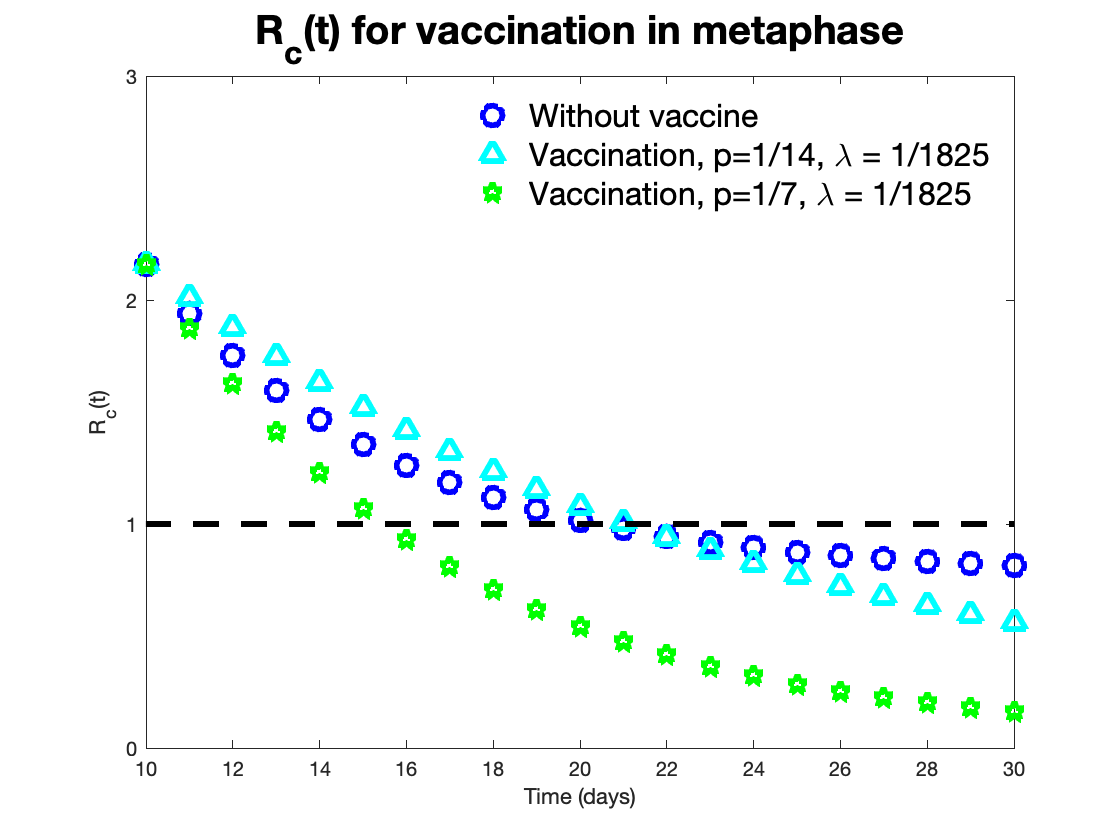}
\end{minipage}}

\subfigure[]{
\begin{minipage}[t]{0.45\linewidth}
\centering
\includegraphics[width=\linewidth]{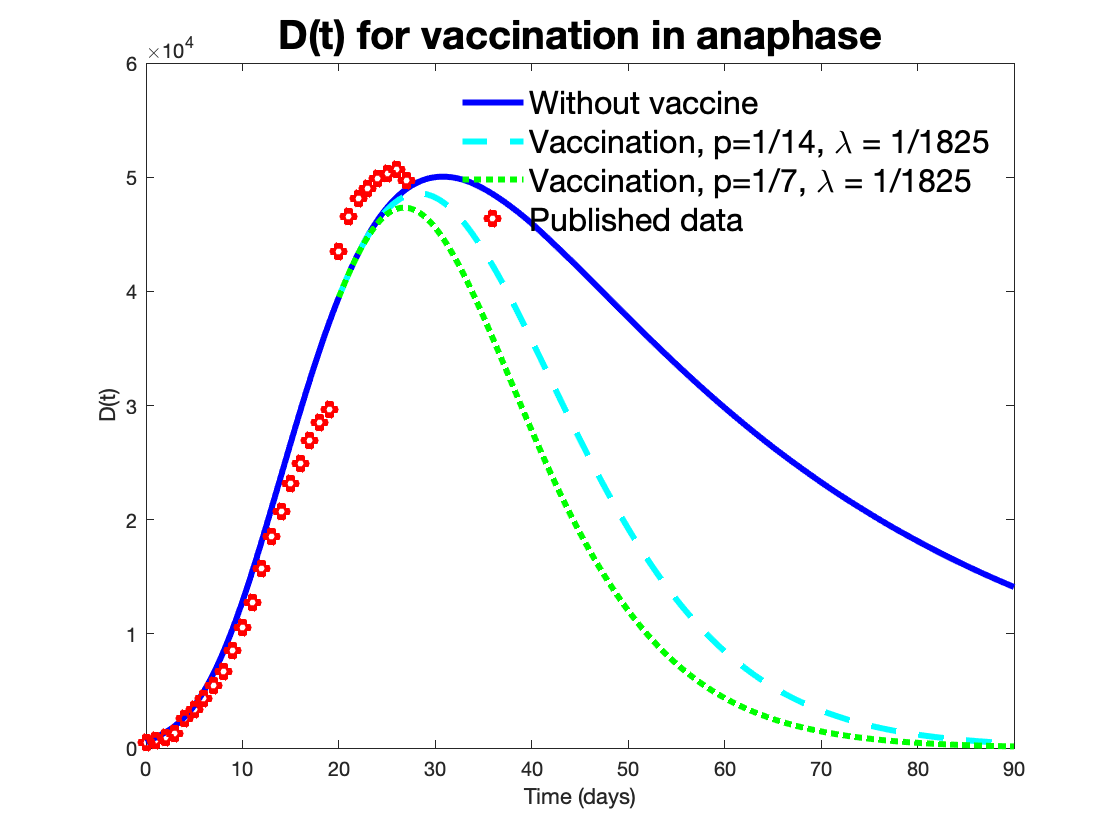}
\end{minipage}}
\subfigure[]{
\begin{minipage}[t]{0.45\linewidth}
\centering
\includegraphics[width=\linewidth]{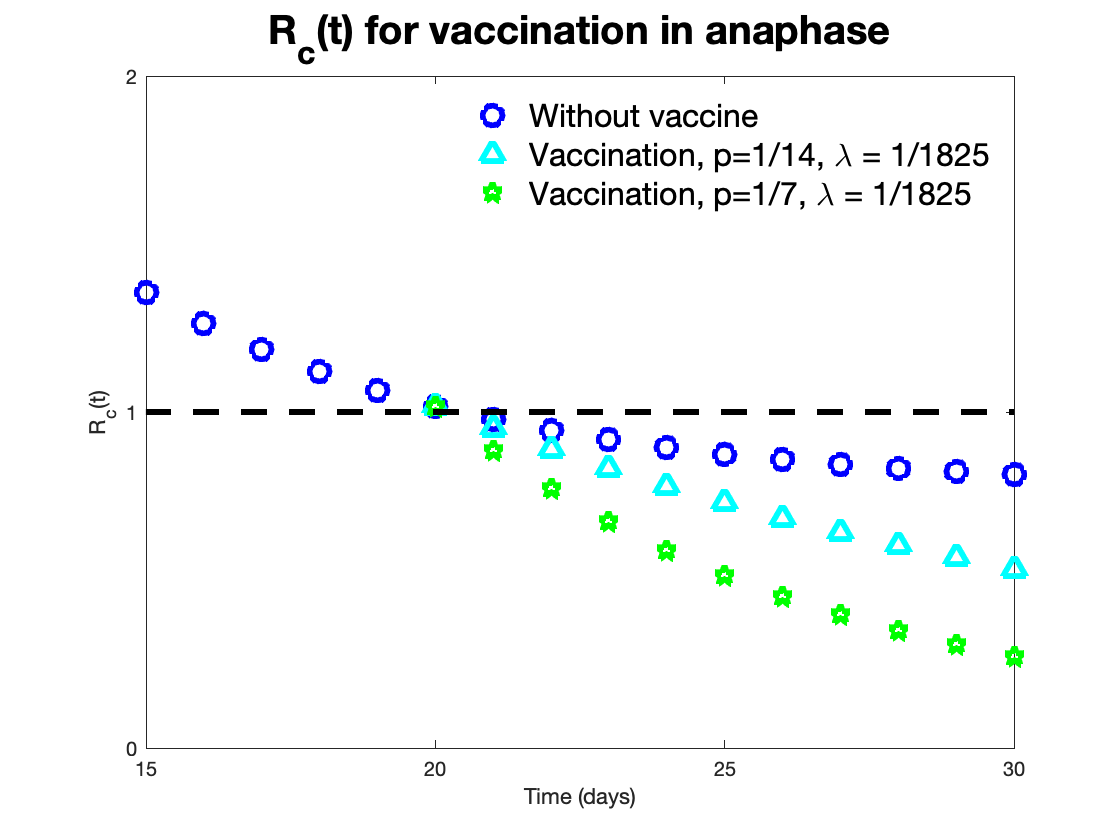}
\end{minipage}}
\caption{Impact of vaccination strategy inside Hubei Province}\label{fig6_in}
\end{figure}

Though the vaccine hasn't come into service, the theoretical analysis is necessary to make better control. In this section, we show the effect of different starting time we begin to use vaccine inside and outside Hubei Province. Considering the three control phases we defined, we take the starting day on January 28 (prophase), February 2 (metaphase) and February 12 (anaphase). We assume the average vaccination  period is five year, it means $\lambda = 1/1825$. We test different parameter $p$ to present the impact of vaccination efficiency, the corresponding numerical results are shown in Figure~\ref{fig6_out} and Figure~\ref{fig6_in}. In the area of outside Hubei Province, we take the parameter $p$ as $1/3$ and $1/7$ to compare with current control strategy ($p=1/3$). We find that the application of vaccine accelerates the epidemic process, bring the peak forward and reduce total amount of diagnostic population. The trends of $\mathcal{R}_{c}(t)   $ with vaccine are much lower than that in current strategy. In the simulation inside Hubei, we take $p$ as $1/7$ and $1/14$, which is smaller than optimal simulation ($p=1/6.2$). Totally speaking, the effect will be better as the parameter $p$ increases. The vaccination strategy does an excellent job of reducing $\mathcal{R}_{c}(t)$ and effective action of taking the vaccine immunity is very important to prevent the growth of diagnostic population. Both inside and outside Hubei, the control of disease in prophase should be paid more attention. If the parameter $p$ is smaller than that in isolation strategy, the peak value of $D(t)$ will be greater than current situation. It means that, in the prophase of control, it is suitable for taking both vaccination and isolation strategy into account. Compared with isolation strategy ,vaccination strategy is more convenience for our daily life and the effect persist long.

\section{Conclusion}

The strictest isolation strategy in China has achieved great success in current control stage. Although the control reproduction number $\mathcal{R}^{0}_{c} $ is high, the effective control reproduction number $\mathcal{R}_{c}(t)   $ decreases sharply under intervention. $\mathcal{R}_{c}(t)   $ of outside Hubei down to the critical value on February 2, 2020, and in our simulation the value will maintain about $0.6$ (see Figure~\ref{fig6_out}b). In Hubei Province, $\mathcal{R}_{c}(t)   $ reaches $1$ on February 13, 2020, we can see that it is  a continued momentum of decline (see Figure~\ref{fig6_in}b). To have a better understanding of the isolation strategy, we simulate for each selected provinces, the goodness of fit shows that the proposed model is suitable to describe the situation of control in China. We find that if the isolation period isn't long enough, the strategy won't work. In Hubei Province, the quarantine period is simulated as $90$ days, our result coincide with Zhong's study \cite{zhongnanshansiyue}. That is, the control period we suggest in Hubei is from January to the end of April. In the area outside Hubei, the trends of COVID-19 in most provinces between $30$ days and $60$ days isolation period are similar. Under careful personal protection, the people can return to their work actively and orderly in these provinces.

After the outbreak of COVID-19, China government pushed a series of policies to ensure the medical care for patients, such as early diagnosis, early isolation, free treatment and etc. We discuss the peak value and peak time of $D(t)$ in many cases. An index called $AMR$ is well defined to measure the needs of medical resources. Sufficient medical resources is the basic of the disease control. So it is necessary to assign medical resources to deal with the shocks caused by booming demand of patients. To study the detail situation of each province, $AMR$ and original medical power are both considered. We find that, the set up of designated hospital decreases the medical burden sharply. It has released the pressure of diagnosis and isolation in each provinces. But the situation of Hubei is crisis, the disease burden is still huge. Medical workers and materials are transferred to Hubei from other provinces to ensure the control of COVID-19.

There are many similarities between SARS and COVID-19, following this characteristic, we explore the relationship between the spread of COVID-19 and meteorological factors, which is considered as a key factor for the disappear of SARS. The impact of meteorological factors is different in high and low migration groups (see Table \ref{table_lr}). Results show that, air index is the most important meteorology factor. It is strongly suggests that for low $MiI$ group, if the air condition is bad with strong wind, please pay more attention to the personal protection. High relative humidity is a positive factor for the COVID-19 control.

In the last part, we do some preparatory work for the coming new vaccine. The situations after the vaccine comes into use are shown. We tested the different starting time in each control phase inside and outside Hubei. Vaccination strategy is more convenient for daily control. But the development cycle of new vaccine is relatively long, the isolation strategy is still necessary in early control. The analysis helps to design the final vaccination strategy once the new vaccine comes out.
All of our study matches the actual control strategy in China and the results are discussed adequately. It could be a guideline for the control of COVID-19 in other countries.

\Acknowledgements{This work was partially supported by National Natural Science Foundation of China (Grant No. 41704116, 11901234, 11926104), Jilin Provincial Excellent Youth Talents Foundation (Grant No. 20180520093JH), Scientific Research Project of Education Department of Jilin Province (Grant No. JJKH20200933KJ).}


\newpage

\section*{Appendix}

\begin{table}[h!]
\centering
    \resizebox{0.9\textwidth}{!}{
    \begin{threeparttable}
\caption{Estimation of peak value, peak time of $D(t)$ and medical resource needed in 90 days.}\label{table_peak_all}
	\begin{tabular}{c|ccc|ccc|ccc|ccc}
    \hline\hline
    $1/\lambda$	&	$Peak$	&	$T_{peak}$	&	$AMR$	&	$Peak$	&	$T_{peak}$	&	$AMR$		&		$Peak$	&	$T_{peak}$	&	$AMR$		&		$Peak$	&	$T_{peak}$	&	$AMR$		\\	\hline
    &\multicolumn{3}{c|}{Anhui}	&\multicolumn{3}{c|}{Beijing}	&	\multicolumn{3}{c|}{Chongqing} &	\multicolumn{3}{c}{Fujian}\\	\hline
    data	&	777	&	20	&	--	&	295	&	21	&	--	&	423	&	20	&	--	&	228	&	19	&	--	\\	\hline
    60	&	770	&	20	&	28759$k$	&	275	&	19	&	10216$k$	&	398	&	18	&	12705$k$	&	227	&	19	&	8866$k$\\	
    30	&	905	&	25	&	46939$k$	&	322	&	24	&	16313$k$	&	436	&	20	&	18344$k$	&	344	&	37	&	25911$k$	\\	
    20	&	1235	&	44	&	93412$k$	&	433	&	39	&	32269$k$	&	495	&	25	&	29820$k$	&	4989	&	*	&	135973$k$	\\	
    10	&	66800	&	*	&	1231648$k$	&	27116	&	*	&	489384$k$	&	6186	&	*	&	181709$k$	&	1736757	&	*	&	18675280$k$	\\	
    5	&	4348428	&	*	&	48700375$k$	&	2067882	&	*	&	24241821$k$	&	215309	&	*	&	2846363$k$	&	7237440	&	74	&	212118841$k$	\\	\hline
    &\multicolumn{3}{c|}{Gansu}	&\multicolumn{3}{c|}{Guangdong}	&	\multicolumn{3}{c|}{Guangxi} &	\multicolumn{3}{c}{Guizhou}\\	\hline
    data	&	66	&	17	&	--	&	1007	&	18	&	--	&	193	&	23	&	--	&	117	&	20	&	--	\\	\hline
    60	&	59	&	15	&	1639$k$	&	988	&	18	&	32278$k$	&	191	&	21	&	8093$k$	&	109	&	21	&	4203$k$	\\	
    30	&	63	&	16	&	2289$k$	&	1192	&	22	&	55932$k$	&	214	&	24	&	10768$k$	&	135	&	27	&	8058$k$	\\	
    20	&	69	&	19	&	3609$k$	&	1692	&	40	&	128580$k$	&	249	&	30	&	15598$k$	&	343	&	*	&	19080$k$	\\	
    10	&	718	&	*	&	23135$k$	&	257422	&	*	&	3953350$k$	&	2062	&	*	&	70888$k$	&	18929	&	*	&	299589$k$	\\	
    5	&	37961	&	*	&	487756$k$	&	16571603	&	*	&	285227509$k$	&	97897	&	*	&	1273544$k$	&	1058173	&	*	&	10532875$k$	\\	\hline
    &\multicolumn{3}{c|}{Hainan}	&\multicolumn{3}{c|}{Hebei}	&	\multicolumn{3}{c|}{Heilongjiang} &	\multicolumn{3}{c}{Henan}\\	\hline
    data	&	126	&	21	&	--	&	211	&	22	&	--	&	370	&	22	&	--	&	901	&	21	&	--	\\	\hline
    60	&	112	&	20	&	4078$k$	&	200	&	20	&	7495$k$	&	362	&	24	&	16787$k$	&	865	&	18	&	30486$k$	\\	
    30	&	138	&	26	&	8172$k$&	257	&	28	&	15477$k$	&	1120	&	*	&	62745$k$	&	1064	&	23	&	53455$k$	\\	
    20	&	427	&	*	&	21573$k$	&	917	&	*	&	43993$k$	&	18180	&	*	&	382084$k$	&	1678	&	60	&	127687$k$	\\	
    10	&	32644	&	*	&	471386$k$	&	102473	&	*	&	1343808$k$	&	3041233	&	*	&	35868972$k$	&	334575	&	*	&	4768513$k$	\\	
    5	&	771931	&	*	&	11927662$k$	&	7477953	&	*	&	77594648$k$	&	7732929$k$	&	74	&	232380351$k$	&	17625745	&	*	&	334589749$k$	\\	\hline
    &\multicolumn{3}{c|}{Hunan}	&\multicolumn{3}{c|}{Inner Mongolia}	&	\multicolumn{3}{c|}{Jiangsu} &	\multicolumn{3}{c}{Jiangxi}\\	\hline
    data	&	698	&	19	&	--	&	66	&	27	&	--	&	456	&	22	&	--	&	712	&	22	&	--	\\	\hline
    60	&	702	&	16	&	21867$k$	&	64	&	28	&	3167$k$	&	445	&	22	&	20035$k$	&	718	&	19	&	24493$k$	\\	
    30	&	797	&	19	&	33027$k$	&	160	&	*	&	9703$k$&	465	&	23	&	22492$k$	&	866	&	24	&	42908$k$	\\	
    20	&	957	&	25	&	58685$k$	&	1882	&	*	&	43991$k$	&	487	&	25	&	25453$k$	&	1251	&	51	&	95956$k$	\\	
    10	&	28641	&	*	&	670425$k$	&	300866	&	*	&	2807802$k$	&	579	&	32	&	37983$k$	&	112385	&	*	&	1870821$k$	\\	
    5	&	3334809	&	*	&	39348405$k$	&	6566392	&	*	&	98519328$k$	&	1479	&	*	&	82541$k$	&	4696779	&	*	&	69937216$k$	\\	\hline
    &\multicolumn{3}{c|}{Jilin}	&\multicolumn{3}{c|}{Liaoning}	&	\multicolumn{3}{c|}{Ningxia} &	\multicolumn{3}{c}{Shaanxi}\\	\hline
    data	&	73	&	17	&	--	&	97	&	17	&	--	&	43	&	22	&	--	&	189	&	20	&	--	\\	\hline
    60	&	62	&	19	&	2086$k$	&	95	&	18	&	3336$k$	&	42	&	18	&	1384$k$	&	187	&	18	&	6093$k$	\\	
    30	&	69	&	21	&	2909$k$	&	126	&	26	&	8210$k$	&	46	&	21	&	1934$k$	&	203	&	19	&	8089$k$	\\	
    20	&	79	&	26	&	4488$k$	&	892	&	*	&	32358$k$	&	51	&	25	&	2960$k$	&	223	&	22	&	11401$k$	\\	
    10	&	780	&	*	&	25419$k$	&	255742	&	*	&	2739032$k$	&	385	&	*	&	13577$k$	&	816	&	*	&	40286$k$	\\	
    5	&	46330	&	*	&	575506$k$	&	5402772	&	84	&	119906271$k$	&	8998	&	*	&	136715$k$	&	22248	&	*	&	391375$k$	\\	\hline
    &\multicolumn{3}{c|}{Shandong}	&\multicolumn{3}{c|}{Shanghai}	&	\multicolumn{3}{c|}{Shanxi} &	\multicolumn{3}{c}{Sichuan}\\	\hline
    data	&	416	&	20	&	--	&	255	&	21	&	--	&	189	&	20	&	--	&	357	&	21	&	--	\\	\hline
    60	&	415	&	20	&	16992$k$	&	232	&	16	&	7036$k$	&	187	&	18	&	6093$k$	&	346	&	22	&	16921$k$	\\	
    30	&	507	&	26	&	26645$k$	&	268	&	19	&	11108$k$	&	203	&	19	&	8089$k$	&	403	&	26	&	22616$k$	\\	
    20	&	720	&	41	&	51630$k$	&	337	&	27	&	22023$k$	&	223	&	22	&	11401$k$	&	504	&	35	&	33486$k$	\\	
    10	&	54739	&	*	&	955181$k$	&	25816	&	*	&	473809$k$	&	816	&	*	&	40286$k$	&	7264	&	*	&	207899$k$	\\	
    5	&	11784044	&	*	&	116251444$k$	&	2900252	&	*	&	42717157$k$	&	22248	&	*	&	391375$k$	&	1016284	&	*	&	10516655$k$	\\	\hline
    &\multicolumn{3}{c|}{Tianjin}	&\multicolumn{3}{c|}{Xinjiang}	&	\multicolumn{3}{c|}{Yunnan} &	\multicolumn{3}{c}{Zhejiang}\\	\hline
    data	&	99	&	21	&	--	&	63	&	23	&	--	&	135	&	22	&	--	&	921	&	16	&	--	\\	\hline
    60	&	82	&	22	&	3660$k$	&	58	&	25	&	2868$k$	&	131	&	18	&	5330$k$	&	905	&	17	&	28601$k$	\\	
    30	&	90	&	25	&	4662$k$	&	77	&	35	&	5010$k$	&	139	&	20	&	6007$k$	&	1089	&	21	&	51192$k$	\\	
    20	&	103	&	30	&	6341$k$	&	196	&	*	&	10834$k$	&	147	&	21	&	6861$k$	&	1642	&	*	&	127746$k$	\\	
    10	&	526	&	*	&	21782$k$	&	10969	&	*	&	168818$k$	&	185	&	29	&	11161$k$	&	374851	&	*	&	5368778$k$	\\	
    5	&	15458	&	*	&	236496$k$	&	1041128	&	*	&	8958716$k$	&	862	&	*	&	39412$k$	&	8395172	&	83	&	210982965$k$	\\	
    \hline\hline
\end{tabular}
\begin{tablenotes}
\item[1]* do not reach peak in 90 days.
\item[2]- not applicable.
\end{tablenotes}
\end{threeparttable}}
\end{table}

\begin{table}[htbp]
\centering
\caption{Simulation parameter Part I}\label{table_para_I}
    \resizebox{0.9\textwidth}{!}{
	\begin{tabular}{c|c|c|c|c|c|c|c}
\hline\hline																
	Parameter	&	Anhui	&	Beijing	&	Chongqing	&	Fujian	&	Gansu	&	Guangdong	&	Guangxi	\\
\hline	$\beta$	&	2.7096E-08	&	9.1997E-08	&	3.6320E-08	&	9.6503E-08	&	6.2034E-08	&	2.7096E-08	&	2.7083E-08	\\
	$\theta$	&	0.1100	&	0.0110	&	0.0800	&	0.0900	&	0.0100	&	0.0050	&	0.0050	\\
	$p$	&	1/4.1	&	 1/4	&	1/6.4	&	 1/3	&	1/6.5	&	 1/3	&	 1/4	\\
	$\lambda$	&	  1/60	&	  1/60	&	  1/60	&	  1/60	&	  1/60	&	  1/60	&	  1/60	\\
	$\sigma$	&	 1/7	&	 1/7	&	 1/7	&	 1/7	&	 1/7	&	 1/7	&	 1/7	\\
	$\rho$	&	0.9000	&	0.9400	&	0.9900	&	0.9300	&	0.9000	&	0.9100	&	0.8800	\\
	$\epsilon_{A}$	&	  1/7 	&	  1/8 	&	  1/9 	&	  1/5 	&	  1/7 	&	  1/10	&	  1/10	\\
	$\epsilon_{I}$	&	 1/5	&	 1/4	&	1/4.5	&	 1/4	&	 1/4	&	1/3.3	&	1/4.5	\\
	$\gamma_{A}$	&	0.0993	&	0.0996	&	0.1190	&	0.0988	&	0.2597	&	0.1075	&	0.0579	\\
	$\gamma_{I}$	&	0.0736	&	0.0766	&	0.0992	&	0.0882	&	0.1998	&	0.0716	&	0.0386	\\
	$\gamma_{D}$	&	0.0883	&	0.0843	&	0.1487	&	0.0970	&	0.2398	&	0.1039	&	0.0560	\\
	$d_{I}$	&	0.0027	&	0.0026	&	0.0040	&	0.0022	&	0.0073	&	0.0024	&	0.0013	\\
	$d_{D}$	&	0.0018	&	0.0017	&	0.0033	&	0.0015	&	0.0049	&	0.0016	&	0.0009	\\
%	population	&	63236000	&	21540000	&	31020000	&	39410000	&	26370000	&	113460000	&	49260000	\\
\hline	$R_{c}$	&	5.0939	&	5.1544	&	3.0885	&	6.4914	&	2.9989	&	6.3126	&	3.7215	\\
\hline	S(0)	&	56912400	&	19601400	&	27918000	&	24434200	&	24524100	&	96441000	&	40885800	\\
	Q(0)	&	6323600	&	1938600	&	3102000	&	14975800	&	1845900	&	17019000	&	8374200	\\
	E(0)	&	591	&	123	&	580	&	80	&	175	&	886	&	95	\\
	A(0)	&	288	&	81	&	90	&	65	&	50	&	37	&	52	\\
	I(0)	&	101	&	58	&	88	&	50	&	5	&	93	&	49	\\
	D(0)	&	15	&	26	&	27	&	10	&	2	&	51	&	13	\\
	R(0)	&	6	&	0	&	0	&	0	&	0	&	2	&	0	\\
\hline\hline																
																
		&	Guizhou	&	Hainan	&	Hebei	&	Heilongjiang	&	Henan	&	Hunan	&	Inner Mongolia	\\
\hline	$\beta$	&	4.9543E-08	&	1.8064E-07	&	2.7096E-08	&	7.9315E-08	&	3.5675E-08	&	2.9823E-08	&	8.8785E-08	\\
	$\theta$	&	0.0500	&	0.1000	&	0.0300	&	0.2000	&	0.2000	&	0.1000	&	0.0130	\\
	$p$	&	1/4.5	&	 1/5	&	 1/5	&	 1/3.5	&	 1/3	&	 1/3	&	 1/5	\\
	$\lambda$	&	  1/60	&	  1/60	&	  1/60	&	  1/60	&	  1/60	&	  1/60	&	   1/60 	\\
	$\sigma$	&	 1/7	&	 1/7	&	 1/7	&	 1/7	&	 1/7	&	 1/7	&	 1/7	\\
	$\rho$	&	0.7000	&	0.9000	&	0.9000	&	0.9500	&	0.8900	&	0.9100	&	0.9700	\\
	$\epsilon_{A}$	&	  1/15	&	  1/10	&	 1/7	&	  1/7 	&	 1/4	&	  1/10	&	  1/5 	\\
	$\epsilon_{I}$	&	 1/9	&	 1/6	&	 1/4	&	 1/5	&	 1/3	&	1/5.1	&	 1/4	\\
	$\gamma_{A}$	&	0.1298	&	0.1495	&	0.1085	&	0.0967	&	0.0993	&	0.1085	&	0.0709	\\
	$\gamma_{I}$	&	0.0998	&	0.0997	&	0.0724	&	0.0744	&	0.0764	&	0.0986	&	0.0473	\\
	$\gamma_{D}$	&	0.1098	&	0.1396	&	0.1049	&	0.0818	&	0.0840	&	0.1085	&	0.0662	\\
	$d_{I}$	&	0.0008	&	0.0045	&	0.0024	&	0.0026	&	0.0026	&	0.0035	&	0.0020	\\
	$d_{D}$	&	0.0006	&	0.0030	&	0.0021	&	0.0018	&	0.0017	&	0.0023	&	0.0014	\\
%	population	&	36000000	&	9340000	&	75560000	&	37730000	&	96050000	&	68990000	&	25340000	\\
\hline	$R_{c}$	&	3.9208	&	5.1065	&	3.8753	&	6.2331	&	6.6235	&	3.5025	&	6.5656	\\
\hline	S(0)	&	23400000	&	8406000	&	51380800	&	22638000	&	83563500	&	37944500	&	22806000	\\
	Q(0)	&	12600000	&	934000	&	24179200	&	15092000	&	12486500	&	31045500	&	2534000	\\
	E(0)	&	280	&	110	&	168	&	308	&	556	&	1500	&	16	\\
	A(0)	&	28	&	45	&	40	&	25	&	211	&	145	&	9	\\
	I(0)	&	8	&	25	&	33	&	15	&	22	&	65	&	4	\\
	D(0)	&	3	&	8	&	1	&	3	&	9	&	24	&	2	\\
	R(0)	&	0	&	0	&	0	&	0	&	0	&	0	&	0	\\
\hline\hline																
\end{tabular}}
\end{table}
%%%%%%%%%%%%%%%%%%%%%%%%%%%%%%%%%%%%%%%%%%%%%%%%%%%%%%%%%%%%%%%%%%%%%%%%%%%%%%%%%%%%%%%%%%%%%%%%%%%%
\begin{table}[htbp]
\centering
\caption{Simulation parameter Part II}\label{table_para_II}
    \resizebox{0.9\textwidth}{!}{
	\begin{tabular}{c|c|c|c|c|c|c|c}
\hline\hline																
	Parameter	&	Jiangsu	&	Jiangxi	&	Jilin	&	Liaoning	&	Ningxia	&	Shaanxi	&	Shandong	\\
\hline	$\beta$	&	5.0000E-09	&	4.8409E-08	&	4.7838E-08	&	6.9683E-08	&	1.5932E-07	&	2.8218E-08	&	2.7094E-08	\\
	$\theta$	&	0.2000	&	0.2000	&	0.0050	&	0.0100	&	0.1000	&	0.0900	&	0.2000	\\
	$p$	&	 1/5	&	1/4.5	&	 1/4	&	 1/3	&	1/7.9	&	1/4.5	&	 1/3	\\
	$\lambda$	&	  1/60	&	  1/60	&	   1/60 	&	  1/60	&	   1/60 	&	  1/60	&	   1/60 	\\
	$\sigma$	&	 1/7	&	 1/7	&	 1/7	&	 1/7	&	 1/7	&	 1/7	&	 1/7	\\
	$\rho$	&	0.9800	&	0.8000	&	0.9600	&	0.9500	&	0.9000	&	0.7000	&	0.9800	\\
	$\epsilon_{A}$	&	  1/10	&	  1/10	&	  1/6 	&	  1/10	&	  1/5 	&	  1/10	&	  1/5 	\\
	$\epsilon_{I}$	&	 1/7	&	 1/3.9	&	 1/5	&	 1/5	&	 1/4	&	 1/9	&	 1/3	\\
	$\gamma_{A}$	&	0.0474	&	0.1296	&	0.1029	&	0.1196	&	0.1462	&	0.1298	&	0.0618	\\
	$\gamma_{I}$	&	0.0379	&	0.0997	&	0.0686	&	0.0997	&	0.0975	&	0.0999	&	0.0515	\\
	$\gamma_{D}$	&	0.0455	&	0.1087	&	0.0960	&	0.1296	&	0.1365	&	0.1098	&	0.0567	\\
	$d_{I}$	&	0.0010	&	0.0037	&	0.0029	&	0.0007	&	0.0042	&	0.0009	&	0.0017	\\
	$d_{D}$	&	0.0007	&	0.0023	&	0.0020	&	0.0007	&	0.0028	&	0.0006	&	0.0012	\\
%	population	&	80507000	&	46480000	&	27040600	&	43590000	&	6880000	&	38640000	&	100470000	\\
\hline	$R_{c}$	&	1.8970	&	4.3156	&	4.2087	&	6.7287	&	2.4398	&	2.2381	&	6.2478	\\
\hline	S(0)	&	70041090	&	37184000	&	24877352	&	30513000	&	5916800	&	23184000	&	90423000	\\
	Q(0)	&	10465910	&	9296000	&	2163248	&	13077000	&	963200	&	15456000	&	10047000	\\
	E(0)	&	452	&	840	&	87	&	44	&	70	&	640	&	165	\\
	A(0)	&	145	&	50	&	6	&	25	&	5	&	36	&	60	\\
	I(0)	&	125	&	67	&	5	&	38	&	4	&	20	&	36	\\
	D(0)	&	9	&	7	&	3	&	4	&	3	&	5	&	15	\\
	R(0)	&	0	&	0	&	0	&	0	&	0	&	0	&	0	\\
\hline\hline																
		&	Shanghai	&	Shanxi	&	Sichuan	&	Tianjin	&	Xinjiang	&	Yunnan	&	Zhejiang	\\
\hline	$\beta$	&	1.2638E-07	&	8.1979E-08	&	2.7018E-08	&	1.4151E-07	&	7.0499E-08	&	2.2519E-08	&	5.7383E-08	\\
	$\theta$	&	0.0100	&	0.1000	&	0.1100	&	0.0050	&	0.0100	&	0.0900	&	0.1000	\\
	$p$	&	  1/3 	&	 1/3	&	 1/3	&	 1/5	&	 1/4	&	1/3.2	&	 1/3	\\
	$\lambda$	&	  1/60	&	  1/60	&	  1/60	&	   1/60 	&	  1/60	&	  1/60	&	  1/60	\\
	$\sigma$	&	 1/7	&	 1/7	&	 1/7	&	 1/7	&	 1/7	&	 1/7	&	 1/7	\\
	$\rho$	&	0.9200	&	0.9000	&	0.9200	&	0.9000	&	0.8900	&	0.9800	&	0.9600	\\
	$\epsilon_{A}$	&	  1/4 	&	 1/3	&	  1/4 	&	  1/10	&	 1/9	&	  1/4 	&	  1/6 	\\
	$\epsilon_{I}$	&	 1/3	&	1/2.5	&	 1/3	&	 1/4	&	 1/5	&	 1/3	&	 1/3	\\
	$\gamma_{A}$	&	0.1213	&	0.0809	&	0.0387	&	0.1254	&	0.0540	&	0.0491	&	0.1190	\\
	$\gamma_{I}$	&	0.0808	&	0.0622	&	0.0322	&	0.0836	&	0.0415	&	0.0327	&	0.0793	\\
	$\gamma_{D}$	&	0.1132	&	0.0685	&	0.0355	&	0.1170	&	0.0457	&	0.0458	&	0.1150	\\
	$d_{I}$	&	0.0035	&	0.0001	&	0.0011	&	0.0036	&	0.0014	&	0.0014	&	0.0017	\\
	$d_{D}$	&	0.0023	&	0.0001	&	0.0007	&	0.0024	&	0.0009	&	0.0009	&	0.0012	\\
%	population	&	24240000	&	37180000	&	83410000	&	15596000	&	24867600	&	48295000	&	57370000	\\
\hline	$R_{c}$	&	5.6066	&	5.4060	&	5.1509	&	5.2949	&	4.2468	&	2.6164	&	6.9822	\\
\hline	S(0)	&	20119200	&	33462000	&	75069000	&	14036400	&	16412616	&	43465500	&	52206700	\\
	Q(0)	&	4120800	&	3718000	&	8341000	&	1559600	&	8454984	&	4829500	&	5163300	\\
	E(0)	&	266	&	48	&	140	&	40	&	40	&	134	&	456	\\
	A(0)	&	22	&	21	&	64	&	26	&	11	&	12	&	145	\\
	I(0)	&	11	&	8	&	28	&	15	&	5	&	10	&	133	\\
	D(0)	&	19	&	1	&	15	&	8	&	2	&	5	&	42	\\
	R(0)	&	3	&	0	&	0	&	0	&	0	&	0	&	1	\\
\hline\hline																
\end{tabular}}
\end{table}

%----------------------------------

\end{document}